\documentclass[sigconf]{acmart}

\setcopyright{none} 
\renewcommand\footnotetextcopyrightpermission[1]{} 
\usepackage{graphicx} 
\usepackage{amsmath}   
\usepackage{amsbsy}    
\usepackage{subfigure}
\usepackage{subcaption}
\usepackage{graphicx}
\usepackage{tabularx} 
\usepackage{booktabs}
\usepackage{multirow}
\usepackage{multicol}
\usepackage{breqn}
\usepackage{extarrows}
\usepackage{enumitem}
\usepackage{bm}
\usepackage{amsmath}
\usepackage{xcolor}
\usepackage{pifont}
\usepackage{changepage}

\theoremstyle{plain}
\newtheorem{theorem}{Theorem}[section]

\newtheorem{lemma}[theorem]{Lemma}

\theoremstyle{definition}

\theoremstyle{remark}

\newcommand{\lc}{\left(} 
\newcommand{\rc}{\right)} 
\newcommand{\ox}{\overline{X}} 
\newcommand{\oy}{\overline{Y}} 
\newcommand{\wh}{\widehat} 
\newcommand{\bl}{\big(} 
\newcommand{\br}{\big)} 
\newcommand{\wt}{\widehat{\theta}} 
\newcommand{\wv}{\widehat{\text{var}}} 
\newcommand{\wc}{\widehat{\text{cov}}} 
\newcommand{\wdd}{\widehat{\delta}} 
\newcommand{\wb}{\widehat{\beta}} 
\newcommand{\wvv}{\widehat{\varepsilon}} 
\newcommand{\p}{\overset{p}{\rightarrow}} 
\AtBeginDocument{%
  }

\acmYear{Preprint. Under review}
\acmDOI{}
\acmConference[Arxiv]{Make sure to enter the correct
conference title from your rights confirmation email}{2025}{Preprint}
\acmISBN{}

\begin{document}

\title{Bridging Control Variates and Regression Adjustment in A/B Testing: From Design-Based to Model-Based Frameworks}

\author{Yu Zhang}
\affiliation{%
  \institution{Shandong University, ByteDance}
  \city{Jinan}
  \country{China}
}
\email{yuzhang.johnny@mail.sdu.edu.cn}

\author{Bokui Wan*, Yongli Qin*}
\thanks{* Corresponding authors.}
\affiliation{%
  \institution{ByteDance}
  \city{Beijing}
  \country{China}}
 \email{{wanbokui, yongli.qin}@bytedance.com}

\renewcommand{\shortauthors}{Zhang et al.}

\begin{abstract}
A/B testing serves as the gold standard for large-scale, data-driven decision-making in online businesses. To mitigate metric variability and enhance testing sensitivity, control variates and regression adjustment have emerged as prominent variance-reduction techniques, leveraging pre-experiment data to improve estimator performance. Over the past decade, these methods have spawned numerous derivatives, yet their theoretical connections and comparative properties remain underexplored. In this paper, we conduct a comprehensive analysis of their statistical properties, establish a formal bridge between the two frameworks in practical implementations, and extend the investigation from design-based to model-based frameworks. Through simulation studies and real-world experiments at ByteDance, we validate our theoretical insights across both frameworks. Our work aims to provide rigorous guidance for practitioners in online controlled experiments, addressing critical considerations of internal and external validity. The recommended method—control variates with group-specific coefficient estimates—has been fully implemented and deployed on ByteDance's experimental platform. 
\end{abstract}




\maketitle

\section{Introduction}

\textbf{Background}. A/B testing, also referred to as online controlled experiments, stands as a ubiquitous tool for major technology companies in the pursuit of maximizing revenue and optimizing user experience \citep{deng2013improving, xie2016improving}. Industry giants conduct hundreds of experiments on millions of users daily, testing modifications across multiple dimensions \citep{larsen2024statistical, xie2016improving}. In practical experiments, to compare different strategies, users of the platform are randomly divided into several groups: one serves as the control, and the others as treatment, with each group exposed to one of the test strategies. The platform incrementally collects the interactions of the users within the site or app for statistical inference \citep{kohavi2009online}. With A/B testing, the causal impact of such changes can be correctly estimated \citep{pearl2009causal, box2005statistics}.

\noindent \textbf{Analyzing experiments}. In the process of A/B testing, the performance of multiple strategies is compared across several key metrics to identify the preferred strategy \citep{siroker2015b}. We first focus on the two-sample $t$-test, due to its wide applicability and ease of implementations \citep{kohavi2020trustworthy}. Suppose we are interested in a specific metric $Y$ (e.g., Gross Merchandise Volume (GMV)). Assume that the observed outcomes in the treatment and control groups are realizations of the random variables \(Y_t\) and \(Y_c\), respectively. The hypothesis testing for the average treatment effect (ATE) is based on the following $t$-statistic: 
\begin{equation}
({\overline{Y}_t - \overline{Y}_c})/{{\widehat{\text{sd}}(\overline{Y}_t - \overline{Y}_c)}}, \notag
\end{equation}
which converges to a normal distribution with a sufficiently large sample size.

\noindent\textbf{Variance reduction}. In online A/B testing, elevated variance reduces the sensitivity of two-sample $t$-test \citep{kohavi2013online}. One technique to mitigate this issue is incorporating relevant pre-experiment covariates as control variates, which reduces the variability of outcome metrics. This technique, also known as CUPED (Controlled Experiments Utilizing Pre-Experiment Data), was first proposed by Deng et al. \cite{deng2013improving}. CUPED leverages the correlation between pre-experiment data and post-experiment metrics to construct unbiased estimators with lower variance. 

Since its introduction, CUPED has been widely implemented across various experimental platforms, including those described by Microsoft \cite{micro}, Netflix \cite{xie2016improving}, and Statsig \cite{Statsig}. Over the past decade, this framework has been widely extended and adapted by major online experimental platforms. Existing literature on variance reduction in A/B testing can be broadly categorized into two strands: 

\begin{itemize}
\item \textbf{Control variates}. This method utilizes a random variable with a known expectation (typically pre-experiment data) to construct an adjusted estimator that preserves the original expectation while reducing variance \citep{deng2013improving, larsen2024statistical, xie2016improving, kohavi2020trustworthy, jin2023toward}. 

\item \textbf{Regression adjustment}. For metrics defined as simple averages, linear regression is often employed to estimate the average treatment effect (ATE) and conduct statistical inference \citep{lin2013agnostic, freedman2008regression, li2020rerandomization, poyarkov2016boosted, hesterberg2024power}. 

\end{itemize}

\noindent\textbf{Design-based and model-based frameworks}. A critical distinction in experimental inference lies in the framework within which analyses are conducted. The design-based perspective treats observed subjects as fixed, with randomness arising solely from treatment assignment, focusing on internal validity for the studied sample \cite{freedman2008regression, lin2013agnostic}. In contrast, the model-based framework considers subjects as a random sample from a larger population, introducing additional randomness from sampling and emphasizing external validity—generalizability to the broader population \cite{deng2013improving}. This distinction has profound implications for variance estimation, estimator performance, and the interpretation of results, yet it has received insufficient attention in the  literature on variance reduction methods. 

\noindent\textbf{Contribution}. Our contributions are fivefold. First, we compile control variates and regression adjustment methods currently deployed in major technology companies or widely studied in the literature, providing a comprehensive review. Second, we elaborate on the estimation approaches for unknown population quantities and analyze the properties of such approximations in real-world settings. Third, we establish a formal bridge between control variates and regression adjustment, revealing their inherent connections in practical implementations. Fourth, we extend the analysis from design-based to model-based frameworks, highlighting key differences in estimator behavior and variance properties. Finally, we validate our theoretical insights through extensive simulations and real-world experiments, offering actionable recommendations for practitioners.

\section{Control Variates}
\label{sec2}

\textbf{Notations}. Unless specified otherwise, all statistical inference in Sections \ref{sec2}–\ref{sec4} is conducted under the design-based framework. The studies on the asymptotic properties of estimator bias and variance estimation are all conducted under the premise of treating the counterfactual outcomes as fixed numbers. Index subjects by $i=1, \ldots, n$. Let \( Y_i^t \) and \( Y_i^c \) denote the potential outcomes of subject $i$ under treatment and control, respectively, with the ATE defined as: 
$$\delta=\frac{1}{n}\sum_{i=1}^nY_i^t-\frac{1}{n}\sum_{i=1}^nY_i^c. $$
Let $T_i$ be the assignment variable: $T_i = 1$ if subject $i$ is assigned to treatment, and $T_i = 0$ if assigned to control. The observed outcome is 
$$Y_i=T_iY_i^t + (1-T_i)Y_i^c, \quad i=1, \ldots, n. $$
We also observe a pre-experiment covariate $X_i$ for each subject. In the design we consider, let $n_t$ denote the number of subjects assigned to treatment, $n_c=n-n_t$ the number assigned to control. Therefore, let $p_t=n_t/n$ and $p_c=n_c/n$ denote the probabilities that subjects are assigned to the treatment and control, respectively. Sample level statistics are defined as: 
$$\overline{Y}=\frac{1}{n}\sum_{i=1}^nY_i,\ \overline{Y}_t=\frac{1}{n_t}\sum_{i=1}^nT_iY_i,\ \overline{Y}_c=\frac{1}{n_c}\sum_{i=1}^n(1-T_i)Y_i, $$
$$\overline{X}=\frac{1}{n}\sum_{i=1}^nX_i,\ \overline{X}_t=\frac{1}{n_t}\sum_{i=1}^nT_iX_i,\ \overline{X}_c=\frac{1}{n_c}\sum_{i=1}^n(1-T_i)X_i. $$


\noindent\textbf{Baseline estimator}. 
We denote the two-sample $t$-test without any control variables as $\widehat\delta_0$ and set it as the baseline estimator: 

$$\widehat\delta_0=\frac{1}{n_t}\sum_{i=1}^nT_iY_i - \frac{1}{n_c}\sum_{i=1}^n(1-T_i)Y_i. $$

\subsection{Overview}

The control variates method is primarily designed to construct a random variable with reduced variance by incorporating a control variable, while preserving the unbiasedness of the response variable expectation \citep{deng2013improving}. Specifically, in addition to the $n$ response variable $Y$, suppose we observe another variable $X$ with a known expectation $\ox$ (Within the design-based framework, once the study subjects are fixed, their respective attributes are treated as constants rather than random variables). Define 
\begin{equation}
\widetilde{Y} = {Y} - \theta {X} + \theta \ox, \notag
\end{equation}
where $\theta$ is an arbitrary constant. This transformation ensures that $\widetilde{Y}$ remains an unbiased estimator of $\oy$ and has the following variance: 
\begin{equation}
\label{var1}
\mathrm{var}(\widetilde{Y}) =  \mathrm{var}(Y) + \theta^2 \mathrm{var}(X) - 2\theta \mathrm{cov}(Y, X), 
\end{equation}
where $\text{cov}(Y, X)$ and $\text{var}(\cdot)$ denote the true population covariance and variance, respectively. The explicit expressions are given by: 
$$
\text{var}(X)=\frac{1}{n-1} \lc \sum_{i=1}^nX_i^2-n\ox^2 \rc, \text{var}(Y)=\frac{1}{n-1} \lc \sum_{i=1}^nY_i^2-n\oy^2 \rc, 
$$
$$
\text{cov}(Y, X)=\frac{1}{n-1}\lc \sum_{i=1}^nY_iX_i-n\oy \ox\rc. 
$$ 
By considering Equation \eqref{var1} as a quadratic function of $\theta$,  $\mathrm{var}(\widetilde Y)$ is minimized when $\theta$ equals to $\mathrm{cov}(Y, X)/\mathrm{var}(X)$. Substituting this $ \theta $, the variance simplifies to: 
\begin{equation}
\mathrm{var}(\widetilde{Y}) = \mathrm{var}({Y})\left(1 - \frac{\text{cov}^2(X, Y)}{\text{var}(Y)\text{var}(X)}\right). \notag
\end{equation}
Consequently, an existence of the covariance of $X$ and $Y$ achieves variance reduction. 
In practical statistical inference, the true value of \(\theta\) remains unknown and therefore necessitates estimation from sample data. When the true \(\theta\) is substituted with its sample-based estimator, the estimator may exhibit bias. Nevertheless, an extensive body of literature \citep{lin2013agnostic, ding2024first} has demonstrated that such bias becomes negligible under large-sample conditions and converges to zero as the sample size tends to infinity. 

In practice, the intuition of the control variates method has been universally recognized by various technology companies. Following the recommendations in Deng et al. \cite{deng2013improving}, they have adopted pre-experimental data as control variates and deployed the method on their experimental platforms. However, discrepancies remain regarding the estimation of $\theta$ in actual deployment. For instance, companies such as Booking \cite{jackson2018booking}, Stasig \cite{Statsig}, and Netflix \citep{xie2016improving} utilize a common $\widehat{\theta}$ for both the treatment and control groups, whereas Meta \cite{meta2024} and Microsoft \citep{micro} employ distinct $\theta$ estimates for each group. In the subsequent discussion, we compile the various $\theta$ estimators currently deployed in major technology companies and elaborate on their asymptotic properties, along with the corresponding ATE estimation, from the perspective of practical implementations.

\subsection{Estimation of $\theta$ and $\widehat\delta$ construction}

\textbf{$\widehat\theta$ shared for two groups}. 
We first consider two approaches that use a common \( \theta \) estimate for both the treatment and control.  
In practical implementations, Booking \citep{jackson2018booking}, Statsig \cite{Statsig}, and Walmart \cite{Walmart} estimate $\theta$ using the full sample covariance and variance: 
\begin{equation}
\label{booking}
\widehat{\theta}_{t,1}=\widehat{\theta}_{c,1}:=\widehat{\theta}_1 = \frac{\widehat{\text{cov}}(\overline Y, \overline X)}{\widehat{\text{var}}(\overline X)}, \notag
\end{equation}
where $\widehat{\text{cov}}(\oy, \ox)$ and $\widehat{\text{var}}(\ox)$ denote the sample covariance and variance, respectively. The explicit expressions are given by: 
$$
\widehat{\text{cov}}(\oy, \ox)=\frac{1}{n(n-1)}\lc \sum_{i=1}^nY_iX_i-n\oy \ox\rc, 
$$
$$
\wv(\ox)=\frac{1}{n(n-1)} \lc \sum_{i=1}^nX_i^2-n\ox^2 \rc. 
$$
The corresponding adjusted estimator is: 
\begin{equation}
\label{delta1}
\widehat \delta_1 =\frac{1}{n_t}\sum_{i=1}^nT_i\lc Y_i-\widehat\theta_{t,1}(X_i - \overline X)\rc - \frac{1}{n_c}\sum_{i=1}^n(1-T_i)\lc Y_i-\widehat\theta_{c,1} (X_i - \overline X)\rc. \nonumber
\end{equation}
As the sample size $n$ increases, \(\widehat\delta_1\) is asymptotically normal: 
$$(\widehat \delta_1 - \delta)/{{\text{sd}}(\widehat{\delta}_1)}\overset{d}{\rightarrow} N(0, 1), $$
where $\text{sd}(\wdd_1)$ denotes the true standard error of $\wdd_1$.

Considering the potential interaction between the treatment and pre-experiment variable, $\text{cov}(Y_t, X_t)$ and $\text{cov}(Y_c, X_c)$ may differ. Consequently, Netflix \citep{xie2016improving} and Airbnb \citep{deng2021improving} account for potential treatment-covariate interactions by pooling within-group covariances and variances: 
\begin{equation}
\label{deng}
\widehat{\theta}_{t, 2}=\widehat{\theta}_{c, 2}:=\widehat{\theta}_2 = \frac{\widehat{\text{cov}}(\overline{Y}_t, \overline{X}_t) + \widehat{\text{cov}}(\overline{Y}_c, \overline{X}_c)}{\widehat{\text{var}}(\overline{X}_t) + \widehat{\text{var}}(\overline{X}_c)}. \notag
\end{equation}

The corresponding $\widehat \delta_2$ is defined as: 
\begin{equation}
\label{delta2}
\widehat \delta_2 =\frac{1}{n_t}\sum_{i=1}^nT_i\lc Y_i-\widehat\theta_{t, 2}(X_i - \overline X)\rc - \frac{1}{n_c}\sum_{i=1}^n(1-T_i)\lc Y_i-\widehat\theta_{c,2} (X_i - \overline X)\rc. \nonumber
\end{equation}
As $n$ grows, $\widehat\delta_2$ remains asymptotically normal: 
$$(\widehat \delta_2 - \delta)/{{\text{sd}}(\widehat{\delta}_2)}\overset{d}{\rightarrow} N(0, 1). $$

\noindent\textbf{$\widehat\theta$ different for two groups}. 
To explicitly allow for heterogeneous treatment effects (HTE) on the covariance between $Y$ and $X$, Meta \cite{meta2024} and Microsoft \cite{micro} employ separate $\theta$ estimates for the treatment and control groups: 
\begin{equation}
\label{the3}
\widehat{\theta}_{t, 3}=\frac{\widehat{\text{cov}}(\overline Y_t, \overline X_t)}{\widehat{\text{var}}(\overline X_t)}, \qquad \widehat{\theta}_{c, 3}=\frac{\widehat{\text{cov}}(\overline Y_c, \overline X_c)}{\widehat{\text{var}}(\overline X_c)}. \notag
\end{equation}
The corresponding $\widehat\delta_3$ is constructed as: 
\begin{equation}
\label{delta3}
\widehat \delta_3 =\frac{1}{n_t}\sum_{i=1}^nT_i\lc Y_i-\widehat\theta_{t, 3} (X_i - \overline X)\rc - \frac{1}{n_c}\sum_{i=1}^n(1-T_i)\lc Y_i-\widehat\theta_{c, 3} (X_i - \overline X)\rc, \nonumber
\end{equation}
from which the following result obtains: 
$$(\widehat \delta_3 - \delta)/{{\text{sd}}(\widehat{\delta}_3)}\overset{d}{\rightarrow} N(0, 1). $$

\subsection{Practical implementations and comparisons}

As can be seen above, all distinct estimators of $\delta$ are asymptotically unbiased and converge to a normal distribution. However, due to the presence of $\widehat{\theta}$, deriving the expression for the true variance is challenging. In practice, we temporarily treat $\widehat{\theta}$ as a constant and use the following variance estimator as a substitute for the true variance: 
\begin{equation}
\begin{aligned}
&\quad \widehat{\text{var}}(\widehat{\delta}_k) \\
&= \frac{1}{n_t(n_t-1)}\sum_{i=1}^n {T_i\Big(Y_i-\widehat{\theta}_{t, k}(X_i-\overline{X}) - \big(\overline{Y}_t-\widehat{\theta}_{t, k}(\overline{X}_t-\overline{X})\big)\Big)^2} + \\
&\quad \frac{1}{n_c(n_c-1)}\sum_{i=1}^n {(1-T_i)\Big(Y_i-\widehat{\theta}_{c, k}(X_i-\overline{X}) - \big(\overline{Y}_c-\widehat{\theta}_{c, k}(\overline{X}_c-\overline{X})\big)\Big)^2}\\
&=\frac{1}{n_t}\Big(\widehat{\text{var}}(Y_t) - 2\widehat{\theta}_{t, k}\widehat{\text{cov}}(Y_t, X_t) + \widehat{\theta}_{t, k}^2\widehat{\text{var}}(X_t)\Big) + \\
&\quad \frac{1}{n_c}\Big(\widehat{\text{var}}(Y_c) - 2\widehat{\theta}_{c, k}\widehat{\text{cov}}(Y_c, X_c) + \widehat{\theta}_{c, k}^2\widehat{\text{var}}(X_c)\Big), \quad k=1, 2, 3 \notag
\end{aligned}
\end{equation}
In the design-based framework, these variance estimations have been proved to be either consistent or conservative \citep{lin2013agnostic, ding2024first}. Nevertheless, such convergence results remain acceptable, as they satisfy the strict industry requirement for A/B testing—specifically, that Type I error rates are not inflated. 



While there are currently three prevalent methods in the industry for estimating $\theta$ and constructing $\widehat\delta$, they are in fact inherently interconnected. The intuition behind $\widehat\delta_2$ and $\widehat\delta_3$ both accounts for the potential impact of treatment $T$ on $\text{cov}(Y_t, X_t)$ and $\text{cov}(Y_c, X_c)$, so they are likely to  exhibit similar asymptotic properties. Additionally, if the treatment does not affect the correlation between pre-experiment data and post-experiment metrics across groups, there should be no distinction in performance among $\widehat\delta_1$, $\widehat\delta_2$, and $\widehat\delta_3$. The following theorem demonstrates that $\widehat{\delta}_2$ and $\widehat{\delta}_3$ are asymptotically equivalent, and $\widehat{\delta}_1$ is asymptotically equivalent to $\widehat{\delta}_3$  under certain conditions. 

\begin{theorem}
\label{equal}
Under the above formulations of $\wdd_1$, $\wdd_2$ and $\wdd_3$, the following asymptotic properties hold:

\noindent i). $\sqrt{n}(\widehat{\delta}_2-\widehat{\delta}_3)$ converges in probability to $0$, meaning that for any $\epsilon>0$, $\lim\limits_{n\rightarrow \infty}\mathbb P(\sqrt n|\widehat{\delta}_2 - \widehat{\delta}_3 |>\epsilon)=0$.

\noindent ii). When $p_t=p_c$ or $\theta_t=\theta_c$, $\sqrt n (\widehat{\delta}_1-\widehat{\delta}_3)$ converges to $0$ in probability, with $\theta_t=\text{cov}(Y_t, X_t)/\text{var}(X_t)$ and $\theta_c=\text{cov}(Y_c, X_c)/\text{var}(X_c)$. 

\end{theorem}

The conclusions of Theorem \ref{equal} provide critical guidance for method selection in practical applications: In testing scenarios where $p_t \neq p_c$ and $\theta_t \neq \theta_c$, the inferential results derived from $\widehat{\delta}_1$ and $\widehat{\delta}_3$ may diverge in a non-negligible manner. Indeed, as explicitly established in Theorem \ref{deltavar}, such discrepancies are ``consistent'', rendering $\widehat{\delta}_3$ distinctly preferable to \(\widehat{\delta}_1\) under these circumstances. 
\begin{theorem}
\label{deltavar}
Building on the foregoing formulations, the following asymptotic relationships among the variance estimators obtain:

\noindent i). The asymptotic estimated variance of $\sqrt{n}\widehat{\delta}_1$ is no less than that of $\sqrt{n}\widehat{\delta}_2$, with strict inequality save when $\theta_t = \theta_c$ or $p_t=p_c$. In formal terms, 
$$\lim_{n\rightarrow \infty}n\left(\widehat{\text{var}}(\widehat{\delta}_2)-\widehat{\text{var}}(\widehat{\delta}_1)\right)=-\frac{1}{p_tp_c}(\theta_t-\theta_c)^2(p_t-p_c)^2\text{var}(X). $$

\noindent ii). The asymptotic estimated variance of $\sqrt{n}\widehat{\delta}_1$ is greater than or equal to that of $\sqrt{n}\widehat{\delta}_3$, with strict inequality except in the case where $\theta_t = \theta_c$. Formally, 
$$\lim_{n\rightarrow \infty}n\left(\widehat{\text{var}}(\widehat{\delta}_3)-\widehat{\text{var}}(\widehat{\delta}_1)\right)=-\lc \frac{p_t^3+p_c^3}{p_tp_c} \rc(\theta_t-\theta_c)^2\text{var}(X).$$

\noindent iii). The asymptotic estimated variance of $\sqrt{n}\widehat{\delta}_2$ is no less than that of $\sqrt{n}\widehat{\delta}_3$, with strict inequality unless $\theta_t = \theta_c$. Precisely, 
$$\lim_{n\rightarrow \infty}n\left(\widehat{\text{var}}(\widehat{\delta}_3)-\widehat{\text{var}}(\widehat{\delta}_2)\right)=-(\theta_t-\theta_c)^2\text{var}(X).$$
\end{theorem}

The above theorem indicates that in practical A/B testing, using $\widehat\delta_3$ for hypothesis testing will necessarily yield statistical power no lower than that of $\widehat\delta_1$ and $\widehat\delta_2$. Since the variance estimates are either consistent or conservative estimators of the true variance, there will be no issue of inflated Type I error rates despite their smaller asymptotic variances.

\section{Regression Adjustment}
\label{sec3}

\subsection{Overview}

The control variate methods described above do not assume a pre-specified functional relationship between the response variable and the control variate; instead, they derive the optimal \(\theta\) based on their covariance structure and substitute $\widehat\theta$ accordingly. Additionally, ATE estimation can be performed via linear regression modeling, a framework generally referred to as regression adjustment. 

\subsection{Statistical models}

\textbf{Simple regression model}. 
We proceed by incrementally incorporating additional covariates into the regression equation. When the regression model includes no covariates other than the grouping variable, we obtain the simple regression model: 
\begin{equation}
\label{eq1}
Y  = \beta_0 + T\cdot\beta_T + \varepsilon, \notag
\end{equation}
where $\beta_0$ denotes the intercept term, $\beta_T$ represents the ATE, and $\varepsilon$ denotes the random noise. Notably, within the design-based framework, the residuals $\varepsilon$ do not satisfy the independence and identical normality assumptions. 

Under the convergence conditions described in Freedman \cite{freedman2008regression}, the simple regression estimator \(\widehat{\beta}_T^{SR}\) obtained via ordinary least squares (OLS) is asymptotically normal: 
\begin{equation}
\label{SR}
(\widehat{\beta}_T^{SR} - \delta)/{\text{sd}(\widehat{\beta}_T^{SR})} \xrightarrow{d}N(0, 1). \notag
\end{equation}
This asymptotic result provides the theoretical foundation for constructing confidence intervals and conducting hypothesis tests regarding the ATE. 

\noindent\textbf{Adjusted regression model}. 
In many practical scenarios, there exists a covariate $X$ that is associated to the outcome variable and can reduce residual variability, thereby enhancing the precision of the ATE estimates. The adjusted regression model specifies a regression of $Y$ on $T$ and the centered covariate $(X-\overline{X})$: 
\begin{equation}
\label{eq2}
Y  = \beta_0 + T\cdot\beta_T + (X-\overline{X})\cdot\beta_X + \varepsilon, \notag
\end{equation}
where $\beta_X$ captures the effect of the centered covariate on the outcome. 

Under the convergence conditions detailed in Freedman \cite{freedman2008regression}, the OLS-based adjusted regression estimator $\widehat{\beta}_T^{AR}$ exhibits asymptotic normality: 
\begin{equation}
\label{AR}
(\widehat{\beta}_T^{AR} - \delta)/{\text{sd}(\widehat{\beta}_T^{AR})} \xrightarrow{d}N(0, 1). \notag
\end{equation}
 
\noindent\textbf{Interactive regression model}. 
In some cases, the treatment effect may not be homogeneous across all units but may instead vary with a covariate. This can be accommodated by incorporating an interaction term between the treatment indicator and the covariate, allowing the model to capture HTE that depends on $X$. 

The resulting interactive regression model is specified as: 
\begin{equation}
\label{eq3}
Y  = \beta_0 + T\cdot\beta_T + (X-\overline{X})\cdot\beta_X + T\cdot(X-\overline{X})\cdot \beta_I+ \varepsilon, \notag
\end{equation}
where \(\beta_I\) denotes the coefficient for the interaction term \(T\cdot(X - \overline{X})\). This term quantifies how the treatment effect varies with the covariate: a statistically significant \(\beta_I \neq 0\) indicates differential treatment effects across subjects with varying $X$ values. Specifically, the treatment effect for a subject with covariate value $X$ is given by \(\beta_T + \beta_I \cdot (X - \overline{X})\), which varies linearly with $X$ in magnitude and direction determined by the sign of \(\beta_I\). 

Under the convergence conditions described in Lin \cite{lin2013agnostic}, the OLS-based ATE estimator $\widehat{\beta}_T^{IR}$ is asymptotically normal: 
\begin{equation}
\label{IR}
(\widehat{\beta}_T^{IR} - \delta)/{\text{sd}(\widehat{\beta}_T^{IR})} \xrightarrow{d}N(0, 1). \notag
\end{equation}

The key advantage of the interaction model lies in its capacity to capture and quantify covariate-dependent treatment effects, offering a more nuanced understanding of the treatment's impact compared to simpler specifications.

\subsection{Comparisons and practical implementations}

Understanding the performance of different regression adjustment estimators is crucial for making informed choices in practical applications. A key aspect of this performance is the variance of the estimators, as it directly affects the precision of the ATE estimates and the power of tests. Lin \cite{lin2013agnostic} has shown that the true variances of the estimators differ in a specific way, establishing that $\sqrt{n}\widehat{\beta}^{SR}_T$ and $\sqrt{n}\widehat{\beta}^{AR}_T$ has at least as much asymptotic variance as $\sqrt{n}\widehat{\beta}^{IR}_T$: 
\begin{equation}
\lim_{n\rightarrow \infty} n\left(\text{var}(\widehat\beta_T^{IR})-\text{var}(\widehat\beta_T^{SR})\right)\le 0, \lim_{n\rightarrow \infty} n\left(\text{var}(\widehat\beta_T^{IR})-\text{var}(\widehat\beta_T^{AR})\right)\le 0. \notag
\end{equation}
This finding aligns with the intuition that incorporating interaction terms can capture additional sources of variability in the treatment effect, thereby reducing the overall variance of the estimator when such effect modification is present. 

Variance estimation remains indispensable in practical implementations, since the true variance of the regression coefficient \(\widehat{\beta}_{T}\) depends on unobservable potential outcomes, which are not directly accessible in real-world data. As a result, we must rely on estimated variances to construct confidence intervals and perform hypothesis tests. However, extensive literature \citep{freedman2008regression, lin2013agnostic} has shown that the standard OLS variance estimators for \(\widehat\beta_T\) are not guaranteed to be consistent. This inconsistency arises because the OLS variance estimator makes assumptions about the structure of the residuals that may not hold in the design-based framework, especially when the relationship between the covariates and the outcome is misspecified or when there is heteroscedasticity. 

Consequently, we discard the OLS variance estimator and instead employ the more robust Eicker-Huber-White (EHW) variance estimator \citep{huber1967behavior, eicker1967limit} for statistical inference. The EHW estimator, often referred to as the ``sandwich'' estimator, is designed to handle heteroscedasticity and does not rely on strong assumptions about the form of the residual variance. It has been shown to be asymptotically consistent or conservative in a wide range of settings \citep{white1980heteroskedasticity, kauermann2000sandwich, ding2017bridging}, making it a reliable choice for variance estimation in regression adjustment models within the design-based framework. Furthermore, when the residual distributions of the treatment and control groups differ substantially and the two groups exhibit marked discrepancies in sample size, the divergence between the EHW variance estimate and the OLS variance estimate becomes non-negligible. 

In practice, clarifying the magnitude relationships among variance estimates of distinct estimators is of critical importance. This is because the variance estimates directly influence the design and interpretation of A/B tests, where researchers need to determine sample sizes and interpret results with confidence intervals. Theorem \ref{reg_com} explicitly specifies the magnitude relationships between the EHW variance estimates of \(\widehat{\beta}_T\), providing substantial guidance for practical A/B testing implementations. 
\begin{theorem}
\label{reg_com}
When employing the EHW method for variance estimation of OLS coefficients, the asymptotic variance of \(\sqrt{n}\widehat{\beta}_T^{IR}\) is no greater than that of \(\sqrt{n}\widehat{\beta}_T^{SR}\) or \(\sqrt{n}\widehat{\beta}_T^{AR}\), formally: 
$$\lim_{n\rightarrow \infty}n\left(\widehat{\text{var}}(\widehat\beta_T^{IR}) - \widehat{\text{var}}(\widehat\beta_T^{SR})\right)\le0, \lim_{n\rightarrow \infty}n\left(\widehat{\text{var}}(\widehat\beta_T^{IR}) - \widehat{\text{var}}(\widehat\beta_T^{AR})\right)\le0. $$

\end{theorem}

Theorem \ref{reg_com} strengthens the theoretical rationale for preferring the interaction regression model in practical applications. It establishes that \(\widehat{\beta}_T^{IR}\) is uniformly superior to \(\widehat{\beta}_T^{AR}\) both theoretically (in terms of true variance) and practically (in terms of EHW-estimated variance). For a fixed significance level, tests based on \(\widehat\beta_T^{IR}\) exhibits  greater power to detect true treatment effects. This is particularly valuable in A/B testing and other experimental settings, where the ability to detect small but meaningful effects can be crucial for making informed decisions.

In practical implementation, these findings underscore the desirability of prioritizing the interactive regression model for ATE estimation within the design-based framework. Adopting this approach leverages the lower variance (both true and estimated) of \(\widehat{\beta}_T^{IR}\), yielding more precise estimates and more powerful statistical tests. Additionally, the use of the EHW variance estimator ensures that the resulting inference is robust to violations of homoscedasticity and other model assumptions, thereby enhancing the reliability of the conclusions drawn from the analysis.

\section{Control Variates \& Regression Adjustments}
\label{sec4}

\begin{table*}[!htb]
    \centering
    \caption{Means and variances at the model-based, design-based, and sample levels.}
    \resizebox{1\textwidth}{!}{
    \begin{tabular}{llllccc}
    \toprule
    & \multicolumn{3}{c}{Means} & \multicolumn{3}{c}{Variances} \\ 
        \cmidrule(r){2-4} \cmidrule(r){5-7} 
    & Treatment & Control & Average treatment effect & Treatment & Control & Individual causal effect \\
    \midrule 
    Model-based level & $\mathbb E\left(W_t(\theta)\right)$ & $\mathbb E\left(W_c(\theta)\right)$ & $\delta=\mathbb E\left(W_t(\theta)-W_c(\theta)\right)$ & $V_t(\theta)$  & $V_c(\theta)$  & $V_\delta$ \\
    Design-based level & $\overline W_t(\theta)$ & $\overline W_c(\theta)$ & $\delta_{\mathcal S }=\overline W_t(\theta)-\overline W_c(\theta)$ & $S_t^2(\theta)$  & $S_c^2(\theta)$  & $S_\delta^2$ \\
    Sample level & $\overline W_t^{obs}(\theta)$ & $\overline W_c^{obs}(\theta)$ & $\widehat\delta=\overline W_t^{obs}(\theta)-\overline W_c^{obs}(\theta)$ & $s_t^2(\theta)$  & $s_c^2(\theta)$  & - \\
    \bottomrule
    \end{tabular}}
    \label{bridge}
\end{table*}

Despite differing intuitions, control variates and regression adjustment yield asymptotically equivalent ATE estimators under specific conditions. Specifically, Deng et al.  \cite{deng2023augmentation} and Tsiatis et al. \cite{tsiatis2008covariate} have shown that the regression adjustment models asymptotically estimate the ATE as \(\overline{Y}_t - \overline{Y}_c - \big( \overline{f(X_t)} - \overline{f(X_c)} \big)\) for some function \(f(\cdot)\). In Theorem \ref{rela1}, we precisely formulate the relationship between the control variates estimator \(\widehat{\delta}\) and the regression adjustment estimator \(\widehat{\beta}_T\). 

\begin{theorem}
\label{rela1}
Adopting the notation introduced in Sections \ref{sec2} and \ref{sec3}, the following asymptotical relationships hold among the estimators under the design-based framework: 

\noindent i). \(\widehat{\delta}_0\) is exactly equal to \(\widehat{\beta}_T^{SR}\). 

\noindent ii).  \(\sqrt n (\widehat{\delta}_1 - \widehat{\beta}_T^{AR})\) converges in probability to $0$. 

\noindent iii). \(\sqrt n (\widehat{\delta}_2 - \widehat{\beta}_T^{IR})\) tends to $0$ in probability.

\noindent iv). \(\widehat{\delta}_3\) coincides exactly with \(\widehat{\beta}_T^{IR}\). 
\end{theorem}

These equivalences reveal that the two sets of estimators, despite originating from distinct methods, yield computationally analogous results. In particular, \(\widehat{\delta}_0\) and \(\widehat{\delta}_3\) are exactly identical to \(\widehat{\beta}_T^{SR}\) and \(\widehat{\beta}_T^{IR}\), respectively. In other words, regardless of the method employed for ATE estimation in practice, the resulting estimates converge to the same value in most scenarios. 

As for variance estimation, the control variates method derives its variance estimate directly from the sample variance with \(\theta\) replaced by \(\widehat{\theta}\), whereas the regression framework employs EHW variance estimation to obtain a robust and conservative variance estimate. While their absolute values may differ marginally in practical implementations, Theorem \ref{finivar} establishes the asymptotic convergence properties of these variance estimators. 

\begin{theorem}
\label{finivar}
Under the design-based framework, the variance estimators satisfy the following asymptotical relationships: 

\noindent i).  $n\widehat{\text{var}}(\widehat\delta_0)$ and $n\widehat{\text{var}}(\widehat{\beta}_{T}^{SR})$ are asymptotically equal in probability, 
$$\lim_{n\rightarrow \infty}n\left(\widehat{\text{var}}(\widehat\delta_0)-\widehat{\text{var}}(\widehat{\beta}_{T}^{SR})\right)=0. $$ 

\noindent ii).  $n\widehat{\text{var}}(\widehat\delta_1)$ and $n\widehat{\text{var}}(\widehat{\beta}_{T}^{AR})$ are asymptotically equivalent in probability, 
$$\lim_{n\rightarrow \infty}n\left(\widehat{\text{var}}(\widehat\delta_1)-\widehat{\text{var}}(\widehat{\beta}_{T}^{AR})\right)=0. $$ 

\noindent iii). $n\widehat{\text{var}}(\widehat\delta_3)$ is asymptotically equal in probability to $n\widehat{\text{var}}(\widehat{\beta}_{T}^{IR})$, 
$$\lim_{n\rightarrow \infty}n\left(\widehat{\text{var}}(\widehat\delta_3)-\widehat{\text{var}}(\widehat{\beta}_{T}^{IR})\right)=0. $$ 

\end{theorem}

Combining Theorems \ref{rela1} and \ref{finivar}, it follows that $\widehat\delta_0$ and $\widehat{\beta}_T^{\text{SR}}$, $\widehat\delta_1$ and $\widehat{\beta}_T^{\text{AR}}$, as well as $\widehat\delta_3$ and $\widehat{\beta}_T^{\text{IR}}$, yield asymptotically equivalent inference. Consequently, within the model-based framework, we center our subsequent analyses on the control variates method. 

\section{Extending to Model-based Framework}
\label{sec5}

In the preceding sections, we have focused on the design-based perspective in randomized experiments, where potential outcomes are treated as fixed quantities, and sampling variability is not considered. Design-based causal inference framework emphasizes experimental design and imposes minimal assumptions on the outcome-generating process. Consequently, while statistical inferences are valid for the specific study sample (internal validity), they may not generalize to broader populations beyond the study sample (external validity). Since experimenters typically seek both internal and external validity, we now discuss the considerations involved in extending control variates methods from the design-based framework to the model-based setting.

\subsection{Design-based and model-based frameworks}

We begin by establishing some notations. Let the random variables $\{Y_t, Y_c, X\}$ represent the triad of potential outcomes and covariate within the model-based framework, from which we take an independent and identically distributed finite population of size $n$: 
$$\mathcal S=\{(Y_i^t, Y_i^c, X_i): i=1, \ldots, n\} \overset{i.i.d.}{\sim}\{Y_t, Y_c, X\}. $$
To distinguish between design-based and model-based quantities, calculations under the finite population are explicitly conditioned on \(\mathcal{S}\), while those under the infinite population omit such conditioning. 

We randomly assign $n_t$ subjects to receive treatment, leaving the remaining $n_c=n-n_t$ subjects to receive control. Define the following transformed variables: 
$$W_i^t(\theta)=Y_i^t-\theta (X_i-\overline X), W_i^c(\theta)=Y_i^c-\theta (X_i-\overline X), W_i^{obs}(\theta)=Y_i-\theta (X_i-\overline X), $$ 
where $\theta$ is a fixed parameter and $\overline X=\sum_{i=1}^n X_i/n$ denotes the sample mean of the covariate. 

\begin{itemize}[leftmargin=10pt]

\item \textbf{Model-based level}

At the model-based level, the average potential outcomes are $\mathbb E\left(W_t(\theta)\right)$ and $\mathbb E\left(W_c(\theta)\right)$, with the ATE given by $\delta=\mathbb E\big(W_t(\theta)-W_c(\theta)\big)$. The population variances of the potential outcomes and individual causal effect are: 
$$V_t=\text{var}\left(W_t(\theta)\right), V_c=\text{var}\left(W_c(\theta)\right), V_{\delta}=\text{var}\left(W_t(\theta)-W_c(\theta)\right). $$

\item \textbf{Design-based level}

Treating the sample $\mathcal S$ as fixed, the average potential outcomes and design-based ATE are: 
$$\overline W_t(\theta)=\frac{1}{n}\sum_{i=1}^n W_i^t(\theta), \overline W_c(\theta)=\frac{1}{n}\sum_{i=1}^n W_i^c(\theta), \delta_{\mathcal S}=\overline W_t(\theta)-\overline W_c(\theta). $$

The corresponding variances of the potential outcomes and individual causal effect are 
$$S_t^2(\theta)=\frac{1}{n-1}\sum_{i=1}^n\big(W_i^t(\theta)-\overline W_t(\theta)\big)^2, S_c^2(\theta)=\frac{1}{n-1}\sum_{i=1}^n\big(W_i^c(\theta)-\overline W_c(\theta)\big)^2,$$
$$S^2_{\delta}=\frac{1}{n-1}\sum_{i=1}^n\left(W_i^t(\theta)-W_i^c(\theta)-\delta_{\mathcal S}\right)^2. $$

\item \textbf{Sample level}

At the observed sample level, we compute the averages of the observed outcomes and the difference-in-means estimator: 
$$\overline W_t^{obs}(\theta)=\frac{1}{n_t}\sum_{i=1}^nT_iW_i^{obs}(\theta), \overline W_c^{obs}(\theta)=\frac{1}{n_c}\sum_{i=1}^n(1-T_i)W_i^{obs}(\theta), $$
$$\widehat \delta=\overline W_t^{obs}(\theta)-\overline W_c^{obs}(\theta), $$
with corresponding sample variances: 
$$s_t^2(\theta)=\frac{1}{n_t-1}\sum_{i=1}^nT_i\left(W_i^{obs}(\theta)-\overline W_t^{obs}(\theta)\right)^2, $$
$$s_c^2(\theta)=\frac{1}{n_c-1}\sum_{i=1}^n(1-T_i)\left(W_i^{obs}(\theta)-\overline W_c^{obs}(\theta)\right)^2. $$

\end{itemize}

No sample analogues of $V_\delta$ or $S^2_{\delta}$ exists, as $W_i^t(\theta)$ and $W_i^c(\theta)$ are never jointly observed for any subject $i$. Table \ref{bridge} summarizes the model-based framework, design-based framework, and sample quantities. 

We distinguish between two key parameters: the model-based average treatment effect \(\delta\) and the design-based average treatment effect \(\delta_{\mathcal{S}}\). The preceding sections focused on inference for \(\delta_{\mathcal{S}}\) under the design-based framework, demonstrating that the sample variance of \(\widehat{\delta}\)—specifically, \(s_t^2(\theta)/n_t + s_c^2(\theta)/n_c\)—provides a conservative estimate of \(\text{var}(\widehat{\delta} \mid \mathcal{S})\). In contrast, we subsequently show that under the model-based framework, the sample variance of $\widehat{\delta}$ may be no longer a consistent or conservative estimate of $\text{var}(\widehat\delta)$. 

\begin{table*}[!htb]
    \centering
    \caption{Distinct ATE estimators, their variance estimation formulas under both the design-based and model-based frameworks, and whether they can guarantee at least the same efficiency as the two-sample $t$-test. }
    \resizebox{1\textwidth}{!}{
    \begin{tabular}{llllcccc}
    \toprule
    &  & \multicolumn{2}{c}{Variance Estimation} & \multicolumn{2}{c}{Guarantee of Variance Reduction} & \multicolumn{2}{c}{Most efficient} \\
    \cmidrule(r){3-4} \cmidrule(r){5-6} \cmidrule(r){7-8} 
     & Expression & Design-based & Model-based & Design-based & Model-based & Design-based & Model-based\\
    \midrule 
    $\widehat\delta_0$ &  $\overline W_t^{obs}(0)-\overline W_c^{obs}(0)$ & {$ s_t^2(0) /{n_t}+  s_c^2(0)/{n_c}$} &  {$ s_t^2(0) /{n_t}+  s_c^2(0)/{n_c}$} & - & - & & \\
    $\widehat\delta_1$ &  $\overline W_t^{obs}(\widehat\theta_1)-\overline W_c^{obs}(\widehat\theta_1)$ &  {$s_t^2(\widehat\theta_1)/{n_t} +  s_c^2(\widehat\theta_1)/{n_c}$} & {$s_t^2(\widehat\theta_1)/{n_t} +  s_c^2(\widehat\theta_1)/{n_c}$} & No & No & & \\
    $\widehat\delta_2$ &  $\overline W_t^{obs}(\widehat\theta_2)-\overline W_c^{obs}(\widehat\theta_2)$ & {$ s_t^2(\widehat\theta_2) /{n_t}+  s_c^2( \widehat\theta_2)/{n_c}$} & {$ s_t^2(\widehat\theta_2) /{n_t}+  s_c^2(\widehat\theta_2)/{n_c}$} & Yes & Yes &  & $\checkmark$\\
    $\widehat\delta_3$ &  $\overline W_t^{obs}(\widehat\theta_t)-\overline W_c^{obs}(\widehat\theta_c)$  & $s_t^2(\widehat\theta_t) /{n_t}+ s_c^2(\widehat\theta_c)/{n_c}$ & $s_t^2(\widehat\theta_t) /{n_t}+ s_c^2( \widehat\theta_c)/{n_c} + (\widehat\theta_t-\widehat\theta_c)^2\widehat{\text{var}}(X)/n$ & Yes & Yes & $\checkmark$ & $\checkmark$\\
    \bottomrule
    \end{tabular}}
    \label{comtable}
\end{table*}

\subsection{Model-based inference}

In statistical inference, the framework within which we operate profoundly shapes how we interpret uncertainty, estimate parameters, and draw conclusions. When transitioning from design-based perspective to model-based framework—where inference is conducted without conditioning on the observed samples—a fundamental transformation occurs in the nature of counterfactual outcomes. In this setting, counterfactuals for each subject are no longer treated as fixed, deterministic values but rather as random variables. This randomness stems from two key sources: sampling variability and treatment assignment variability. This dual source of randomness complicates variance estimation and necessitates a reevaluation of the properties of $\widehat{\delta}$. Theorem \ref{supervar} formalizes the behavior of variance estimates in the model-based context, revealing a critical limitation: $\widehat{\text{var}}(\widehat{\delta}_3)$ is not guaranteed to be a consistent estimator of the true variance. Instead, it may systematically underestimate the true variability, rending it ``aggressive''. 
\begin{theorem}
\label{supervar}
Within the model-based framework, the asymptotic properties of \(\text{var}(\widehat{\delta}_3)\) and its estimator are as follows:

\noindent i). $n\text{var}(\widehat\delta_3)$ converges in probability to 
$$ \frac{\text{var}(Y_t)}{p_t} + \frac{\text{var}(Y_c)}{p_c} - \left(\frac{p_c}{p_t}\theta_t^2 + 2\theta_t\theta_c + \frac{p_t}{p_c}\theta_c^2\right)\text{var}(X). $$

\noindent ii). $n\widehat{\text{var}}(\widehat\delta_3)$ converges in probability to 
$$ \frac{\text{var}(Y_t)}{p_t} + \frac{\text{var}(Y_c)}{p_c} - \left(\frac{1}{p_t}\theta_t^2+ \frac{1}{p_c}\theta_c^2\right)\text{var}(X). $$
Their asymptotic difference satisfies 
$$\lim_{n\rightarrow \infty}n\left(\text{var}(\widehat\delta_3)-\widehat{\text{var}}(\widehat\delta_3)\right)=(\theta_t-\theta_c)^2\text{var}(X). $$

\end{theorem}

Theorem \ref{supervar} highlights the critical discrepancy between true and estimated variances: as the sample size \(n\) grows large, the difference scales with \((\theta_t - \theta_c)^2\text{var}(X)\). This term is non-negative, implying the estimated variance is always less than or equal to the true variance. Importantly, the discrepancy grows with two factors: the degree of heterogeneity in treatment effects (captured by \((\theta_t - \theta_c)^2\)) and the variability in the covariate \(X\) (captured by \(\text{var}(X)\)). When \(\theta_t = \theta_c\), the difference vanishes, and the estimated variance becomes consistent with the true variance. However, when treatment effects are heterogeneous (\(\theta_t \neq \theta_c\)), the estimated variance systematically underestimates the true variance. 

Theorem \ref{supervar} thus carries profound implications for statistical inference in the presence of HTE. When HTE exists, \(\theta_t\) and \(\theta_c\) diverge, and the estimated variance of \(\widehat\delta\) may become ``aggressive''. Such underestimated variance threatens the external validity of inferences; results based on these estimates may fail to generalize to the model-based framework, as underlying uncertainty is misrepresented.

The shift in the variance estimate of $\delta_3$ from design-based to model-based frameworks motivates an investigation of the remaining $\widehat\delta$. Notably, in contrast to $\widehat\delta_3$, the variance estimates of $\widehat\delta_0$, $\widehat\delta_1$, and $\widehat\delta_2$ within the model-based framework are consistent estimators of the true variance, as established in Theorem \ref{supervarelse}. 

\begin{theorem}
\label{supervarelse}
Within the model-based framework, we establish: 

\noindent i). For $k=0, 1, 2$, $n\widehat{\text{var}}(\widehat\delta_k)$ converges in probability to 
$$ \frac{\text{var}(Y_t)}{p_t} + \frac{\text{var}(Y_c)}{p_c} - \left(\frac{2\theta_k\theta_t-\theta_k^2}{p_t}+ \frac{2\theta_k\theta_c-\theta_k^2}{p_c}\right)\text{var}(X), $$
with $\theta_0=0, \theta_1=p_t\theta_t+p_c\theta_c$ and $\theta_2=p_t\theta_c+p_c\theta_t$. 

\noindent ii). For each $k=0, 1, 2$, $n\widehat{\text{var}}(\widehat\delta_k)$ and $n\text{var}(\widehat\delta_k)$ are asymptotically equal in probability, 
$$ \lim_{n\rightarrow \infty} n\left(\widehat{\text{var}}(\widehat\delta_k)-\text{var}(\widehat\delta_k)\right)=0. $$

\end{theorem}

These results underscore the robustness of \(\widehat\delta_0\), \(\widehat\delta_1\), and \(\widehat\delta_2\) across inferential frameworks. Their variance estimates maintain consistency in the model-based setting, avoiding the aggressive underestimation exhibited by \(\widehat\delta_3\) under treatment effect heterogeneity.

\noindent \textbf{Variance correction}. While the variance estimate of $\widehat\delta_3$ may become underestimated when extended to the model-based framework, the incremental term is actually identifiable. Therefore, a correction can be achieved by incorporating an additional term: 
$$\widehat{\text{var}}(\widehat{\delta}_3)_{\text{correct}}=\frac{1}{n_t}s_t^2(\widehat\theta_{t, 3})+\frac{1}{n_c}s_c^2(\widehat\theta_{c, 3}) + \underbrace{\textcolor{black}{\frac{1}{n}(\widehat{\theta}_t-\widehat{\theta}_c)^2\widehat{\text{var}}(X)}}_{\text{correction term}}. $$

The analysis in Section \ref{sec2} shows that within the design-based framework, inference using $\widehat\delta_3$ is at least as efficient as the two-sample $t$-test. In the model-based framework, however, the corrected variance of $\widehat\delta_3$ increases commensurately due to the inclusion of the non-negative correction term. This raises natural questions: Does the corrected variance of $\widehat\delta_3$ still achieve variance reduction relative to $\widehat\delta_0$? Furthermore, does the corrected $\widehat\delta_3$ retain its advantage when compared to other $\delta$ estimators? These questions are addressed in Theorem \ref{supercom}.

\begin{theorem}
\label{supercom}
Under the model-based framework, the following asymptotic relationships hold: 

\noindent i). $n\widehat{\text{var}}(\widehat{\delta}_3)_{\text{correct}}$ and $n\widehat{\text{var}}(\widehat{\delta}_2)$ are asymptotically equal in probability. 

\noindent ii). Asymptotically, $n\widehat{\text{var}}(\widehat{\delta}_3)_{\text{correct}}$ and $n\widehat{\text{var}}(\widehat{\delta}_2)$ are no larger than $n\widehat{\text{var}}(\widehat{\delta_0})$, with strict inequality unless $p_c\theta_t+p_t\theta_c=0$. 

\noindent iii). Asymptotically, $n\widehat{\text{var}}(\widehat{\delta}_3)_{\text{correct}}$ and $n\widehat{\text{var}}(\widehat{\delta}_2)$ are less than or equal to $n\widehat{\text{var}}(\widehat{\delta_1})$, with strict inequality unless $p_t=p_c$ or $\theta_t=\theta_c$. 

\noindent iv). Asymptotically, $n\widehat{\text{var}}(\widehat{\delta_1})$ is not necessarily smaller than $n\widehat{\text{var}}(\widehat{\delta_0})$. 
\end{theorem}

Theorem \ref{supercom} establishes that within the model-based framework, $\widehat\delta_2$ and $\widehat\delta_3$ exhibit theoretically identical statistical power and outperform \(\widehat{\delta}_0\) and \(\widehat{\delta}_1\) under most practical conditions. Consistent with conclusions drawn under the design-based framework, $\widehat\delta_1$ still fails to guarantee superiority over $\widehat\delta_0$; instead, it may even yield counterproductive results. 

To conclude this section, Table \ref{comtable} presents the expressions for all $\delta$ estimators, their variance estimation formulas under both the design-based and model-based frameworks, and their guaranteed efficiency relative to the two-sample $t$-test.

\section{Numerical Experiments}
\label{sec6}

\subsection{Simulation studies}
In this section, we conducted extensive simulations to evaluate the aforementioned estimation methods, examining their asymptotic properties, variance estimates, type I error rates, and statistical power. Specifically, we considered $16$ experimental configurations, encompassing design-based versus model-based settings, varying group assignment probabilities($p_t=0.5, 0.4$), null versus alternative hypothesis($\text{ATE}=0, 0.1$), and the presence or absence of heterogeneous treatment effects ($\text{HTE}=0, 0.5$). Across all experiments, the total sample size for treatment and control groups was fixed at $3,000$, with noise following a mean-zero normal distribution with variance $1$. The data-generating model was specified as: 
\begin{equation}
\left\{
\begin{aligned}
&Y_t = 1 + \text{ATE} + X + \text{HTE}\cdot X + \varepsilon_t, \\
&Y_c = 1 + X  + \varepsilon_c, \\
&Y = T\cdot Y_t + (1-T)\cdot Y_c, \notag
\end{aligned}
\right.
\end{equation}
where $X \sim N(0, 2)$.

\begin{table*}[!htb]
    \centering
   \caption{Type I error rates and statistical power of estimators across experimental configurations. }
    \resizebox{0.99\textwidth}{!}{
    \begin{tabular}{lcccccccccc|ccccccc}
    \toprule
     & & & & \multicolumn{7}{c}{Design-based} & \multicolumn{7}{c}{Model-based} \\
     \midrule
    Metric  & ATE & $p_t$ & HTE & $\widehat\delta_0$ & $\widehat\delta_1$ & $\widehat\delta_2$ & $\widehat\delta_3$ & $\widehat\beta_{T}^{SR}$ & $\widehat\beta_{T}^{AR}$ & $\widehat\beta_{T}^{IR}$ & $\widehat\delta_0$ & $\widehat\delta_1$ & $\widehat\delta_2$ & $\widehat\delta_3$ & $\widehat\beta_{T}^{SR}$ & $\widehat\beta_{T}^{AR}$ & $\widehat\beta_{T}^{IR}$  \\
    \midrule
    \multirow{4}{*}{Type I Error}  & \multirow{4}{*}{0}  & \multirow{2}{*}{0.5} & 0  &  0.0462 & 0.0376 & 0.0376 & 0.0377 & 0.0463 & 0.0376 & 0.0379 &  0.0482 & 0.0487 & 0.0489  &0.0490  &0.0483 & 0.0490 & 0.0490   \\
      &   & & 0.5  & 0.0441 & 0.0222 & 0.0223 & 0.0404 & 0.0441 & 0.0223 & 0.0404 & 0.0466 & 0.0497 & 0.0497 & 0.0494 & 0.0467  & 0.0497 & 0.0494 \\
\cmidrule(r){3-18}
       &  & \multirow{2}{*}{0.4} & 0  &  0.0497 & 0.0379 & 0.0377 & 0.0377 & 0.0498 & 0.0379 & 0.0378  & 0.0492 & 0.0479 & 0.0476 & 0.0477 & 0.0492 & 0.0480 & 0.0477 \\
      &   & & 0.5  & 0.0465 & 0.0231 & 0.0220 & 0.0405 & 0.0466 & 0.0231 & 0.0405 & 0.0481 & 0.0481&  0.0491 & 0.0491&  0.0483 & 0.0483 & 0.0493 \\
    \midrule
        \multirow{4}{*}{Power} & \multirow{4}{*}{0.1}  & \multirow{2}{*}{0.5} & 0  &  0.2494 & \textbf{0.8350} & \textbf{0.8352} & \textbf{0.8352} & 0.2495 & \textbf{0.8353} & \textbf{0.8354} & 0.2389 & \textbf{0.7799} & \textbf{0.7801} & \textbf{0.7800} & 0.2392 & \textbf{0.7799} & \textbf{0.7799 } \\
      &   & & 0.5  & 0.1829 & 0.7847 & 0.7847 & \textbf{0.8511} & 0.1830 & 0.7852 & \textbf{0.8512} & 0.1716 & \textbf{0.6867} & \textbf{0.6871} & \textbf{0.6873} & 0.1716 & \textbf{0.6872} & \textbf{0.6873} \\	
\cmidrule(r){3-18}
       &  & \multirow{2}{*}{0.4} & 0  &  0.2390 & \textbf{0.8174} & \textbf{0.8177} & \textbf{0.8176} & 0.2394 & \textbf{0.8175} & \textbf{0.8177}  & 0.2157 & \textbf{0.7634} & \textbf{0.7638} & \textbf{0.7638} & 0.2157 & \textbf{0.7634} & \textbf{0.7639} \\
      &   & & 0.5  & 0.1691 & 0.7549 & 0.7701 & \textbf{0.8381} & 0.1696 & 0.7552 & \textbf{0.8384} & 0.1583 & 0.6613 & \textbf{0.6729} & \textbf{0.6729} & 0.1586 & 0.6615 & \textbf{0.6731} \\
    \bottomrule
    \end{tabular}}
    \label{tab}
\end{table*}

In the design-based setup, counterfactual outcomes and noise terms were pre-generated for all samples and held fixed across repeated experiments; randomness arose solely from the group assignment variable. In contrast, under the model-based framework, counterfactual outcomes, noise terms, and group assignments were generated independently and identically in each replication. All experiments were replicated $10,000$ times. 

\begin{figure}[!h]
  \centering 
  \subfigure[Design-based \& $p=0.5$ \& $\text{ATE}=0.1$ \&  $\text{HTE}=0$]
  {  \label{cona}
   \includegraphics[width=0.225\textwidth]{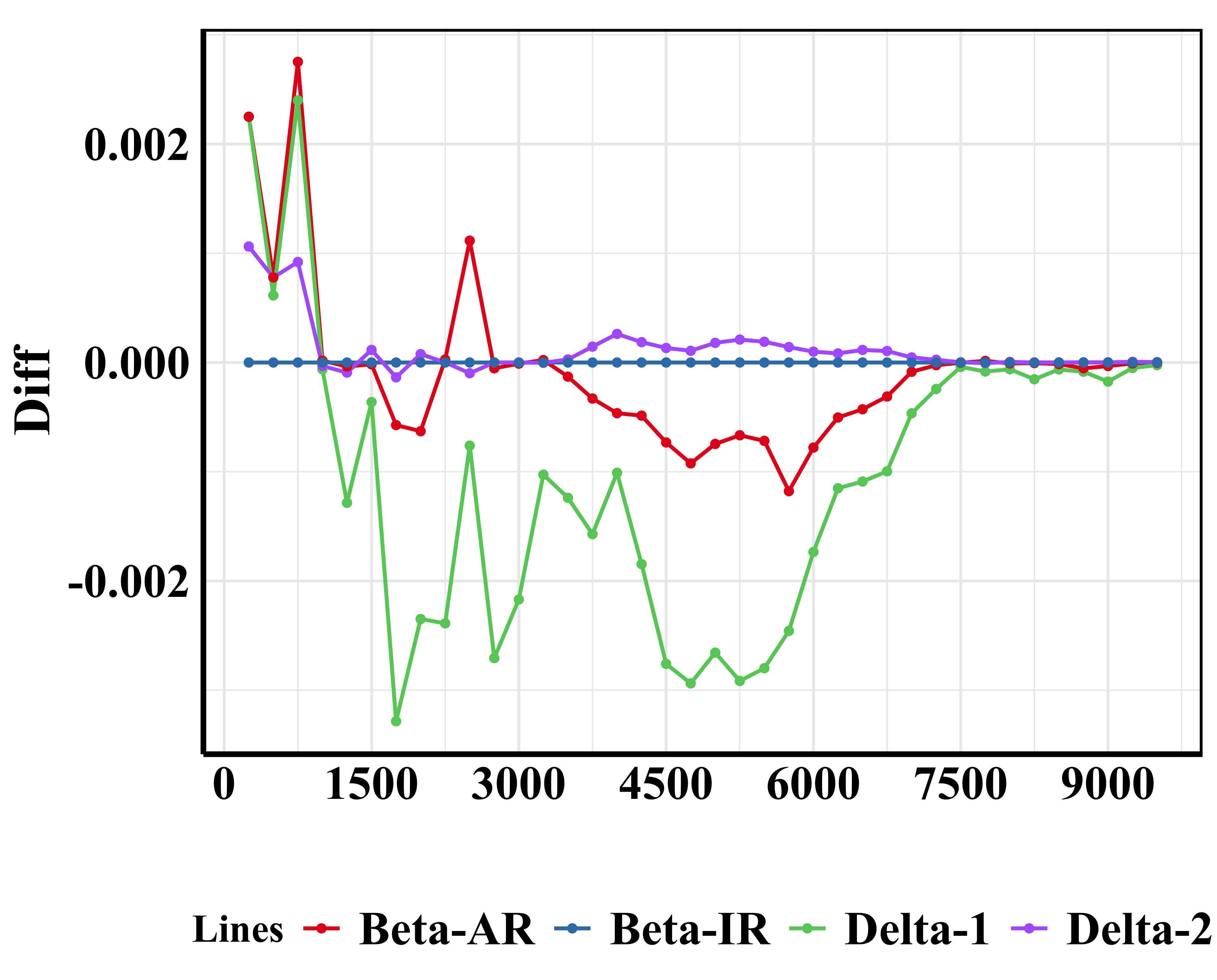}
  }
    \subfigure[Design-based \& $p=0.5$ \& $\text{ATE}=0.1$ \& $\text{HTE}=0.5$]
  {  \label{conb}
   \includegraphics[width=0.225\textwidth]{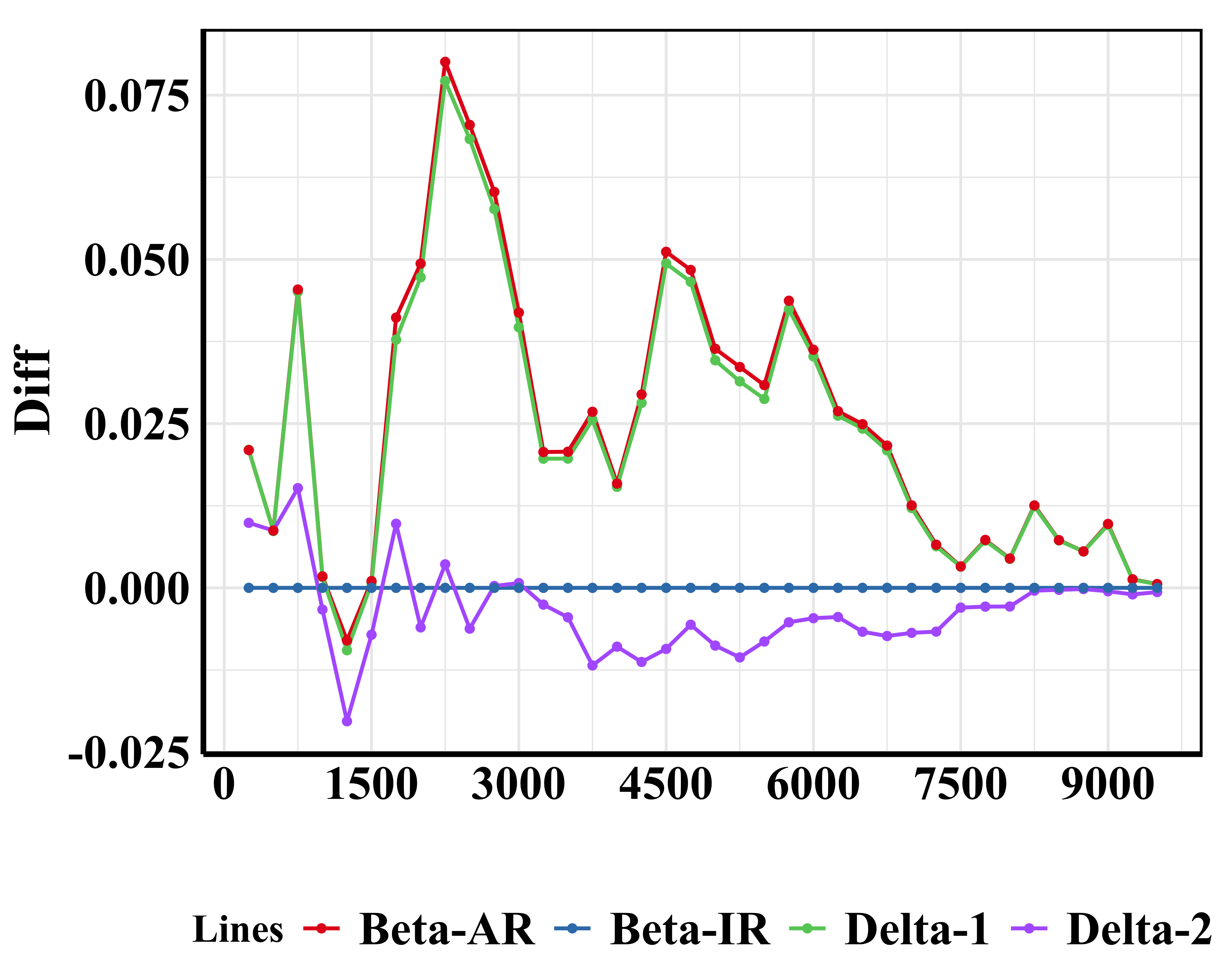}
  }
    \subfigure[Design-based \& $p=0.4$ \& $\text{ATE}=0.1$ \& $\text{HTE}=0$]
  {  \label{conc}
   \includegraphics[width=0.225\textwidth]{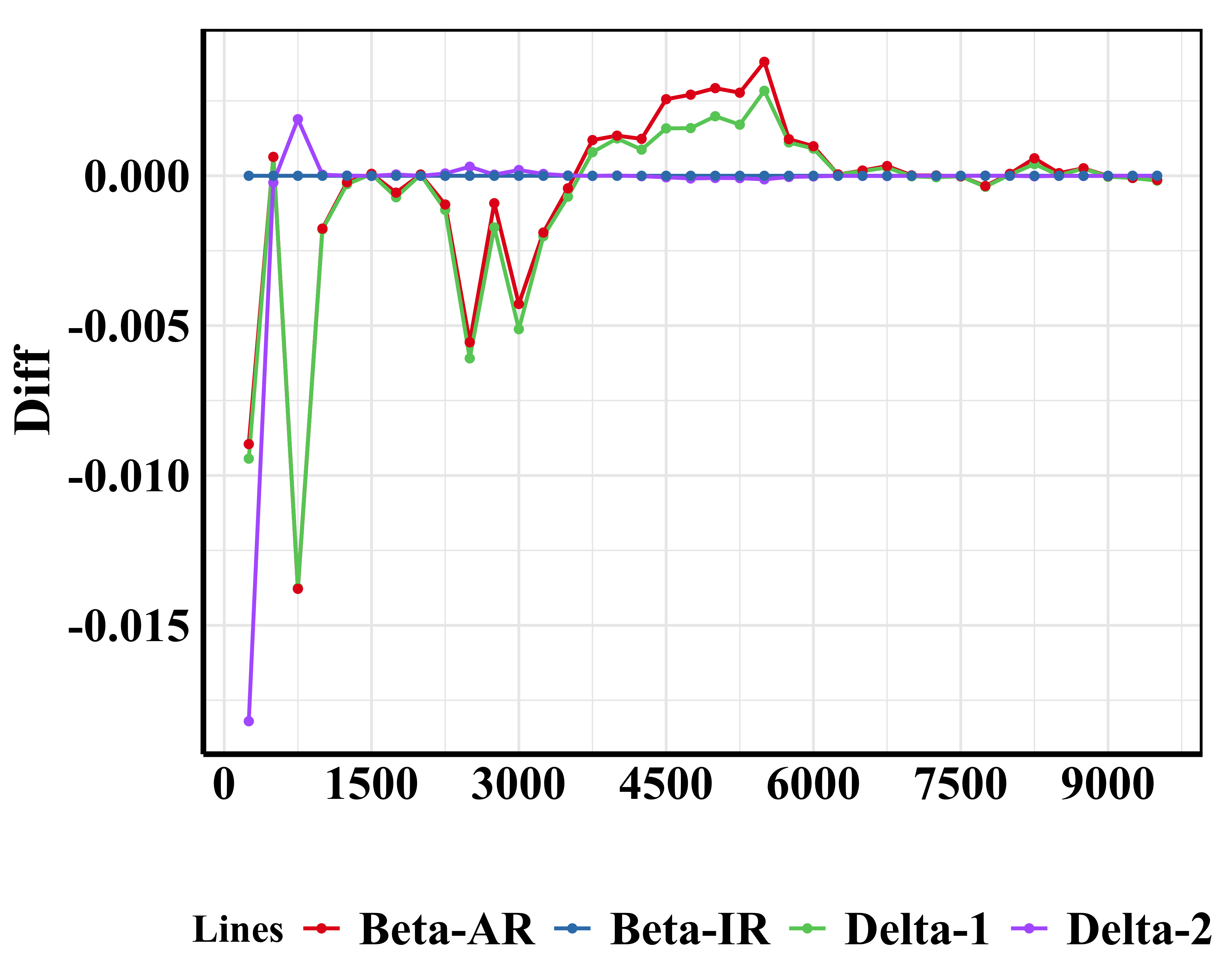}
  }
      \subfigure[Design-based \& $p=0.4$ \& $\text{ATE}=0.1$ \& $\text{HTE}=0.5$]
  {  \label{cond}
   \includegraphics[width=0.225\textwidth]{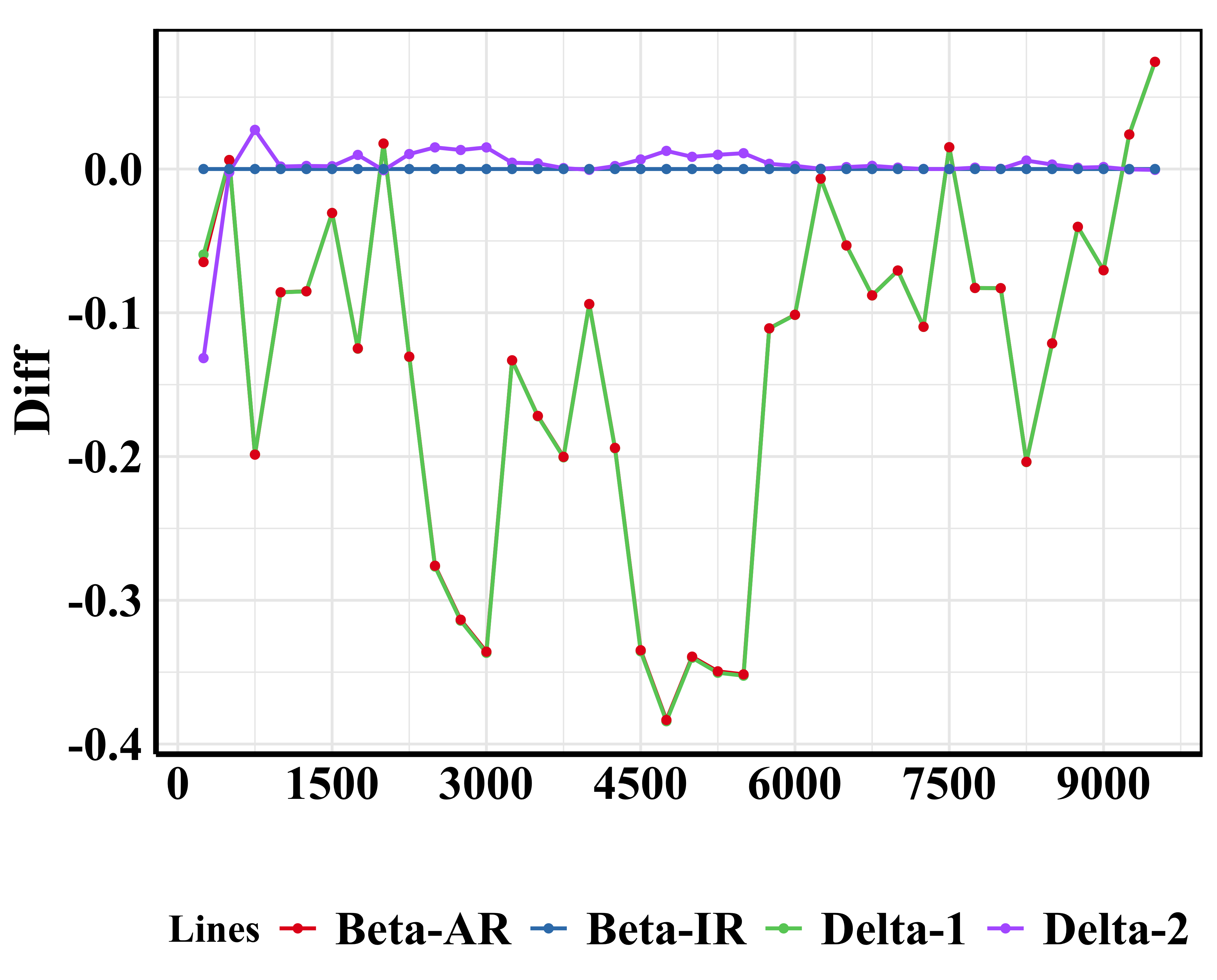}
  }

  \caption{Asymptotically discrepancies between \(\widehat{\delta}_3\) and other estimators, scaled as \(\sqrt n\cdot(\widehat{\text{ATE}} - \widehat{\delta}_3)\). }
  \label{convergence}
\end{figure}

Figure \ref{convergence} illustrates the convergence properties of ATE estimators under design-based setting and other specific configurations. Using \(\widehat{\delta}_3\) as the baseline, we examine differences between \(\widehat{\delta}_3\) and other estimators, with results scaled by \(\sqrt n\) (i.e., $\sqrt n\cdot(\widehat{\text{ATE}}-\widehat\delta_3)$). $\widehat\delta_0$ and $\widehat\beta_T^{SR}$ are omitted due to their excessive variance. Additional results are provided in Appendix \ref{appB}.

Figure \ref{convergence} confirms our earlier findings: $\widehat{\beta}_T^{IR}$ is exactly equal to $\widehat{\delta}_3$, and $\widehat{\delta}_2$ is asymptotically equivalent to $\widehat{\delta}_3$. Furthermore, Figures \ref{cona} and \ref{conb} demonstrate that when $p_t = p_c$ or $\text{HTE} = 0$, both $\widehat{\beta}_T^{AR}$ and $\widehat{\delta}_1$ are asymptotically equivalent to $\widehat{\delta}_3$. However, when neither conditions holds, $\widehat{\beta}_T^{AR}$ and $\widehat{\delta}_1$ remain asymptotically equivalent to each other but diverge from $\widehat{\delta}_3$ (see Figure \ref{cond}).

\begin{figure}[!h] 
  \centering 
  \subfigure[Design-based \& $p=0.5$ \& $\text{ATE}=0.1$ \& $\text{HTE}=0.5$]
  {
   \includegraphics[width=0.22\textwidth]{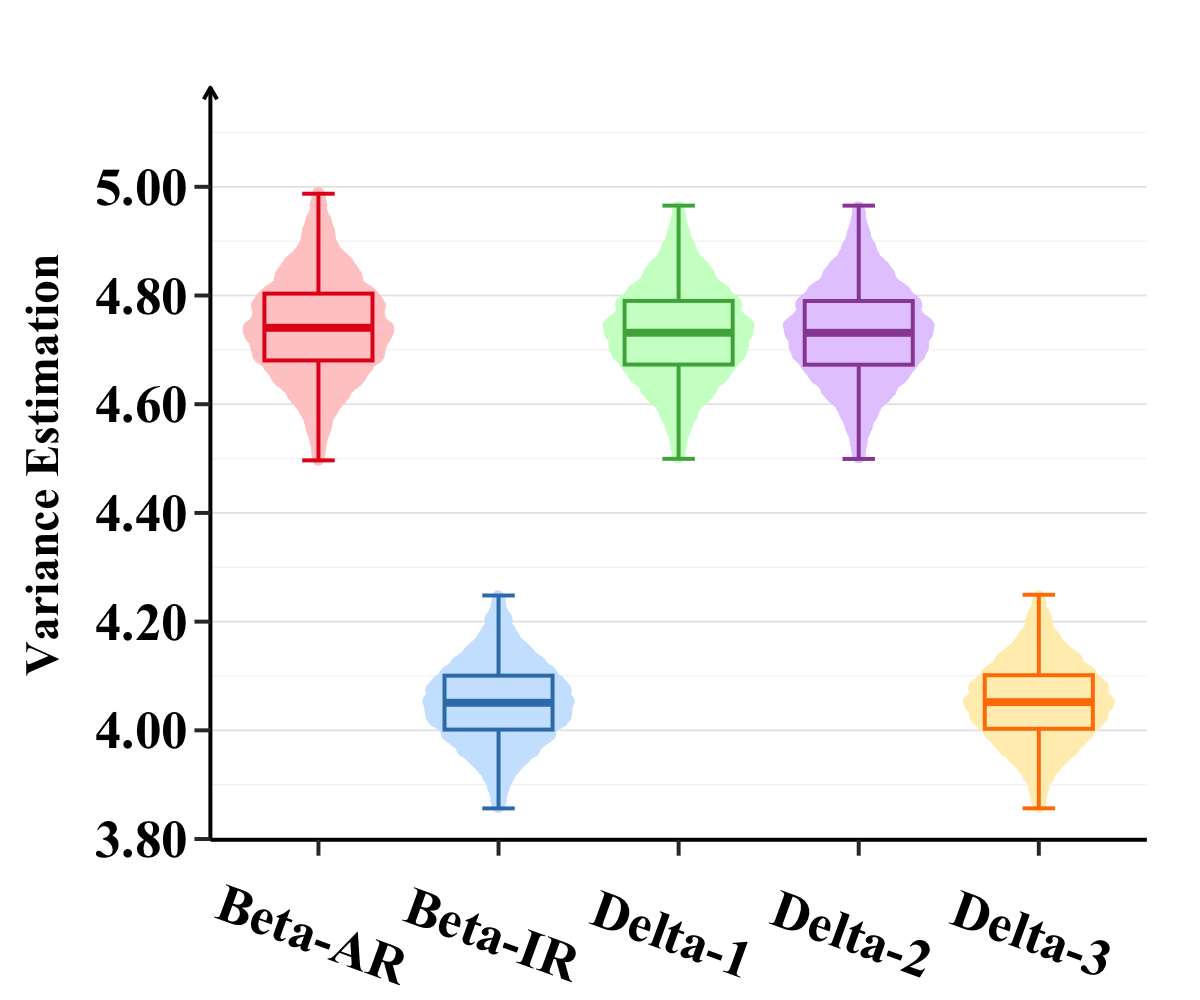}
  }
    \subfigure[Design-based \& $p=0.4$ \& $\text{ATE}=0.1$ \& $\text{HTE}=0.5$]
  {
   \includegraphics[width=0.22\textwidth]{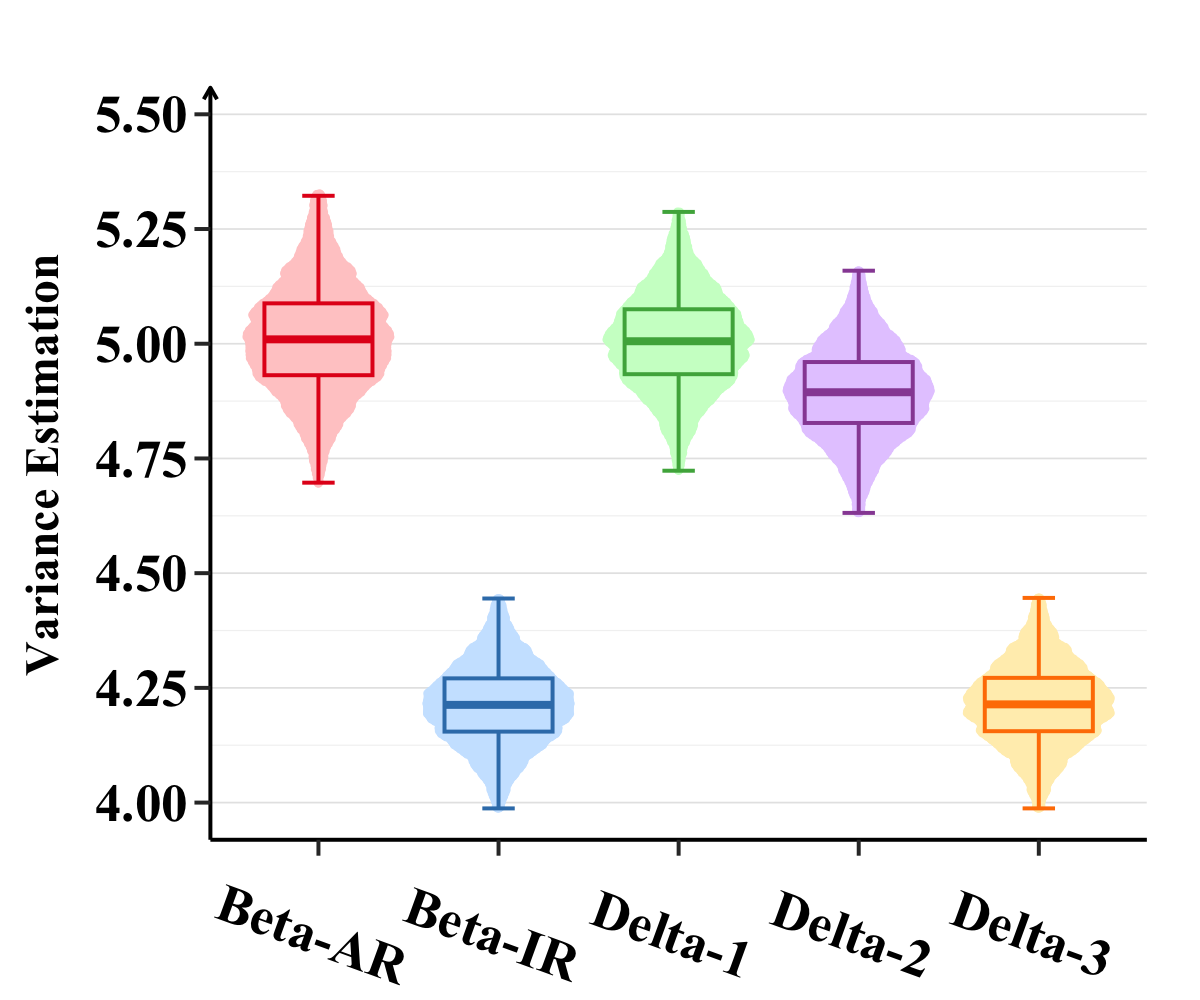}
  }
      \subfigure[Model-based \& $p=0.5$ \& $\text{ATE}=0.1$ \& $\text{HTE}=0.5$]
  {
   \includegraphics[width=0.22\textwidth]{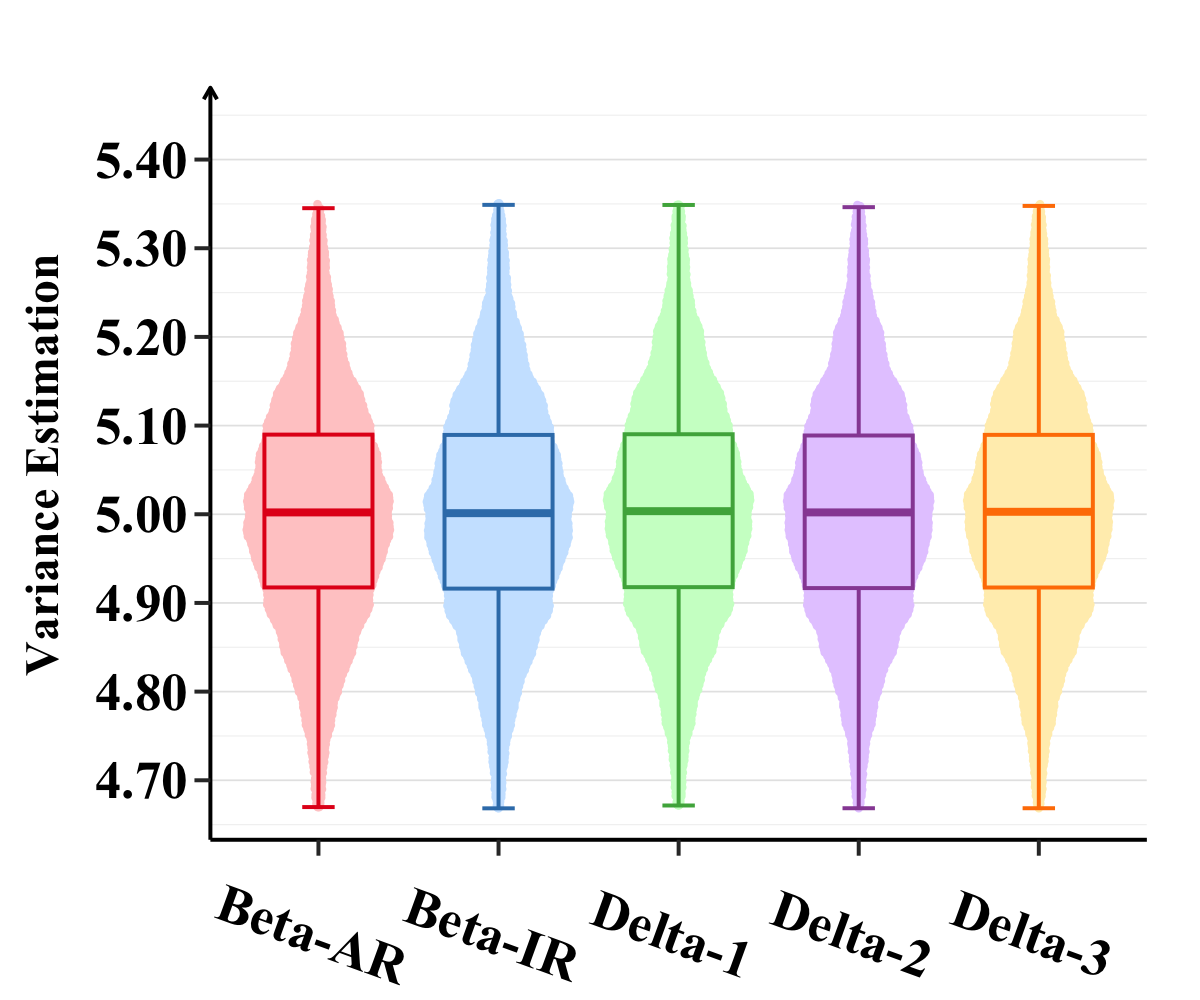}
  }
      \subfigure[Model-based \& $p=0.4$ \& $\text{ATE}=0.1$ \& $\text{HTE}=0.5$]
  {
   \includegraphics[width=0.22\textwidth]{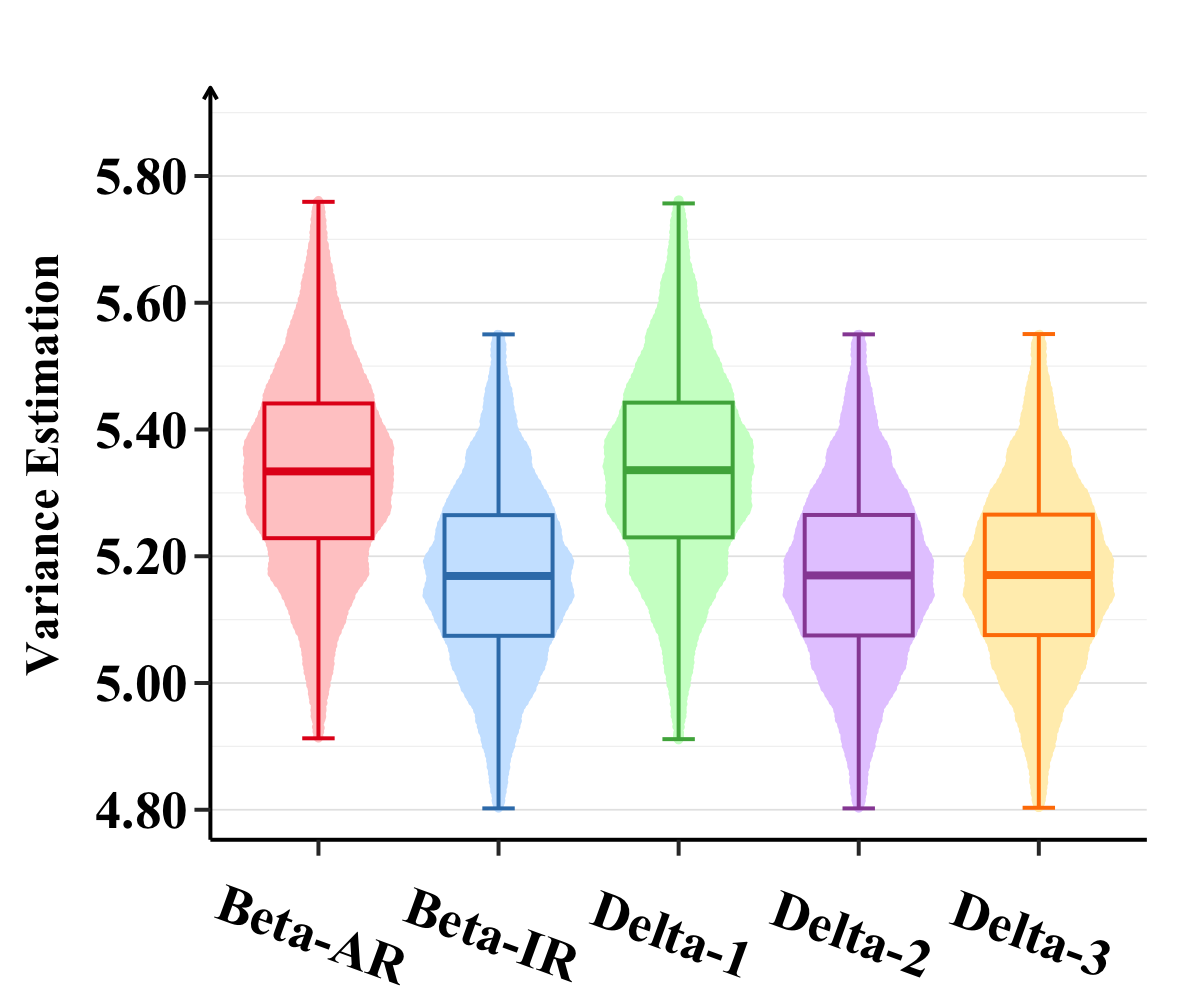}
  }

  \caption{Box plots of the variance estimates of ATE estimators, scaled by \(n \times \widehat{\text{var}}  \). }
  \label{hatvar}
\end{figure}

Figure \ref{hatvar} represents box plots of variance estimates for different ATE estimators. The height of each boxplot (from top to bottom) reflects the stability of the variance estimators, while the median indicates conservatism. Overall, $\widehat{\delta}_3$ and $\widehat{\beta}_T^{IR}$ exhibit the smallest median variance estimates, outperforming other estimators. Specifically, under the design-based framework with HTE, $\widehat{\beta}_T^{IR}$ and $\widehat{\delta}_3$ yield smaller variance estimates regardless of group assignment probabilities. Focusing on \(\widehat{\delta}_1\), \(\widehat{\delta}_2\), and \(\widehat{\delta}_3\), we observe: under the design-based framework, \(\widehat{\delta}_2\) outperforms \(\widehat{\delta}_1\) but lags behind \(\widehat{\delta}_3\); under the model-based framework, \(\widehat{\delta}_2\) and \(\widehat{\delta}_3\) (with variance correction) perform identically, consistent with our theoretical results.

Finally, we validated the Type I error rates and statistical power using simulated datasets, with results in Table \ref{tab}. In the design-based setting, all estimators exhibit relative conservatism, with Type I error rates strictly controlled below $0.05$; in the model-based setup, they demonstrate consistency in error rate control. 
In terms of statistical power, $\widehat{\delta}_3$ and $\widehat{\beta}_T^{IR}$ yield comparable performance across all experimental configurations, irrespective of whether the framework is design-based or model-based. For control variates estimators, their power properties are consistent with theoretical predictions: 
\begin{itemize}[leftmargin=10pt]

\item Within the design-based framework: \(\widehat{\delta}_1\), \(\widehat{\delta}_2\), and \(\widehat{\delta}_3\) are asymptotically equivalent under the conditions of no HTE, regardless of whether group assignment probabilities are equal. In the presence of HTE, \(\widehat{\delta}_3\) outperforms the other two estimators significantly. 

\item Within the model-based framework: \(\widehat{\delta}_2\) and \(\widehat{\delta}_3\) maintain asymptotic equivalence across all scenarios, while \(\widehat{\delta}_1\) achieves asymptotic equivalence with them under conditions of equal group assignments or the absence of HTE. 

\end{itemize}

\subsection{Real data analysis}
We evaluated the performance of control variates estimators via simulation studies leveraging real-world datasets from ByteDance’s experimental platform. To simulate a design-based scenario, we repeatedly partition the original data under two group assignment probabilities: equal assignment and \( p_t = 0.4 \), generating $1,000$ datasets for each scenario.

\begin{figure}[!h]
  \centering 
  \subfigure[Design-based \& $p=0.4 $ \& $ \text{ATE}=9$]
  {  \label{realplot1}
   \includegraphics[width=0.225\textwidth]{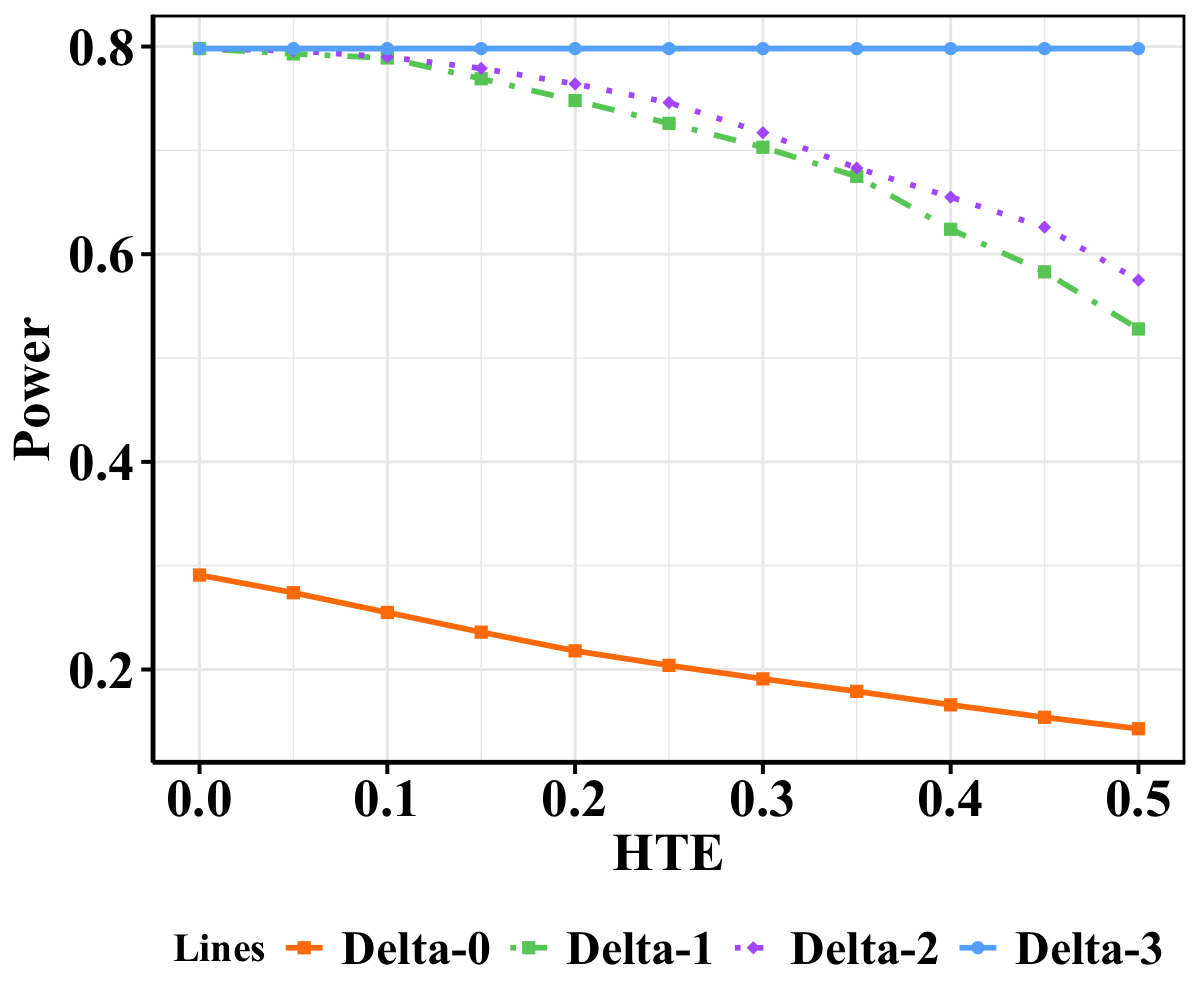}
  }
    \subfigure[Design-based \& $p=0.4 $ \& $ \text{HTE}=0.4$]
  {  \label{realplot2}
   \includegraphics[width=0.225\textwidth]{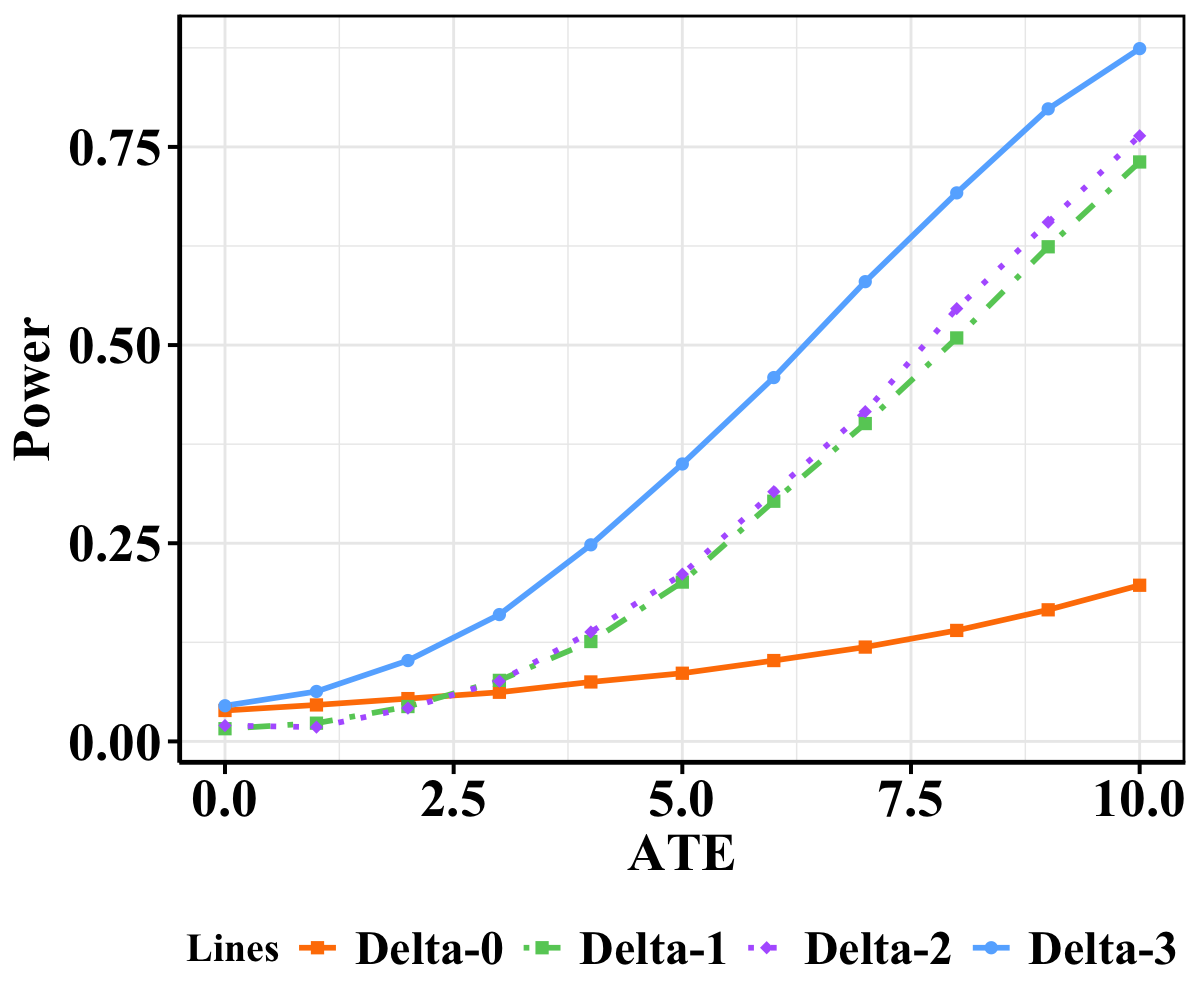}
  }
      \subfigure[Design-based \& $p=0.5 $ \& $ \text{ATE}=9$]
  {  \label{realplot3}
   \includegraphics[width=0.225\textwidth]{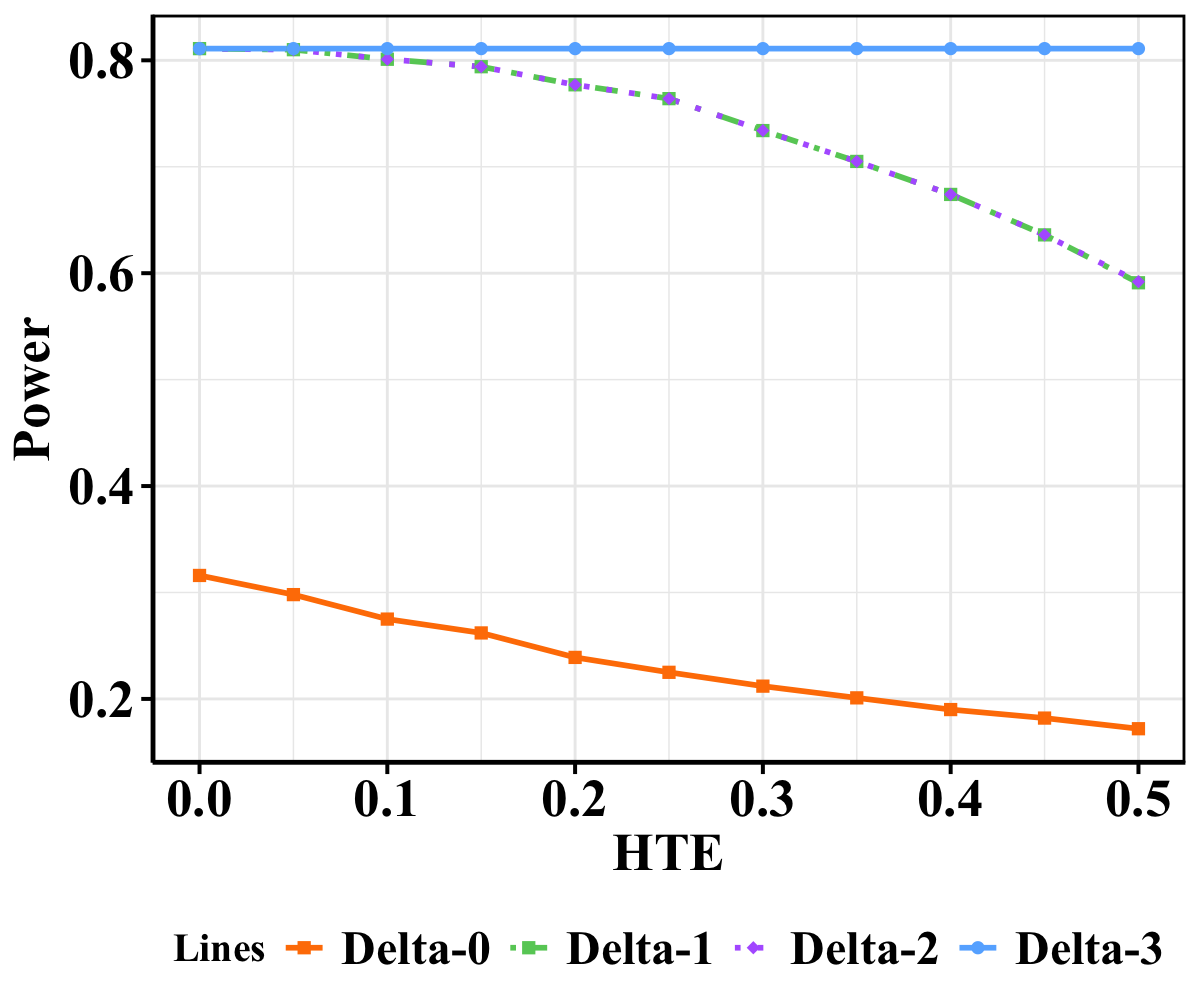}
  }
      \subfigure[Design-based \& $p=0.5 $ \& $ \text{HTE}=0.4$]
  {  \label{realplot4}
   \includegraphics[width=0.225\textwidth]{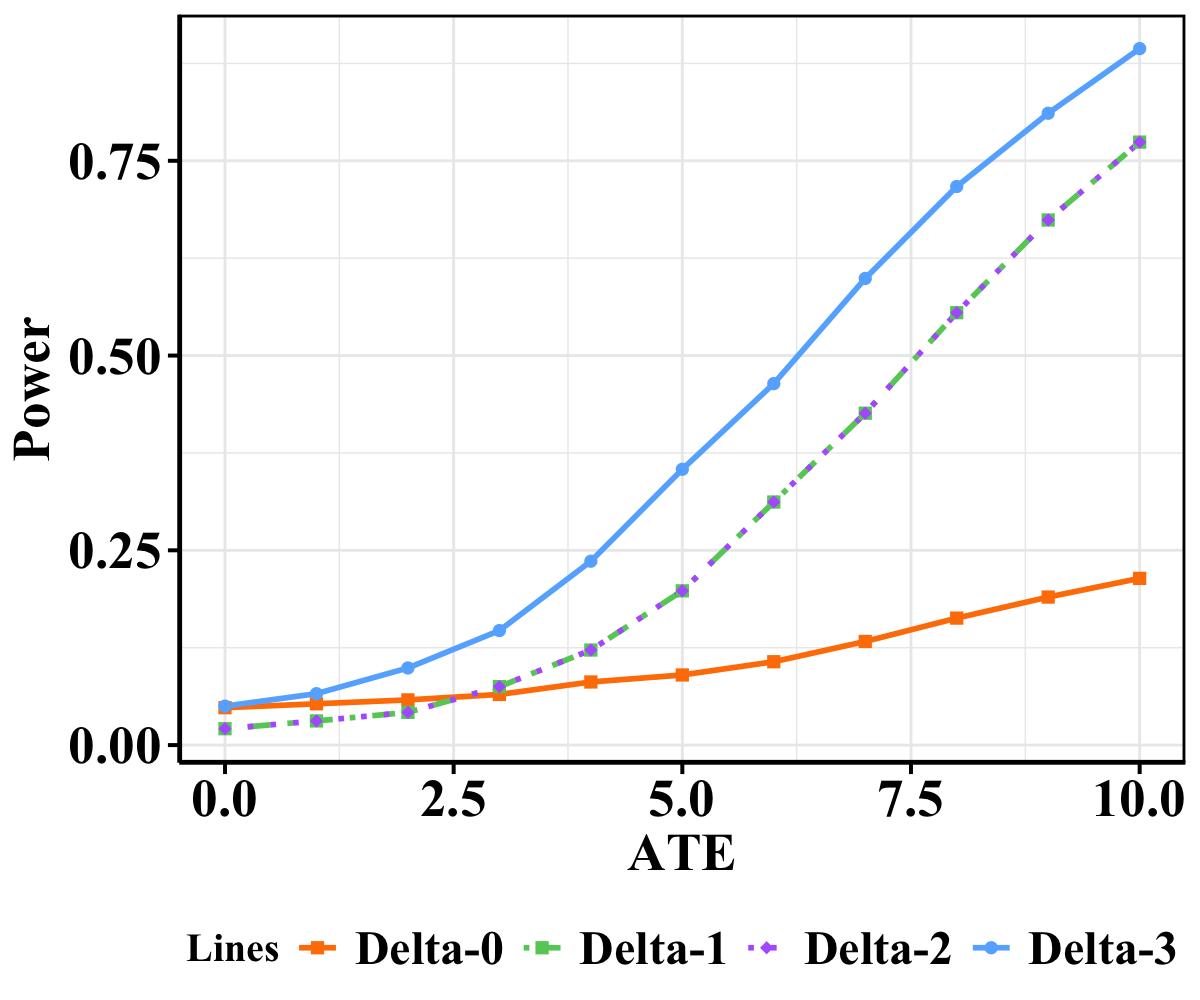}
  }
  \caption{Power - ATE/HTE curves for estimators under varying group assignment probabilities. }
  \label{realplot}
\end{figure}

\noindent \textbf{Type I error}. Consistent with the simulations, all estimators maintained Type I error rates at $0.05$, regardless of treatment allocation or the presence of HTE. 

\noindent \textbf{Statistical power}. To evaluate power, we artificially introduced specific ATE values into the A/A datasets to approximate alternative scenarios, with results in Figure \ref{realplot}. As shown in Figures \ref{realplot1} and \ref{realplot3}, HTE reduces the power of all estimators except $\widehat{\delta}_3$; stronger HTE exacerbates this power loss, particularly under unequal allocation. When HTE is fixed, $\widehat{\delta}_3$ exhibits consistently higher power across all ATE levels. Notably, $\widehat{\delta}_1$ and $\widehat{\delta}_2$ exhibit identical power under the equal group assignments, a property that vanishes under unequal assignments, as established in Theorem \ref{deltavar}.

\begin{table}[!htb]
    \centering
   \caption{Coverage rates and statistical power of $\widehat\delta$ in the model-based simulations using real-world data. }
    \resizebox{0.48\textwidth}{!}{
    \begin{tabular}{lcccccc}
    \toprule
    Metric & $p_t$ & $\widehat\delta_0$ & $\widehat\delta_1$ & $\widehat\delta_2$ & $\widehat\delta_3^{\text{correct}}$ & $\widehat\delta_3^{\text{wrong}}$ \\
    \midrule
    \multirow{2}{*}{Coverage} & 0.5  &  94.78\% & 94.74\% & 94.74\% & 94.76\% & 86.01\% \\
    \cmidrule(r){2-7}
     & 0.3  & 94.90\% & 94.61\% & 94.88\% & 94.86\% & 88.13\%  \\
    \midrule
    \multirow{2}{*}{Power} & 0.5 & 0.6274 & \textbf{0.7816} & \textbf{0.7816} & \textbf{0.7859} & -  \\
    \cmidrule(r){2-7}
    & 0.3 & 0.4655  & 0.6321 & \textbf{0.7407} & \textbf{0.7452} & -\\
    \bottomrule
    \end{tabular}}
    \label{tabreal}
\end{table}

To simulate the model-based scenario, we treated a large real dataset as the super population and repeatedly drew small samples with replacement. We compared coverage rates and power for \(\widehat{\delta}_3\) with (\(\widehat{\delta}_3^{\text{correct}}\)) and without (\(\widehat{\delta}_3^{\text{wrong}}\)) variance correction, alongside other estimators. Results (with artificially introduced HTE) are in Table \ref{tabreal}. Key findings from the model-based experiments include:  

\begin{itemize}[leftmargin=10pt]

\item \(\widehat{\delta}_3^{\text{wrong}}\) underestimated true variance, leading to inflated Type I error rates (via reduced coverage). 

\item \(\widehat{\delta}_2\) and \(\widehat{\delta}_3^{\text{correct}}\) were asymptotically equivalent regardless of group assignment or HTE.

\item Under equal group assignment, \(\widehat{\delta}_1\) was asymptotically equivalent to \(\widehat{\delta}_2\) and \(\widehat{\delta}_3^{\text{correct}}\), which aligns with the results of Theorem \ref{supercom}. 

\end{itemize}

\section{Conclusions and Recommendations}
\label{sec7}

This paper establishes a formal connection between control variates and regression adjustment in A/B testing, with a focus on their performance across design-based and model-based frameworks. First, control variates with group-specific \(\theta\) estimates (\(\widehat{\delta}_3\)) are asymptotically equivalent to regression adjustment with interaction terms, outperforming other estimators in the presence of HTE or unequal allocation. Second, the unadjusted variance estimator for \(\widehat{\delta}_3\) is conservative or consistent in the design-based framework but systematically underestimates the true variance in the model-based setting. 

For practitioners, we recommend that prioritize control variates with group-specific \(\theta\) estimates (\(\widehat{\delta}_3\)), as they balance efficiency and computational feasibility for large-scale online experiments. Regarding variance estimation, the choice of estimator depends on whether the primary focus is on internal validity or external validity. The recommended control variates method has been fully implemented and deployed on ByteDance's experimental platform, yielding consistent improvements in testing sensitivity. 



\begin{acks}

\end{acks}

\bibliographystyle{ACM-Reference-Format}
\bibliography{KDD_template}

\appendix
\onecolumn

\section{Technical Proofs}
\label{appA}

\begin{lemma}[Finite-population version of the Weak Law of Large Numbers]\label{lem1}
Assume some regularity conditions mentioned in \citet{lin2013agnostic}, the means over treatment or control of $Y_i^t, Y_i^c, X_i, (Y_i^t)^2, (Y_i^c)^2, X_i^2, Y_i^tY_i^c, Y_i^tX_i, Y_i^cX_i$ converge in probability to the limits of the population means. 
\end{lemma}
\noindent\textbf{Proof}. See \citet{Lin2012SupplementT} Lemma 1 for details. 

\subsection{Proof of Theorem \ref{equal}}

i). We start by expanding the expression for \(\sqrt{n}(\widehat{\delta}_2-\widehat{\delta}_3)\): 
\begin{equation}
\begin{aligned}
 \sqrt{n}(\widehat{\delta}_2-\widehat{\delta}_3) & = \sqrt{n} \lc\bl \oy_t - \wh{\theta}_{t, 2}( \ox_t-\ox) \br - \bl \oy_c - \wh{\theta}_{c, 2}( \ox_c-\ox) \br - \bl \oy_t - \wh{\theta}_{t, 3}( \ox_t-\ox) \br + \bl \oy_c - \wh{\theta}_{c, 3}( \ox_c-\ox) \br\rc \\
&= \sqrt n \lc (\wt_{t, 3} - \wt_{t, 2})(\ox_t-\ox) + (\wt_{c, 2} - \wt_{c, 3})(\ox_c-\ox) \rc \\
&= \sqrt n (\ox_t-\ox_c)\lc p_c\wt_{t, 3} + p_t\wt_{c, 3}-\frac{p_c\wt_{t, 3}\wv(X_t)+p_t\wt_{c, 3}\wv(X_c)}{p_c\wv(X_t) + p_t\wv(X_c)} \rc.  \nonumber
\end{aligned}
\end{equation}

From Lemma \ref{lem1}, 
$$
p_c\wt_{t, 3} + p_t\wt_{c, 3}-\frac{p_c\wt_{t, 3}\wv(X_t)+p_t\wt_{c, 3}\wv(X_c)}{p_c\wv(X_t) + p_t\wv(X_c)}\p 0 \text{ and } \sqrt n (\ox_t-\ox_c) \overset{d}{\rightarrow} N(0, 2\text{var}(X)). 
$$

Hence, we conclude that $ \sqrt{n}(\widehat{\delta}_2-\widehat{\delta}_3)\overset{p}{\rightarrow}0. $

\noindent ii). Similarly, consider the difference \(\sqrt{n}(\widehat{\delta}_1-\widehat{\delta}_3)\): 
\begin{equation}
\begin{aligned}
 \sqrt{n}(\widehat{\delta}_1-\widehat{\delta}_3) & = \sqrt{n} \lc\bl \oy_t - \wh{\theta}_{t, 1}( \ox_t-\ox) \br - \bl \oy_c - \wh{\theta}_{c, 1}( \ox_c-\ox) \br - \bl \oy_t - \wh{\theta}_{t, 3}( \ox_t-\ox) \br + \bl \oy_c - \wh{\theta}_{c, 3}( \ox_c-\ox) \br\rc \\
&= \sqrt n \lc (\wt_{t, 3} - \wt_{t, 1})(\ox_t-\ox) + (\wt_{c, 1} - \wt_{c, 3})(\ox_t-\ox) \rc \\
&= \sqrt n (\ox_t-\ox_c)\lc p_c\wt_{t, 3} + p_t\wt_{c, 3}-\frac{\wc(Y, X)}{\wv(X)} \rc  \nonumber
\end{aligned}
\end{equation}

From Lemma \ref{lem1}, 
$$
\frac{\wc(Y, X)}{\wv(X)}\p p_t\theta_t+p_c\theta_c \text{ and } p_c\wt_{t, 3} + p_t\wt_{c, 3}\p p_c\theta_t+p_t\theta_c. 
$$

Therefore, $ \sqrt{n}(\widehat{\delta}_1-\widehat{\delta}_3)\overset{p}{\rightarrow}0$ holds when \(p_t=p_c\) or \(\theta_t=\theta_c\).

\subsection{Proof of Theorem \ref{deltavar}}

i). We begin by examining the difference in variances: 
\begin{equation}
\begin{aligned}
n\lc\wv(\wdd_1) -\wv(\wdd_2)\rc&=\frac{1}{p_t} \lc 2(\wt_{t, 2}-\wt_{t, 1})\wc(Y_t, X_t) + (\wt_{t, 1}^2-\wt_{t, 2}^2)\wv(X_t)\rc + \frac{1}{p_c} \lc 2(\wt_{c, 2}-\wt_{c, 1})\wc(Y_c, X_c) + (\wt_{c, 1}^2-\wt_{c, 2}^2)\wv(X_c)\rc\\
&=\frac{\wv(X_t)}{p_t} \lc 2(\wt_{t, 2}-\wt_{t, 1})\wt_{t, 3} + \wt_{t, 1}^2-\wt_{t, 2}^2\rc + \frac{ \wv(X_c)}{p_c} \lc 2(\wt_{c, 2}-\wt_{c, 1})\wt_{c, 3} + \wt_{c, 1}^2-\wt_{c, 2}^2\rc.  \nonumber
\end{aligned}
\end{equation}

Then, 
\begin{equation}
\begin{aligned}
\lim_{n\rightarrow \infty} n\lc\wv(\wdd_1) -\wv(\wdd_2)\rc&=\lc\frac{\theta_{1}+\theta_2-2\theta_t}{p_t} + \frac{\theta_{1}+\theta_2-2\theta_c}{p_c}\rc(\theta_1-\theta_2)\text{var}(X)\\
&=\lc\frac{\theta_c-\theta_t}{p_t} + \frac{\theta_t-\theta_c}{p_c}\rc(\theta_1-\theta_2)\text{var}(X)\\
&=\frac{1}{p_tp_c} (\theta_1-\theta_2)^2\text{var}(X). \nonumber
\end{aligned}
\end{equation}

This implies: 
\begin{equation}
\begin{aligned}
\lim_{n\rightarrow \infty} n\lc\wv(\wdd_1) -\wv(\wdd_2)\rc=0 
\iff \theta_1=\theta_2 
\iff p_t\theta_t+p_c\theta_c=p_c\theta_t+p_t\theta_c \iff (p_t-p_c)(\theta_t-\theta_c)=0 \iff p_t=p_c \text{ or } \theta_t=\theta_c. \nonumber
\end{aligned}
\end{equation}

\noindent ii). Next, consider the difference between \(\wdd_1\) and \(\wdd_3\): 
\begin{equation}
\begin{aligned}
n\lc\wv(\wdd_1) -\wv(\wdd_3)\rc&=\frac{1}{p_t} \lc 2(\wt_{t, 3}-\wt_{t, 1})\wc(Y_t, X_t) + (\wt_{t, 1}^2-\wt_{t, 3}^2)\wv(X_t)\rc + \frac{1}{p_c} \lc 2(\wt_{c, 3}-\wt_{c, 1})\wc(Y_c, X_c) + (\wt_{c, 1}^2-\wt_{c, 3}^2)\wv(X_c)\rc\\
&=\frac{\wv(X_t)}{p_t} \lc 2(\wt_{t, 3}-\wt_{t, 1})\wt_{t, 3} + \wt_{t, 1}^2-\wt_{t, 3}^2\rc + \frac{ \wv(X_c)}{p_c} \lc 2(\wt_{c, 3}-\wt_{c, 1})\wt_{c, 3} + \wt_{c, 1}^2-\wt_{c, 3}^2\rc\\
&=\frac{\wv(X_t)}{p_t} \lc \wt_{t, 1}-\wt_{t, 3}\rc^2 + \frac{ \wv(X_c)}{p_c} \lc \wt_{c, 1}-\wt_{c, 3}\rc^2. \nonumber
\end{aligned}
\end{equation}

Then, 
$$
\lim_{n\rightarrow \infty} n\lc\wv(\wdd_1) -\wv(\wdd_3)\rc=\lc\frac{(\theta_{1}-\theta_t)^2}{p_t} + \frac{(\theta_{1}-\theta_c)^2}{p_c}\rc\text{var}(X). 
$$

Hence, 
$$
\lim_{n\rightarrow \infty} n\lc\wv(\wdd_1) -\wv(\wdd_3)\rc\ge 0, \text{with strict inequality unless } \theta_t=\theta_c. 
$$

\noindent iii). Finally, consider the difference between \(\wdd_2\) and \(\wdd_3\): 
\begin{equation}
\begin{aligned}
n\lc\wv(\wdd_2) -\wv(\wdd_3)\rc&=\frac{1}{p_t} \lc 2(\wt_{t, 3}-\wt_{t, 2})\wc(Y_t, X_t) + (\wt_{t, 2}^2-\wt_{t, 3}^2)\wv(X_t)\rc + \frac{1}{p_c} \lc 2(\wt_{c, 3}-\wt_{c, 2})\wc(Y_c, X_c) + (\wt_{c, 2}^2-\wt_{c, 3}^2)\wv(X_c)\rc\\
&=\frac{\wv(X_t)}{p_t} \lc 2(\wt_{t, 3}-\wt_{t, 2})\wt_{t, 3} + \wt_{t, 2}^2-\wt_{t, 3}^2\rc + \frac{ \wv(X_c)}{p_c} \lc 2(\wt_{c, 3}-\wt_{c, 2})\wt_{c, 3} + \wt_{c, 2}^2-\wt_{c, 3}^2\rc\\
&=\frac{\wv(X_t)}{p_t} \lc \wt_{t, 2}-\wt_{t, 3}\rc^2 + \frac{ \wv(X_c)}{p_c} \lc \wt_{c, 2}-\wt_{c, 3}\rc^2. \nonumber
\end{aligned}
\end{equation}

Then, 
$$
\lim_{n\rightarrow \infty} n\lc\wv(\wdd_2) -\wv(\wdd_3)\rc=\lc\frac{(\theta_{2}-\theta_t)^2}{p_t} + \frac{(\theta_{2}-\theta_c)^2}{p_c}\rc\text{var}(X)=(\theta_t-\theta_c)^2\text{var}(X).
$$

Thus, 
$$
\lim_{n\rightarrow \infty} n\lc\wv(\wdd_2) -\wv(\wdd_3)\rc\ge 0, \text{with strict inequality unless } \theta_t=\theta_c. 
$$

\subsection{Proof of Theorem \ref{reg_com}}
The results presented in Theorem \ref{reg_com} follow as a direct corollary of Theorem \ref{deltavar} and Theorem \ref{finivar}.

\subsection{Proof of Theorem \ref{rela1}}

i). First, consider the simple linear model without covariates, let $Y=\beta_0+T\cdot\beta_T+\varepsilon$, then we have: 

$$\wb^{SR}_T=\oy_t-\oy_c=\frac{1}{n_t}\sum_{i=1}^n T_iY_i - \frac{1}{n_c} \sum_{i=1}^n (1-T_i)Y_i=\wdd_0. $$

\noindent ii). Next, extend to the model with an additional covariate $X - \ox$, let $Y = \beta_0 + T\cdot\beta_T + (X-\ox)\cdot\beta_X + \varepsilon$. 

The design matrix product \(\mathcal X'\mathcal X\) takes the form: 
$$
\mathcal X'\mathcal X= \begin{pmatrix}
n & n_t & 0 \\
n_t & n_t & n_t(\ox_t-\overline{X}) \\
0 & n_t(\ox_t-\overline{X}) & (X-\overline{X})' (X-\overline{X})
\end{pmatrix}. 
$$

The estimator for \(\wb_T^{AR}\) is derived as: 

$$
\wb_T^{AR}=\frac{1}{|\mathcal X'\mathcal X|}\lc-n_tn(n-1)\wv(X)\oy + n_tn(n-1)\wv(X)\oy_t - n_tn(\ox_t-\ox)(X-\ox)'Y\rc, 
$$
where $|\mathcal X'\mathcal X|=n_tn_c(n-1)\wv(X) - nn_t^2(\ox_t-\ox)^2$. 

Rewriting $\wdd_1=\oy_t-\oy_c-\wt_1(\ox_t-\ox_c)$, we now show the asymptotic equivalence between \(\wb_T^{AR}\) and \(\wdd_1\): 

\begin{equation}
\begin{aligned}
\sqrt{n} (\wb_T^{AR} - \wdd_1)&=\frac{\sqrt{n}}{|\mathcal X' \mathcal X|/(n_tn(n-1))} \lc p_c(\oy_t-\oy_c)\wv(X)-(\ox_t-\ox)\wc(X, Y) - \Big((\oy_t-\oy_c) - \wt_1(\ox_t - \ox_c)\Big)\Big(p_c\wv(X) - \frac{n_t^2n}{n_tn_c(n-1)}(\ox_t-\ox)^2\Big)\rc \\
&=\frac{1}{|\mathcal X' \mathcal X|/(n_tn(n-1))} \Big((\oy_t-\oy_c) - \wt_1(\ox_t - \ox_c)\Big)\frac{n_t^2n}{n_tn_c(n-1)} \cdot \sqrt{n}(\ox_t-\ox)^2 \overset{p}{\rightarrow} 0. \nonumber
\end{aligned}
\end{equation}

\noindent iv). Finally, consider the interactive model $Y=(1-T)\cdot \beta_0 + T\cdot \beta_1 + (X-\ox)\cdot\beta_X + T\cdot(X-\ox)\cdot\beta_I + \varepsilon$. 

Define the sum of squared residuals function: 

Let $f(\beta_0, \beta_1, \beta_X, \beta_I) = \left\|Y-\Big((1-T)\cdot \beta_0 + T\cdot \beta_1 + (X-\ox)\cdot\beta_X + T\cdot(X-\ox)\cdot\beta_I\Big)\right\|^2 $. 

Taking partial derivatives with respect to the parameters and setting them to zero gives the first-order conditions: 

\begin{equation}
\left\{
\begin{aligned}
& \frac{\partial f}{\partial \beta_0}(\beta_0, \beta_1, \beta_X, \beta_I)=-2\sum\limits_{i=1}^n(1-T_i)\lc Y_i-\Big((1-T_i)\cdot\beta_0 + T_i\beta_1 + (X_i-\ox)\beta_X + T_i(X_i-\ox)\beta_I\Big)\rc =0, \\
& \frac{\partial f}{\partial \beta_1}(\beta_0, \beta_1, \beta_X, \beta_I)= -2\sum\limits_{i=1}^n T_i\lc Y_i-\Big((1-T_i)\cdot\beta_0 + T_i\beta_1 + (X_i-\ox)\beta_X + T_i(X_i-\ox)\beta_I\Big)\rc =0, \\
& \frac{\partial f}{\partial \beta_X}(\beta_0, \beta_1, \beta_X, \beta_I)=-2\sum\limits_{i=1}^n(X_i-\ox)\lc Y_i-\Big((1-T_i)\cdot\beta_0 + T_i\beta_1 + (X_i-\ox)\beta_X + T_i(X_i-\ox)\beta_I\Big)\rc =0, \\
& \frac{\partial f}{\partial \beta_I}(\beta_0, \beta_1, \beta_X, \beta_I)= -2\sum\limits_{i=1}^n T_i(X_i-\ox)\lc Y_i-\Big((1-T_i)\cdot\beta_0 + T_i\beta_1 + (X_i-\ox)\beta_X + T_i(X_i-\ox)\beta_I\Big)\rc =0. \nonumber
\end{aligned}
\right.
\end{equation}

\begin{equation}
\iff \left\{
\begin{aligned}
&\sum\limits_{i=1}^n(1-T_i)\lc Y_i-\beta_0 - (X_i-\ox)\beta_X\rc =0, \\
&\sum\limits_{i=1}^nT_i\lc Y_i-\beta_1 - (X_i-\ox)(\beta_X + \beta_I)\rc =0, \\
&\sum\limits_{i=1}^n(1-T_i)(X_i-\ox)\lc Y_i-\beta_0 - (X_i-\ox)\beta_X\rc =0, \\ 
&\sum\limits_{i=1}^nT_i(X_i-\ox) \lc Y_i-\beta_1 - (X_i-\ox)(\beta_X + \beta_I)\rc =0. \nonumber
\end{aligned}
\right.
\end{equation}

Solving this system and computing the difference between \(\wb_1\) and \(\wb_0\) leads to: 
 
$$
\wb_1-\wb_0=\lc\oy_t-\wt_{t, 3}(\ox_t-\ox)\rc - \lc \oy_c - \wt_{c, 3}(\ox_c-\ox)\rc=\wdd_3. 
$$

\subsection{Proof of Theorem \ref{finivar}}

i). First, consider the EHW estimator for the variance of the simple regression model, where the design matrix \(\mathcal X = \left(\bm{1}\quad T\right)\): 

$$
\begin{aligned}
\widehat{\Sigma}&=(\mathcal X'\mathcal X)^{-1}(\mathcal X' \text{diag}(\widehat{\varepsilon}_1^2, \ldots, \widehat{\varepsilon}_n^2)\mathcal X)(\mathcal X'\mathcal X)^{-1}, \\
&=
\begin{pmatrix}
\frac{1}{n_c^2}\sum\limits_{i=1}^n (1-T_i)\wvv_i^2 & -\frac{1}{n_c^2}\sum\limits_{i=1}^n (1-T_i)\wvv_i^2  \\
-\frac{1}{n_c^2}\sum\limits_{i=1}^n (1-T_i)\wvv_i^2  & \frac{1}{n_t^2}\sum\limits_{i=1}^n T_i\wvv_i^2 + \frac{1}{n_c^2}\sum\limits_{i=1}^n (1-T_i)\wvv_i^2
\end{pmatrix}. 
\end{aligned}
$$

Extracting the variance of \(\wb_T^{SR}\) and scaling by n gives: 

\begin{equation*}
\begin{aligned}
n\wv(\wb_T^{SR})&=\frac{1}{p_t n_t}\sum\limits_{i=1}^n T_i\wvv_i^2 + \frac{1}{p_c n_c}\sum\limits_{i=1}^n (1-T_i)\wvv_i^2 \\
&=\frac{1}{p_t n_t}\sum\limits_{i=1}^n T_i(Y_i-\oy_t)^2 + \frac{1}{p_c n_c}\sum\limits_{i=1}^n (1-T_i)(Y_i-\oy_c)^2 \\
&=n\wv(\wdd_0). 
\end{aligned}
\end{equation*}

\noindent ii). Next, extend to the adjusted regression model with covariate $(X-\ox)$, where \(\mathcal X = \left(\bm{1}\quad T \quad (X-\ox)\right)\). The EHW variance estimator becomes: 

$$
\begin{aligned}
\widehat{\Sigma}&=(\mathcal X'\mathcal X)^{-1}(\mathcal X' \text{diag}(\widehat{\varepsilon}_1^2, \ldots, \widehat{\varepsilon}_n^2)\mathcal X)(\mathcal X'\mathcal X)^{-1} \\
&=(\mathcal X'\mathcal X)^{-1} 
\begin{pmatrix}
\sum\limits_{i=1}^n\wvv_i^2 & \sum\limits_{i=1}^nT_i\wvv_i^2 & \sum\limits_{i=1}^n(X_i-\ox)\wvv_i^2  \\
\sum\limits_{i=1}^nT_i\wvv_i^2 & \sum\limits_{i=1}^nT_i\wvv_i^2 & \sum\limits_{i=1}^nT_i(X_i-\ox)\wvv_i^2 \\
\sum\limits_{i=1}^n(X_i-\ox)\wvv_i^2 &\sum\limits_{i=1}^nT_i(X_i-\ox)\wvv_i^2 & \sum\limits_{i=1}^nT_i(X_i-\ox)^2\wvv_i^2
\end{pmatrix} (\mathcal X'\mathcal X)^{-1} , 
\end{aligned}
$$

where 
$$
\begin{aligned}
(\mathcal X'\mathcal X)^{-1}=\frac{1}{|\mathcal X'\mathcal X|}
\begin{pmatrix}
n_t(n-1)\wv(X) - n_t^2(\ox_t-\ox)^2 & -n_t(n-1)\wv(X) & n_t^2(\ox_t-\ox)  \\
-n_t(n-1)\wv(X) & n(n-1)\wv(X) & -n_tn(\ox_t-\ox) \\
n_t^2(\ox_t-\ox) & -n_tn(\ox_t-\ox) & n_tn_c
\end{pmatrix}. 
\end{aligned}
$$

The scaled variance of \(\wb_T^{AR}\) is derived as:
\begin{equation*}
\begin{aligned}
n\wv(\wb_T^{AR})&=\frac{1}{p_tn_t}\sum\limits_{i=1}^nT_i\wvv_i^2 + \frac{1}{p_cn_c}\sum\limits_{i=1}^n(1-T_i)\wvv_i^2 \\
&=\frac{1}{p_tn_t}\sum\limits_{i=1}^nT_i\lc Y_i-\wb_0-\wb_T^{AR}-(X_i-\ox)\wb_X\rc^2 + \frac{1}{p_cn_c}\sum\limits_{i=1}^n(1-T_i)\lc Y_i-\wb_0-(X_i-\ox)\wb_X\rc^2, 
\end{aligned}
\end{equation*}

where 
\begin{equation*}
\begin{aligned}
\begin{pmatrix}
\wb_0 \\
\wb_T^{AR} \\
\wb_X
\end{pmatrix} =\frac{1}{|\mathcal X'\mathcal X|}
\begin{pmatrix}
n_tn_c(n-1)\oy_c\wv(X) - n_t^2n\oy(\ox_t-\ox)^2 + n_t^2(n-1)(\ox_t-\ox)\wc(X, Y)\\
n_tn_c(n-1)(\oy_t-\oy_c)(\ox_t-\ox)^2-n_tn(n-1)(\ox_t-\ox)\wc(X, Y) \\
n_t^2n_c(\ox_t-\ox)(\oy_c-\oy_t) + n_tn_c(n-1)\wc(X, Y)
\end{pmatrix}. 
\end{aligned}
\end{equation*}

Hence, 
\begin{equation*}
\begin{aligned}
&\quad \lim_{n\rightarrow \infty} n\lc\wv(\wb_T^{AR}) - \wv(\wdd_1)\rc \\
&= \lim_{n\rightarrow \infty} \lc\frac{1}{p_tn_t}\sum_{i=1}^nT_i \lc (\oy_t-\wb_0-\wb_T^{AR}) + (\wt_1-\wb_X)(X_i-\ox)\rc\lc 2Y_i-\oy_t-\wb_0-\wb_T^{AR}-(\wb_X + \wt_1)(X_i-\ox) + \wt_1(\ox_t-\ox)\rc \right.\\
&\quad + \left. \frac{1}{p_cn_c}\sum_{i=1}^n(1-T_i) \lc (\oy_c-\wb_0) + (\wt_1-\wb_X)(X_i-\ox)\rc \lc 2Y_i-\oy_c-\wb_0-(\wb_X + \wt_1)(X_i-\ox) + \wt_1(\ox_c-\ox) \rc\rc \\
& = 0. 
\end{aligned}
\end{equation*}

\noindent iii). Finally, analyze the interactive regression model with design matrix \(\mathcal X = \left(1-T\quad T \quad (X-\ox) \quad T(X-\ox)\right)\). 

$$
\begin{aligned}
\lim_{n\rightarrow \infty}n\widehat{\Sigma}&=\lim_{n\rightarrow \infty}n(\mathcal X'\mathcal X)^{-1}(\mathcal X' \text{diag}(\widehat{\varepsilon}_1^2, \ldots, \widehat{\varepsilon}_n^2)\mathcal X)(\mathcal X'\mathcal X)^{-1} \\
&=\lim_{n\rightarrow \infty}n(\mathcal X'\mathcal X)^{-1}\lc\frac{1}{n}\mathcal X' \text{diag}(\widehat{\varepsilon}_1^2, \ldots, \widehat{\varepsilon}_n^2)\mathcal X\rc n(\mathcal X'\mathcal X)^{-1}\\
&=\lc\lim_{n\rightarrow \infty}n(\mathcal X'\mathcal X)^{-1}\rc \lc\lim_{n\rightarrow \infty}\frac{1}{n}\mathcal X' \text{diag}(\widehat{\varepsilon}_1^2, \ldots, \widehat{\varepsilon}_n^2)\mathcal X\rc \lc\lim_{n\rightarrow \infty}n(\mathcal X'\mathcal X)^{-1}\rc. 
\end{aligned}
$$

The asymptotic limit of the scaled design matrix product is: 

$$
\begin{aligned}
\lim_{n\rightarrow \infty}\frac{1}{n}\mathcal X'\mathcal X&=
\frac{1}{n}\begin{pmatrix}
p_c & 0 & 0 & 0 \\ 
0& p_t & 0 & 0 \\ 
0 & 0 & \text{var}(X) & \lim\limits_{n\rightarrow \infty} \sum\limits_{i=1}^n(X_i-\overline{X})^2T_i \\ 
0 & 0 & \lim\limits_{n\rightarrow \infty}\sum\limits_{i=1}^n(X_i-\overline{X})^2T_i & \lim\limits_{n\rightarrow \infty} \sum\limits_{i=1}^n(X_i-\overline{X})^2T_i 
\end{pmatrix}. 
\end{aligned}
$$

The scaled weighted design matrix product (with squared residuals) is: 

$$
\begin{aligned}
&\quad \frac{1}{n}\mathcal X' \text{diag}(\widehat{\varepsilon}_1^2, \ldots, \widehat{\varepsilon}_n^2)\mathcal X\\
&=\frac{1}{n}
\begin{pmatrix}
(1-T)' \\ 
T' \\ 
(X-\ox)' \\ 
T(X-\ox)' 
\end{pmatrix}
\begin{pmatrix}\widehat{\varepsilon}_1^2&0&\cdots&0 \\ 0&\widehat{\varepsilon}_2^2&\cdots&0 \\ \vdots&\vdots&&\vdots \\ 0&0&\cdots&\widehat{\varepsilon}_n^2\end{pmatrix}
\begin{pmatrix}
1-T \quad T  \quad X-\ox \quad T(X-\ox)
\end{pmatrix}\\
&=\frac{1}{n}
\begin{pmatrix}
\sum\limits_{i=1}^{n}(1-T_i)\widehat{\varepsilon}_i^2&0&\sum\limits_{i=1}^{n}(1-T_i)(X_i-\overline{X})\widehat{\varepsilon}_i^2&\sum\limits_{i=1}^{n}(1-T_i)(X_i-\overline{X})\widehat{\varepsilon}_i^2 \\ 
0&\sum\limits_{i=1}^{n}T_i\widehat{\varepsilon}_i^2&\sum\limits_{i=1}^{n}T_i(X_i-\overline{X})\widehat{\varepsilon}_i^2&0 \\ 
\sum\limits_{i=1}^{n}(1-T_i)(X_i-\overline{X})\widehat{\varepsilon}_i^2&\sum\limits_{i=1}^{n}T_i(X_i-\overline{X})\widehat{\varepsilon}_i^2&\sum\limits_{i=1}^{n}(X_i-\overline{X})^2\widehat{\varepsilon}_i^2&\sum\limits_{i=1}^{n}(1-T_i)(X_i-\overline{X})^2\widehat{\varepsilon}_i^2 \\ 
\sum\limits_{i=1}^{n}(1-T_i)(X_i-\overline{X})\widehat{\varepsilon}_i^2&0&\sum\limits_{i=1}^{n}(1-T_i)(X_i-\overline{X})^2\widehat{\varepsilon}_i^2&\sum\limits_{i=1}^{n}(1-T_i)(X_i-\overline{X})^2\widehat{\varepsilon}_i^2
\end{pmatrix}. 
\end{aligned}
$$

Combining these results, the asymptotic estimated variance of \(\wb_T^{IR}\) matches that of \(\wdd_3\): 
$$
\begin{aligned}
\lim_{n\rightarrow \infty} n\wv(\wb_T^{IR}) &= \lim_{n\rightarrow \infty}\lc\frac{1}{p_tn_t}\sum_{i=1}^nT_i\wvv_i^2 + \frac{1}{p_cn_c}\sum_{i=1}^n(1-T_i)\wvv_i^2\rc \\
& =\lim_{n\rightarrow \infty}\lc\frac{1}{p_tn_t}\sum_{i=1}^n(Y_i-\oy_t-\widehat{\theta}_t(X_i-\ox))^2+\frac{1}{p_cn_c}\sum_{i=1}^n(Y_i-\oy_c-\widehat{\theta}_c(X_i-\ox))^2\rc\\
& =\lim_{n\rightarrow \infty} n\wv(\wdd_3). 
\end{aligned}
$$

\begin{lemma}\label{lem2}
Within the model-based framework, the randomness of sampling has no impact on the variance of $\wdd_0$, i.e., 
$$\text{var}(\wdd_0)=\frac{\text{var}(Y_t)}{n_t} + \frac{\text{var}(Y_c)}{n_c}. $$
\end{lemma}

\noindent \textbf{Proof}. 
$$
\begin{aligned}
\text{var}(\wdd_0) &= \mathbb E(\wdd_0^2) - (\mathbb E(\wdd_0))^2\\
&=\mathbb E\lc\sum_{i=1}^n\sum_{j=1}^n\Big(\frac{T_i}{n_t}Y_i^t-\frac{1-T_i}{n_c}Y_i^c\Big)\Big(\frac{T_j}{n_t}Y_j^t-\frac{1-T_j}{n_c}Y_j^c\Big)\rc - (\mu_t-\mu_c)^2\\
&=n(n-1)\lc\frac{1}{n_t^2}\Big(p_t^2-\frac{p_tp_c}{n-1}\Big)\mu_t^2 - \frac{2}{n_tn_c}\Big(p_t-p_t^2+\frac{p_tp_c}{n-1}\Big)\mu_t\mu_c + \frac{1}{n_c^2}\Big(1-2p_t+p_t^2-\frac{p_tp_c}{n-1}\Big)\mu_c^2\rc \\&\quad + n\lc\frac{1}{n_t^2}p_t(\text{var}(Y_t) + \mu_t^2) + \frac{1}{n_c^2} p_c (\text{var}(Y_c) + \mu_c^2) \rc - (\mu_t-\mu_c)^2\\
&=\frac{\text{var}(Y_t)}{n_t} + \frac{\text{var}(Y_c)}{n_c}. 
\end{aligned}
$$

\subsection{Proof of Theorem \ref{supervar}}

\noindent i). We first define that   
$$
\begin{aligned}
\widetilde{\delta}_3&=\lc\oy_t - \theta_t(\ox_t-\ox)\rc - \lc\oy_c - \theta_c(\ox_c-\ox)\rc \\
&=\lc\oy_t - \theta_t(\ox_t-p_t\ox_t-p_c\ox_c)\rc - \lc\oy_c - \theta_c(\ox_c-p_t\ox_t-p_c\ox_c)\rc \\
&=\oy_t-\oy_c-\lc(p_c\theta_t + p_t\theta_c)(\ox_t-\ox_c)\rc. 
\end{aligned}
$$

Introduce the transformed variable $Z_i^t=Y_i^t - (p_c\theta_t + p_t\theta_c) X_i$ and $Z_i^c=Y_i^c - (p_c\theta_t + p_t\theta_c) X_i$, which allows us to rewrite \(\widetilde{\delta}_3\) as: 

$$\widetilde{\delta}_3=\frac{1}{n_t}\sum\limits_{i=1}^nT_iZ_i^t - \frac{1}{n_c}\sum\limits_{i=1}^n(1 - T_i)Z_i^c. $$

The variances of $Z$ for the treatment and control groups are: 

$$\text{var}(Z_t)=\text{var}(Y_t) + \lc(p_t^2-1)\theta_t^2 - 2p_t^2\theta_t\theta_c + p_t^2\theta_c^2\rc\text{var}(X),\quad \text{var}(Z_c)=\text{var}(Y_c) + \lc p_c^2\theta_t^2 - 2p_c^2\theta_t\theta_c + (p_t^2-2p_t)\theta_c^2\rc\text{var}(X). $$

From Lemma \ref{lem2}, we have: 
$$
\begin{aligned}
\lim_{n\rightarrow \infty}n\text{var}(\widetilde{\delta}_3)&=\frac{\text{var}(Z_t)}{p_t} + \frac{\text{var}(Z_c)}{p_c}\\
&=\frac{\text{var}(Y_t)}{p_t} + \frac{\text{var}(Y_c)}{p_c} - \lc\frac{p_c}{p_t}\theta_t^2 + 2\theta_t\theta_c + \frac{p_t}{p_c}\theta_c^2\rc\text{var}(X). 
\end{aligned}
$$

We next proceed to show that 
$$
\lim_{n\rightarrow \infty}n\lc\text{var}(\wdd_3)-\text{var}(\widetilde{\delta}_3)\rc=0. 
$$

It suffices to show that
$$
\lim_{n\rightarrow \infty}\sqrt{n}\lc\wdd_3-\widetilde{\delta}_3\rc=0. 
$$

$$
\begin{aligned}
\lim_{n\rightarrow \infty}\sqrt{n}\lc\wdd_3-\widetilde{\delta}_3\rc= \lim_{n\rightarrow \infty}\sqrt{n}\lc (\theta_t-\wt_{t, 3})(\ox_t-\ox) - (\theta_c-\wt_{c, 3})(\ox_c-\ox) \rc=0. 
\end{aligned}
$$

\noindent ii). Next, consider the asymptotic limit of the estimated variance \(\wv(\wdd_3)\): 

$$
\begin{aligned}
\lim_{n\rightarrow\infty} n \wv(\wdd_3) &= \lim_{n\rightarrow\infty} \frac{1}{p_t}\lc \wv(Y_t) - 2\wt_{t, 3}\wc(Y_t, X_t) + \wt_{t, 3}^2\wv(X_t)\rc + \frac{1}{p_c}\lc \wv(Y_c) - 2\wt_{c, 3}\wc(Y_c, X_c) + \wt_{c, 3}^2\wv(X_c)\rc \\
&=\frac{1}{p_t}\lc \text{var}(Y_t) - 2\theta_{t}\text{cov}(Y_t, X_t) + \theta_{t}^2\text{var}(X_t)\rc + \frac{1}{p_c}\lc \text{var}(Y_c) - 2\theta_{c}\text{cov}(Y_c, X_c) + \theta_{c}^2\text{var}(X_c)\rc \\
&=\frac{\text{var}(Y_t)}{p_t} + \frac{\text{var}(Y_c)}{p_c} - \lc\frac{1}{p_t}\theta_t^2 + \frac{1}{p_c}\theta_c^2\rc\text{var}(X). 
\end{aligned}
$$

Comparing the results from i) and ii), the asymptotic difference between the true variance and its estimator is: 
$$\lim_{n\rightarrow \infty}n\left(\text{var}(\widehat\delta_3)-\widehat{\text{var}}(\widehat\delta_3)\right)=(\theta_t-\theta_c)^2\text{var}(X). $$

\subsection{Proof of Theorem \ref{supervarelse}}

i). First define the general form of the estimator \(\widetilde{\delta}_k\) for \(k = 0, 1, 2\): 
$$
\begin{aligned}
\widetilde{\delta}_k&=\lc\oy_t - \theta_k(\ox_t-\ox)\rc - \lc\oy_c - \theta_k(\ox_c-\ox)\rc \\
&=\lc\oy_t - \theta_k(\ox_t-p_t\ox_t-p_c\ox_c)\rc - \lc\oy_c - \theta_k(\ox_c-p_t\ox_t-p_c\ox_c)\rc \\
&=\oy_t-\oy_c-\theta_k(\ox_t-\ox_c). 
\end{aligned}
$$

Introduce the transformed variable 
$$
Z_i^t = Y_i^t - \theta_k X_i, Z_i^c = Y_i^t - \theta_k X_i,  \text{ for each } k = 0, 1, 2, and i = 1, \ldots, n
$$

and rewrite that 
$$
\widetilde{\delta}_k=\frac{1}{n_t}\sum\limits_{i=1}^nT_iZ_i^t - \frac{1}{n_c}\sum\limits_{i=1}^n(1 - T_i)Z_i^c. 
$$

The variances for the treatment $Z_t$ and control $Z_c$ are derived as: 

$$
\text{var}(Z_t)=\text{var}(Y_t) + \lc\theta_k^2-2\theta_k\theta_t\rc\text{var}(X), \quad \text{var}(Z_c)=\text{var}(Y_c) + \lc \theta_k^2-2\theta_k\theta_c\rc\text{var}(X). 
$$

From Lemma \ref{lem2}, we have: 
$$
\begin{aligned}
\lim_{n\rightarrow \infty}n\text{var}(\wdd_k)&=\lim_{n\rightarrow \infty}n\text{var}(\widetilde\delta_k)\\
&=\frac{\text{var}(Z_t)}{p_t} + \frac{\text{var}(Z_c)}{p_c}\\
&= \frac{\text{var}(Y_t)}{p_t} + \frac{\text{var}(Y_c)}{p_c} - \left(\frac{2\theta_k\theta_t-\theta_k^2}{p_t}+ \frac{2\theta_k\theta_c-\theta_k^2}{p_c}\right)\text{var}(X). 
\end{aligned}
$$

\noindent ii). Next, we calculate the asymptotic limit of the estimated variance \(\wv(\wdd_k)\) for $k=0, 1, 2$: 

$$
\lim_{n\rightarrow \infty}n\wv(\wdd_k) = \frac{1}{p_t}\lc\text{var}(Y_t) + \lc\theta_k^2-2\theta_k\theta_t\rc\text{var}(X)\rc + \frac{1}{p_c}\lc\text{var}(Y_c) + \lc\theta_k^2-2\theta_k\theta_c\rc\text{var}(X)\rc. 
$$
Comparing the above results, for each $k=0, 1, 2$, $n\widehat{\text{var}}(\widehat\delta_k)$ and $n\text{var}(\widehat\delta_k)$ are asymptotically equal in probability.

\subsection{Proof of Theorem \ref{supercom}}

i). This conclusion is a direct corollary of Theorem \ref{deltavar} and Theorem \ref{supervar}. 

\noindent ii). Consider the asymptotic difference between the estimated variances of \(\wdd_0\) and \(\wdd_2\): 
$$
\lim_{n\rightarrow \infty} n\lc\wv(\wdd_0)-\wv(\wdd_2)\rc = \lc\frac{p_c}{p_t}\theta_t^2 + 2\theta_t\theta_c + \frac{p_t}{p_c}\theta_c^2\rc\text{var}(X)=\lc\sqrt{\frac{p_c}{p_t}}\theta_t + \sqrt{\frac{p_t}{p_c}}\theta_c\rc^2\text{var}(X)\ge 0, 
$$
with strict inequality unless $p_c\theta_t+p_t\theta_c=0$. 

\noindent iii). Next, we analyze the estimated variance difference between \(\wdd_1\) and \(\wdd_2\): 
$$
\begin{aligned}
\lim_{n\rightarrow \infty} n\lc\wv(\wdd_1)-\wv(\wdd_2)\rc &= \lc\frac{p_c}{p_t}\theta_t^2 + 2\theta_t\theta_c + \frac{p_t}{p_c}\theta_c^2\rc\text{var}(X)-\left(\frac{2\theta_1\theta_t-\theta_1^2}{p_t}+ \frac{2\theta_1\theta_c-\theta_1^2}{p_c}\right)\text{var}(X) \\
&=\lc \frac{(\theta_1-\theta_t)^2}{p_t} + \frac{(\theta_1-\theta_c)^2}{p_c} - (\theta_t-\theta_c)^2\rc\text{var}(X) \\
&= (\theta_t-\theta_c)^2\lc \frac{p_c^2}{p_t}+ \frac{p_t^2}{p_c} - 1 \rc\text{var}(X) \\
&= (\theta_t-\theta_c)^2\frac{(2p_t-1)^2}{p_t(1-p_t)}\text{var}(X), 
\end{aligned}
$$
where strict inequality holds except when $p_t=p_c$ or $\theta_t=\theta_c$. 

\noindent iv). Finally, we examine the asymptotic difference between \(\wdd_0\) and \(\wdd_1\): 
$$
\begin{aligned}
\lim_{n\rightarrow \infty} n\lc\wv(\wdd_0)-\wv(\wdd_1)\rc &= \left(\frac{2\theta_1\theta_t-\theta_1^2}{p_t}+ \frac{2\theta_1\theta_c-\theta_1^2}{p_c}\right)\text{var}(X) \\
&=\frac{1}{p_tp_c}\lc 2\theta_1\theta_tp_c-\theta_1^2p_c+2\theta_1\theta_cp_t-\theta_1^2p_t\rc \\
&=\frac{1}{p_tp_c}\lc2\theta_1\theta_2-\theta_1^2\rc \\
&=\frac{1}{p_tp_c}\lc \theta_2^2-\lc\theta_1-\theta_2\rc^2\rc. 
\end{aligned}
$$

Therefore, asymptotically, $n\wv(\wdd_1)$ is not necessarily smaller than $n\wv(\wdd_0)$.

\newpage

\section{Supplementary Experiments}
\label{appB}

\begin{figure}[htbp]
  \centering 
  \subfigure[Design-based \& $p=0.5$ \& $\text{ATE}=0$ \& $\text{HTE}=0$]
  {  \label{cona}
   \includegraphics[width=0.225\textwidth]{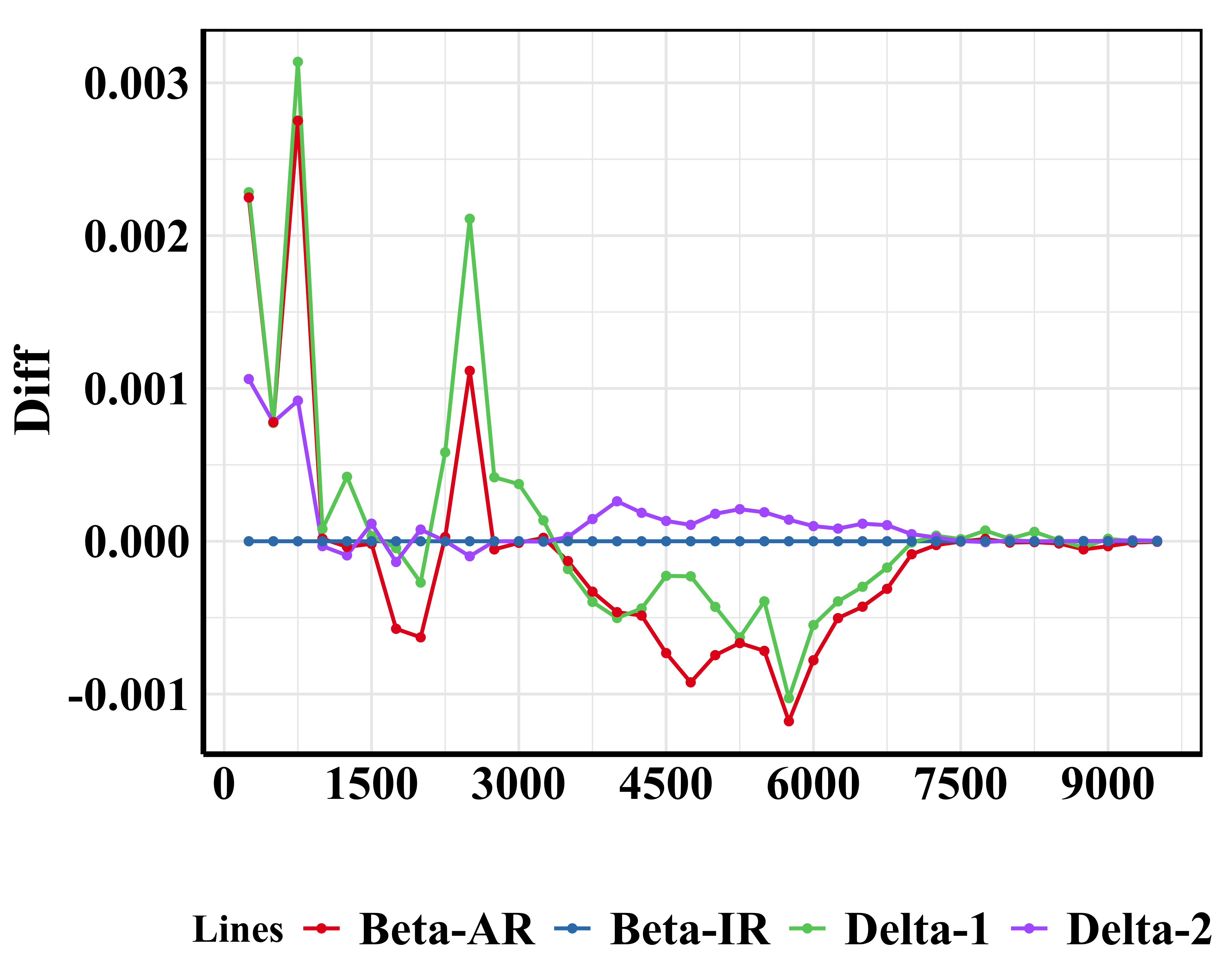}
  }
    \subfigure[Design-based \& $p=0.5$ \& $\text{ATE}=0$ \& $\text{HTE}=0.5$]
  {  \label{conb}
   \includegraphics[width=0.225\textwidth]{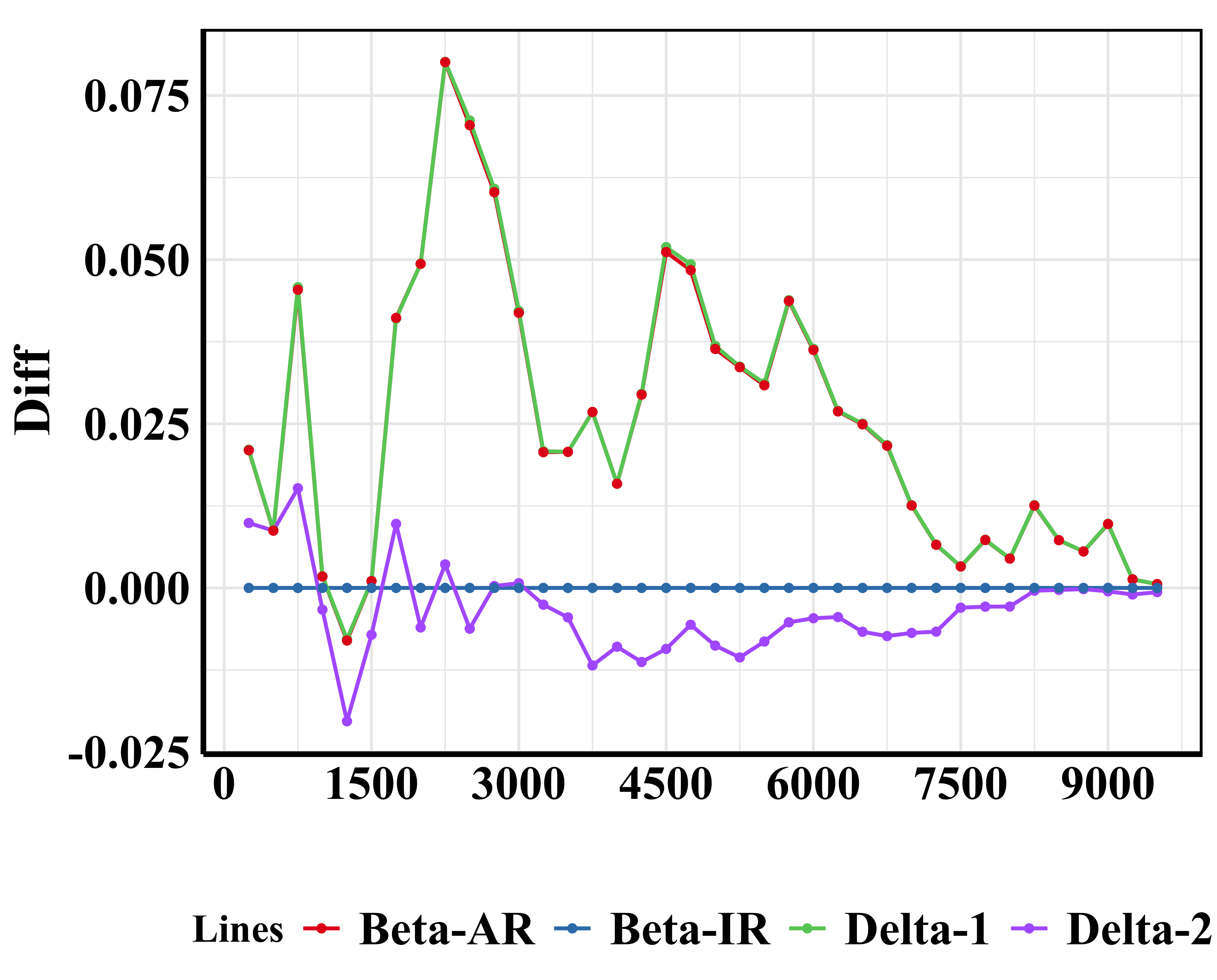}
  }
    \subfigure[Design-based \& $p=0.5$ \& $\text{ATE}=0.1$ \& $\text{HTE}=0$]
  {  \label{conc}
   \includegraphics[width=0.225\textwidth]{lineplot1121.png}
  }
      \subfigure[Design-based \& $p=0.5$ \& $\text{ATE}=0.1$ \& $\text{HTE}=0.5$]
  {  \label{cond}
   \includegraphics[width=0.225\textwidth]{lineplot1122.png}
  }
  \subfigure[Design-based \& $p=0.4$ \& $\text{ATE}=0$ \& $\text{HTE}=0$]
  {  \label{cone}
   \includegraphics[width=0.225\textwidth]{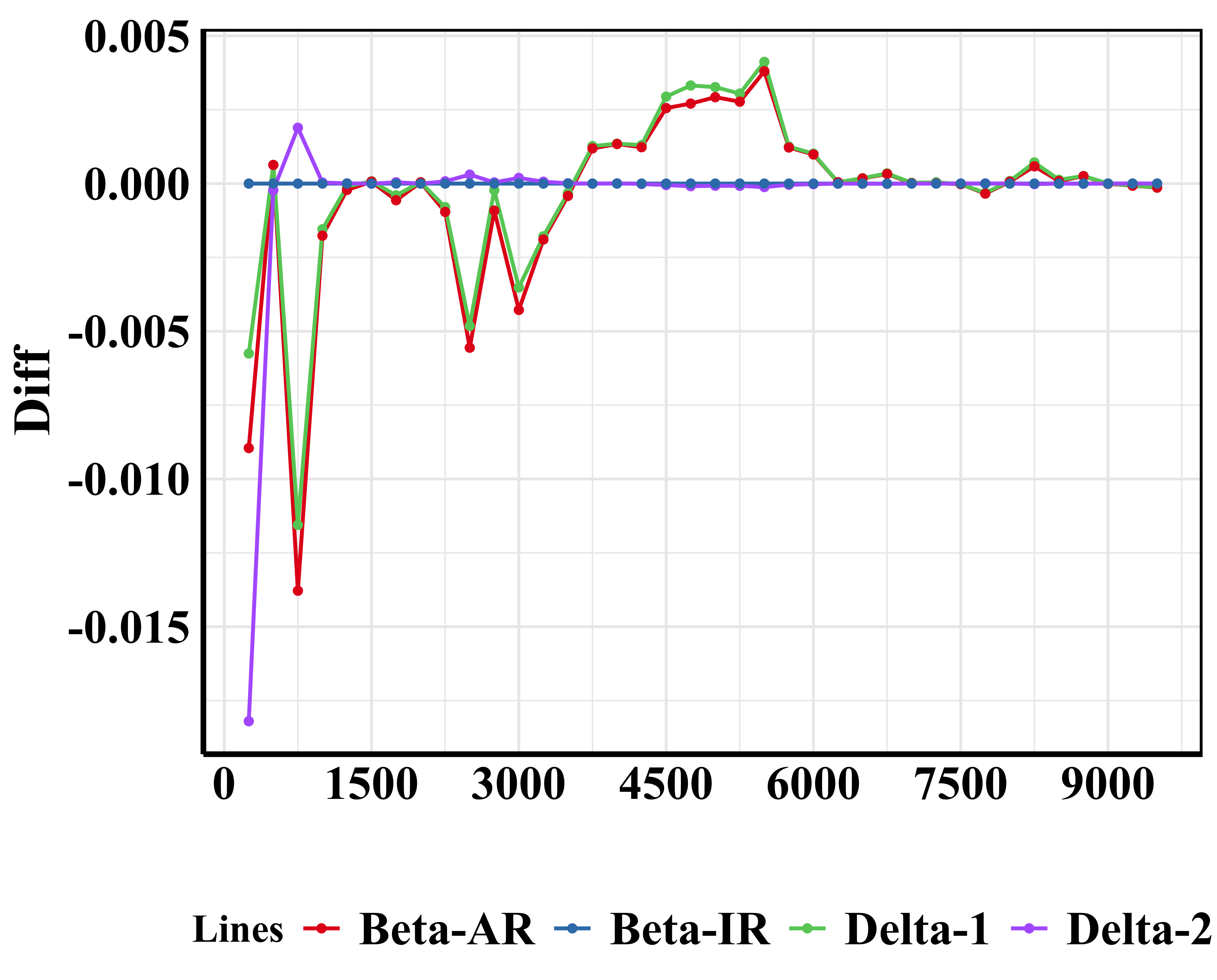}
  }
    \subfigure[Design-based \& $p=0.4$ \& $\text{ATE}=0$ \& $\text{HTE}=0.5$]
  {  \label{conf}
   \includegraphics[width=0.225\textwidth]{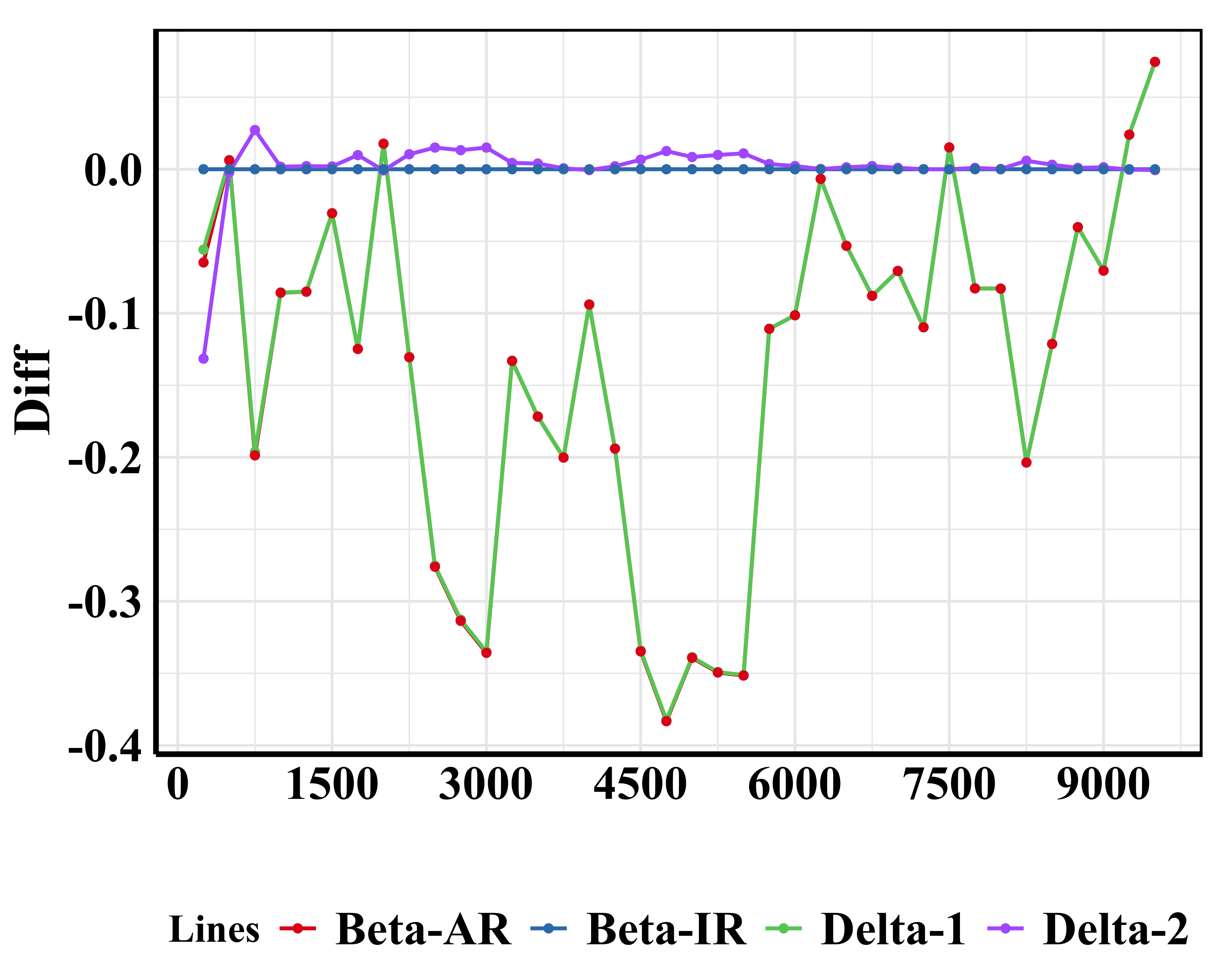}
  }
    \subfigure[Design-based \& $p=0.4$ \& $\text{ATE}=0.1$ \& $\text{HTE}=0$]
  {  \label{cong}
   \includegraphics[width=0.225\textwidth]{lineplot1221.png}
  }
      \subfigure[Design-based \& $p=0.4$ \& $\text{ATE}=0.1$ \& $\text{HTE}=0.5$]
  {  \label{conh}
   \includegraphics[width=0.225\textwidth]{lineplot1222.png}
  }
  \subfigure[Model-based \& $p=0.5$ \& $\text{ATE}=0$ \& $\text{HTE}=0$]
  {  \label{coni}
   \includegraphics[width=0.225\textwidth]{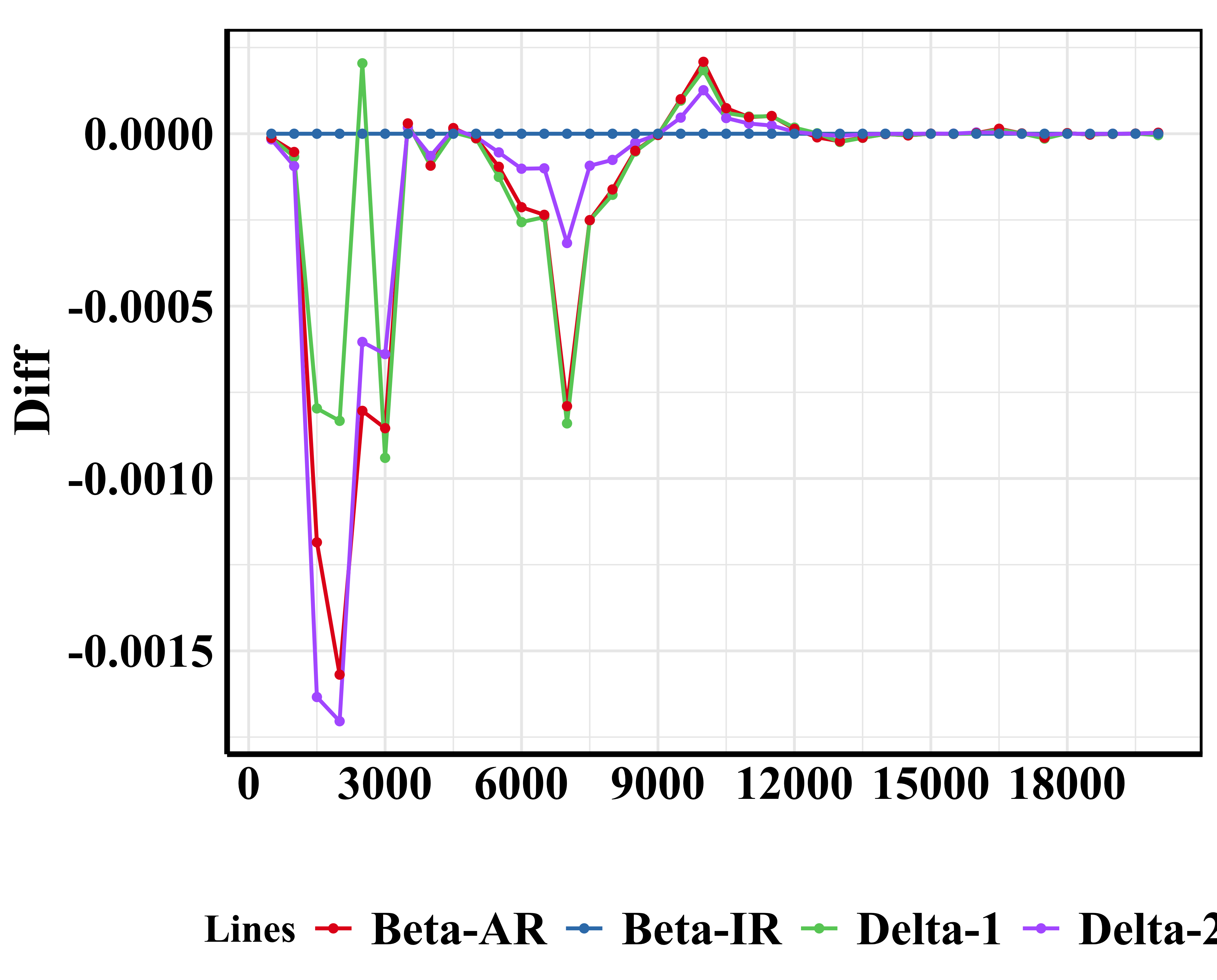}
  }
    \subfigure[Model-based \& $p=0.5$ \& $\text{ATE}=0$ \& $\text{HTE}=0.5$]
  {  \label{conj}
   \includegraphics[width=0.225\textwidth]{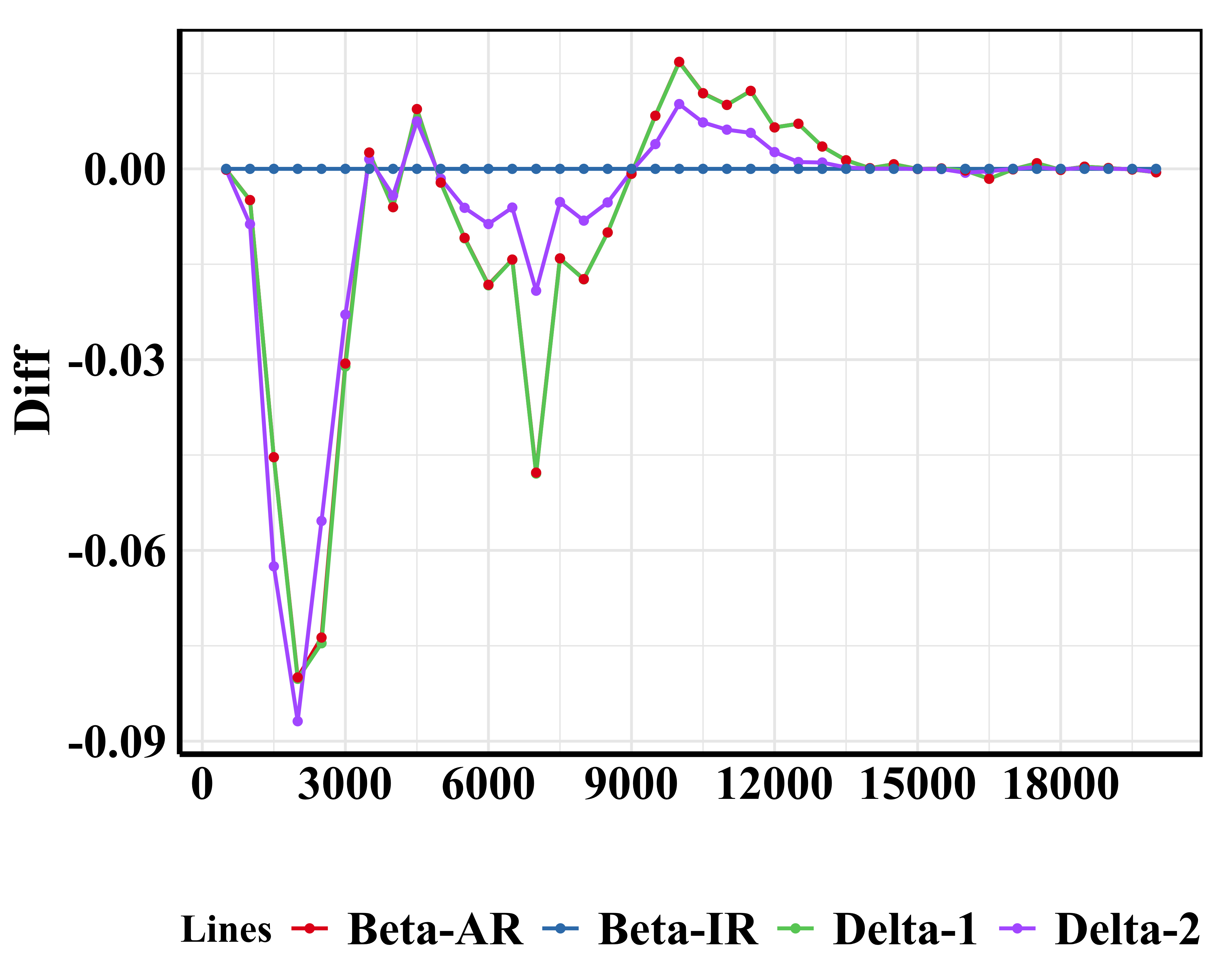}
  }
    \subfigure[Model-based \& $p=0.5$ \& $\text{ATE}=0.1$ \& $\text{HTE}=0$]
  {  \label{conk}
   \includegraphics[width=0.225\textwidth]{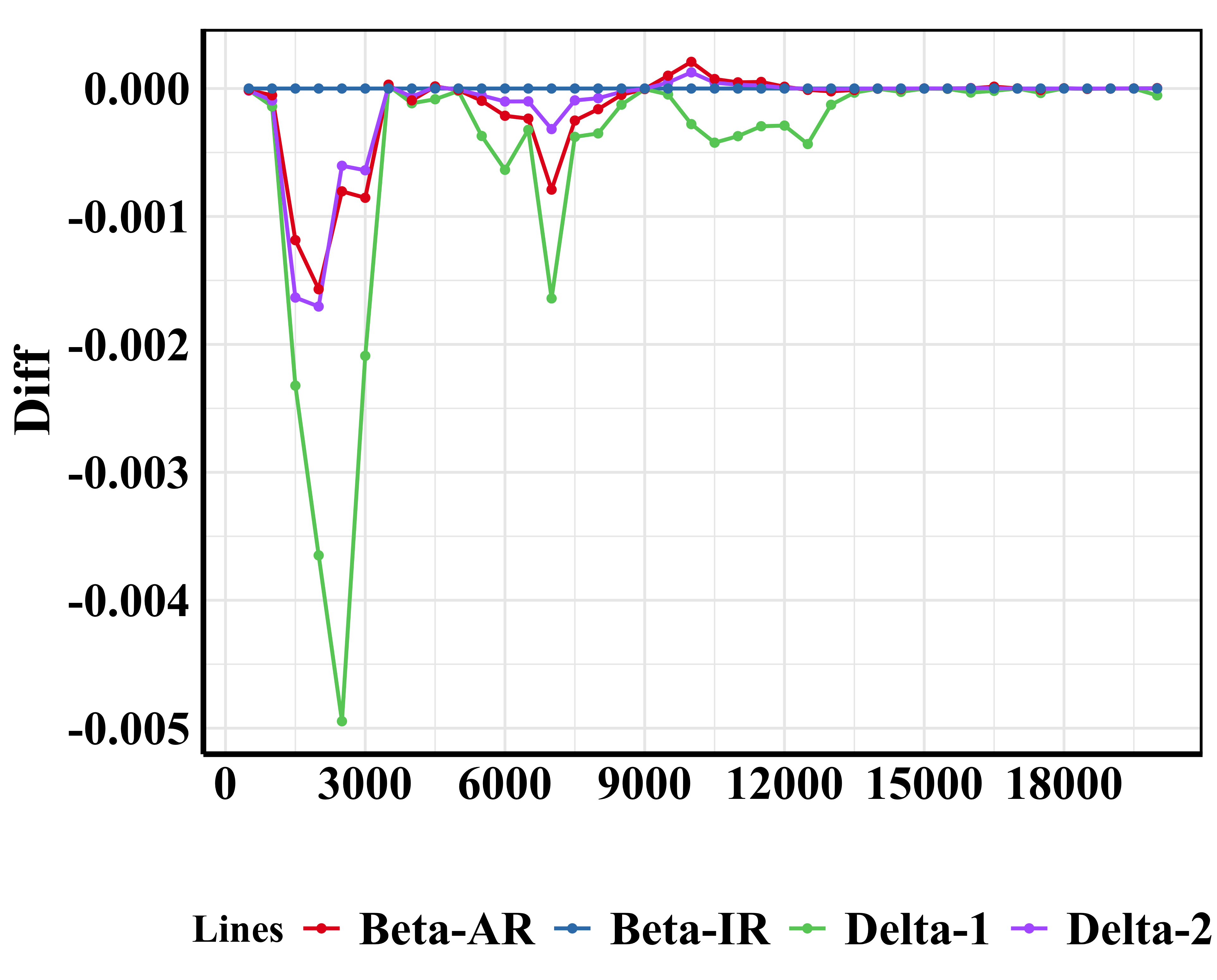}
  }
      \subfigure[Model-based \& $p=0.5$ \& $\text{ATE}=0.1$ \& $\text{HTE}=0.5$]
  {  \label{conl}
   \includegraphics[width=0.225\textwidth]{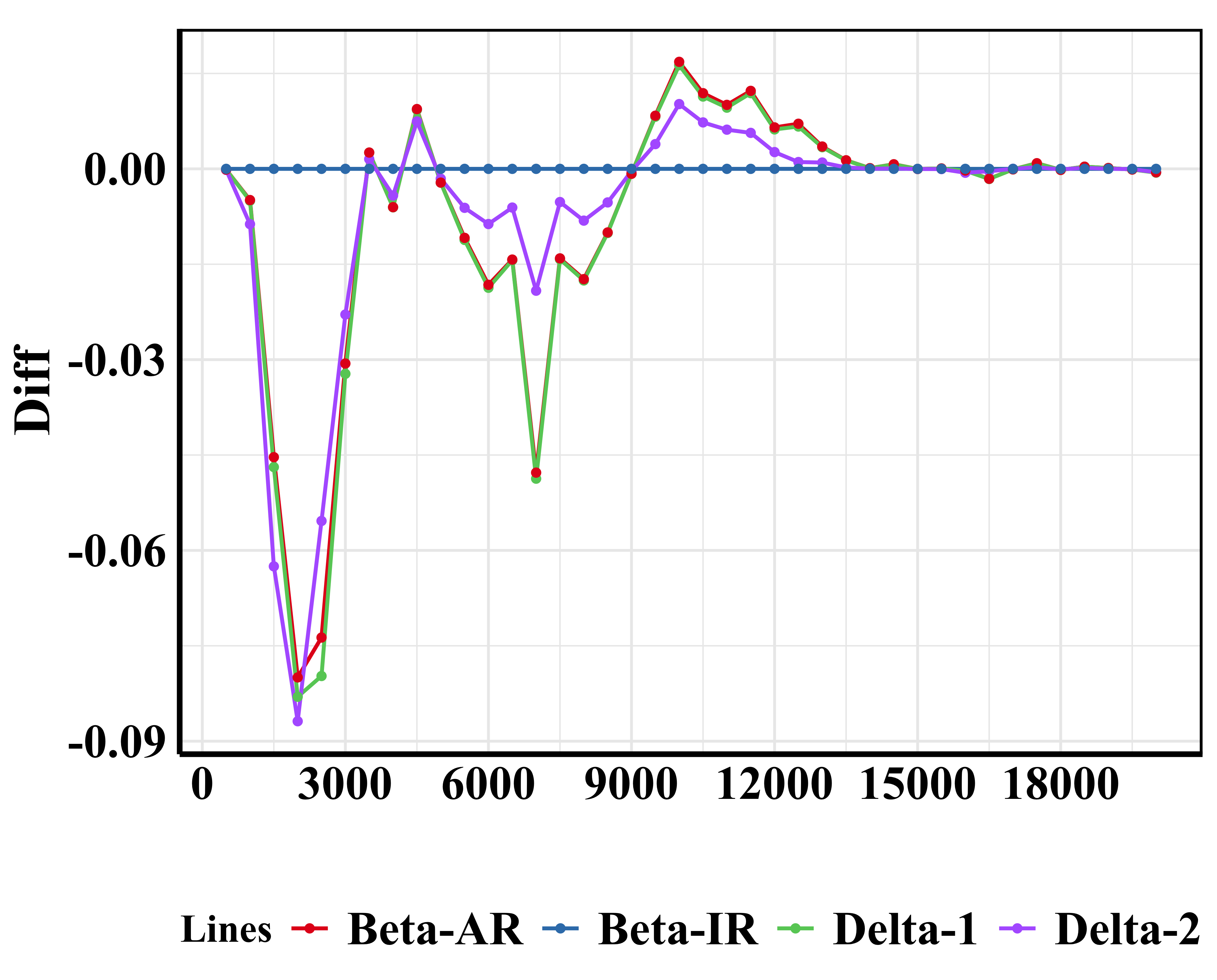}
  }
  \subfigure[Model-based \& $p=0.4$ \& $\text{ATE}=0$ \& $\text{HTE}=0$]
  {  \label{conm}
   \includegraphics[width=0.225\textwidth]{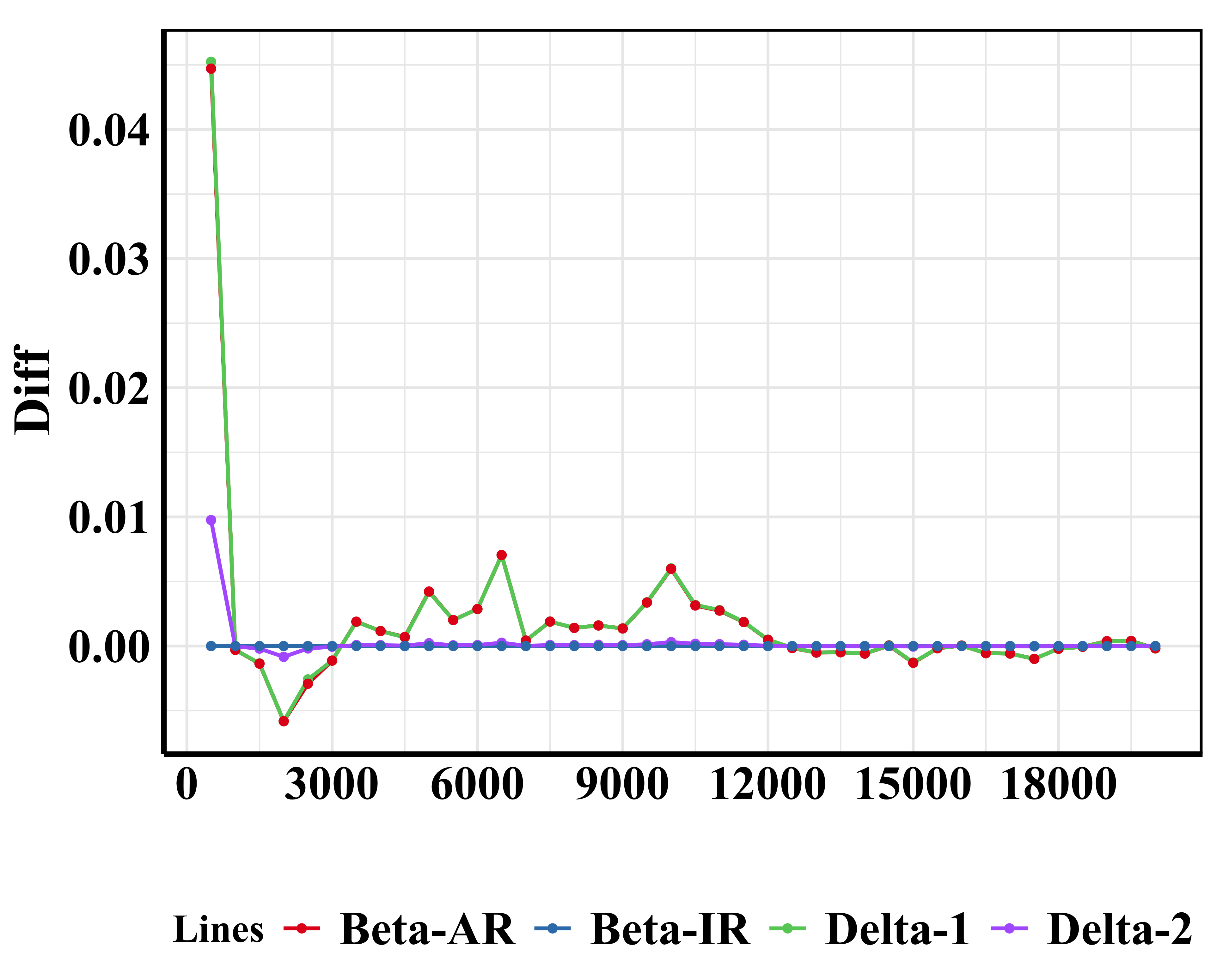}
  }
    \subfigure[Model-based \& $p=0.4$ \& $\text{ATE}=0$ \& $\text{HTE}=0.5$]
  {  \label{conn}
   \includegraphics[width=0.225\textwidth]{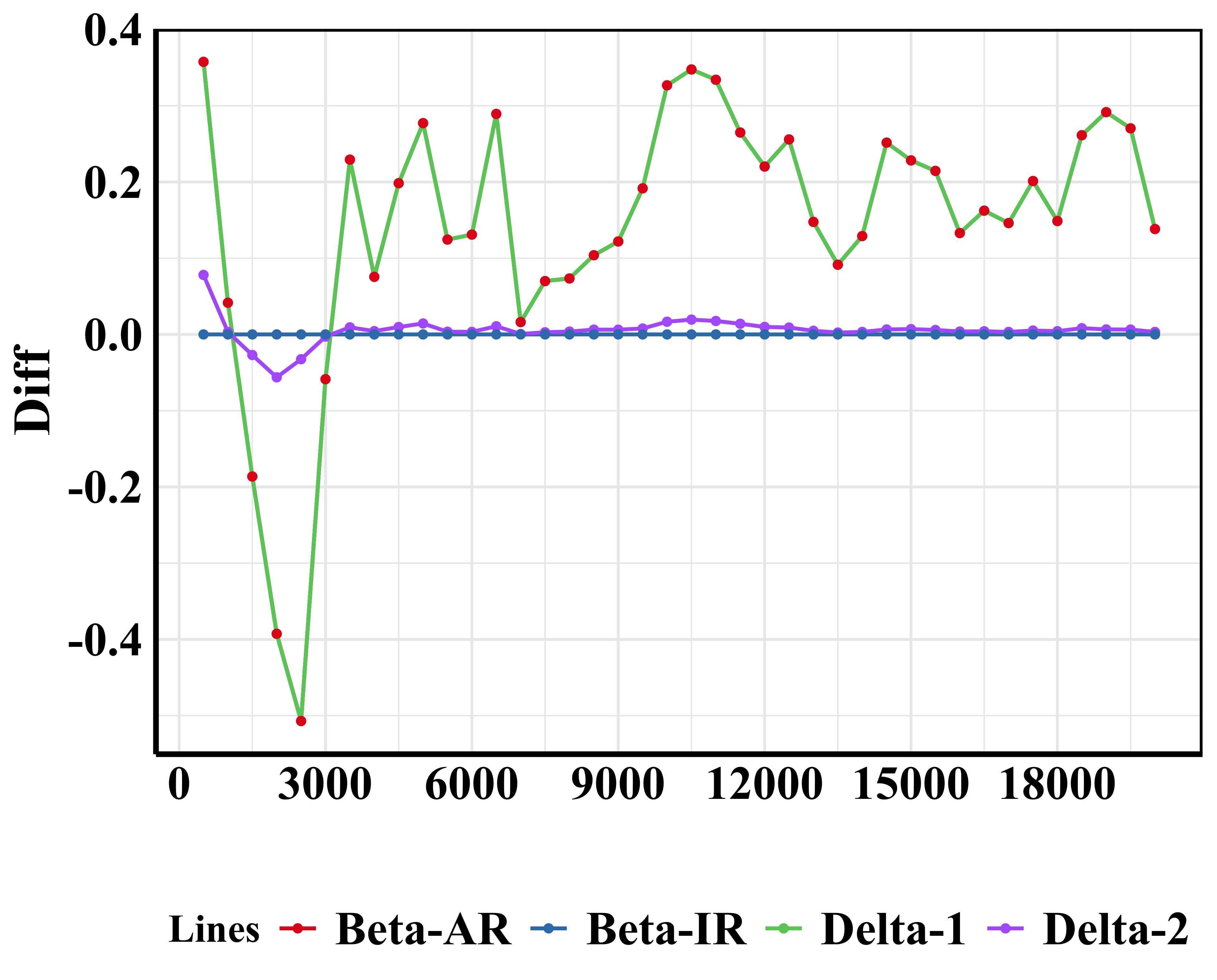}
  }
    \subfigure[Model-based \& $p=0.4$ \& $\text{ATE}=0.1$ \& $\text{HTE}=0$]
  {  \label{cono}
   \includegraphics[width=0.225\textwidth]{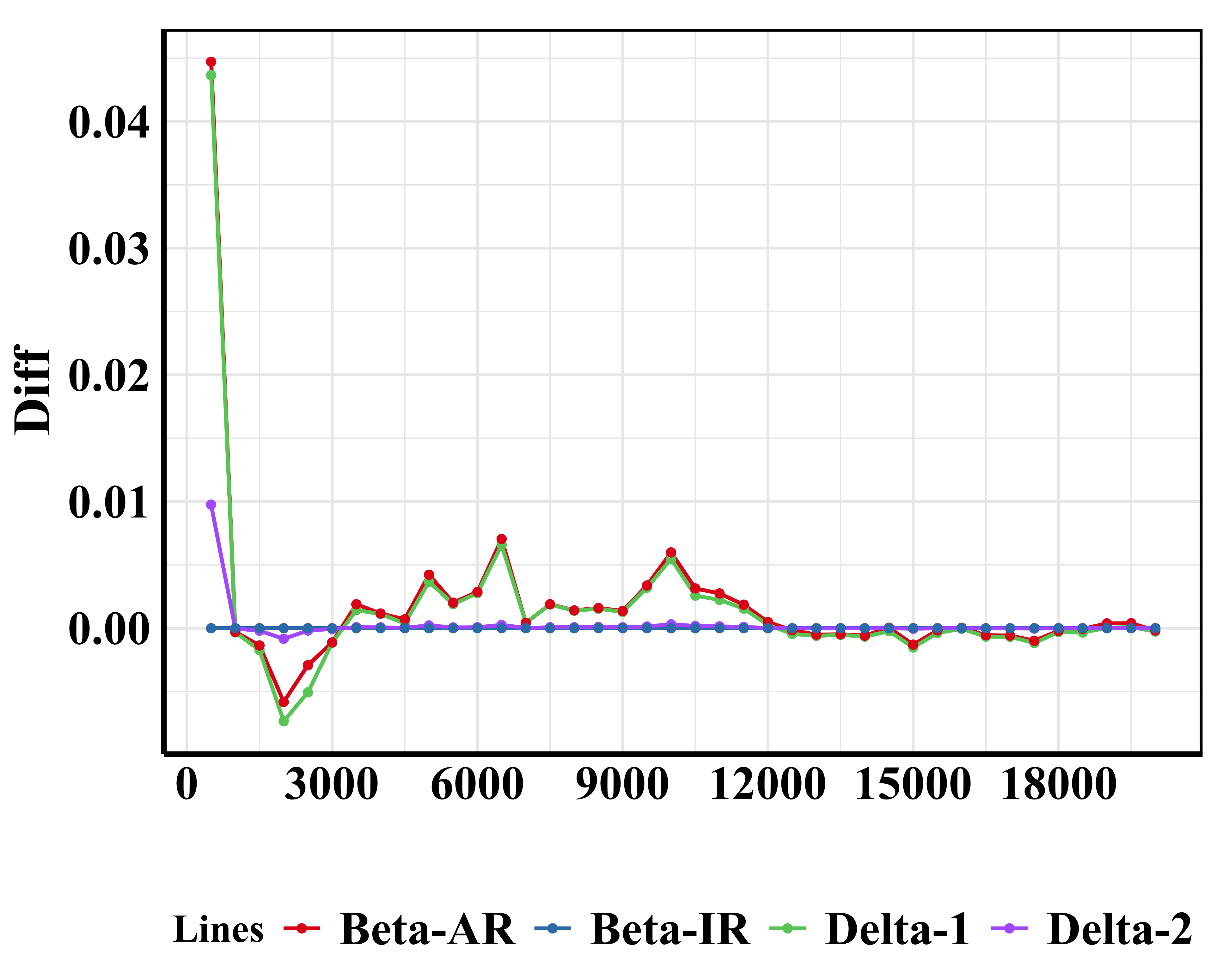}
  }
      \subfigure[Model-based \& $p=0.4$ \& $\text{ATE}=0.1$ \& $\text{HTE}=0.5$]
  {  \label{conp}
   \includegraphics[width=0.225\textwidth]{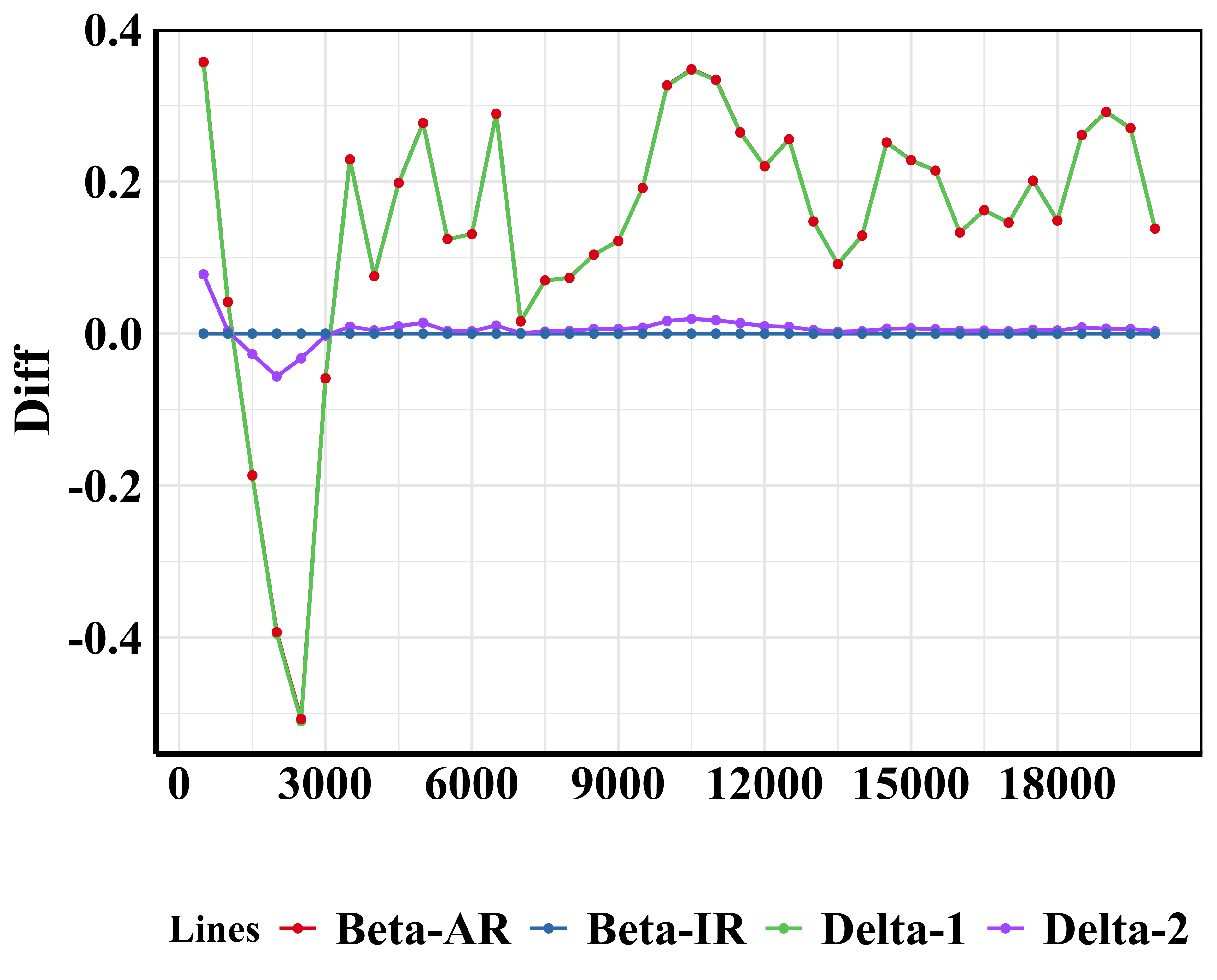}
  }

  \caption{Asymptotically discrepancies between \(\widehat{\delta}_3\) and other estimators, scaled as \(\sqrt n\cdot(\widehat{\text{ATE}} - \widehat{\delta}_3)\). }
  \label{convergence_all}
\end{figure}

We present convergence properties of distinct estimators across all $16$ configurations in Figure \ref{convergence_all}. These findings demonstrate that, irrespective of the inferential framework or the presence of an ATE, the difference between any pair of \(\sqrt{n}\widehat{\delta}_1\), \(\sqrt{n}\widehat{\delta}_2\), \(\sqrt{n}\widehat{\delta}_3\), \(\sqrt{n}\widehat{\beta}_T^{{AR}}\), and \(\sqrt{n}\widehat{\beta}_T^{{IR}}\) converges in probability to $0$, provided either that group assignment probabilities are equal or that no HTE exists in the population. Conversely, when group assignment probabilities are unequal and HTE is present simultaneously, the difference between \(\sqrt{n}\widehat{\delta}_1\) and \(\sqrt{n}\widehat{\beta}_T^{{AR}}\) still converges in probability to $0$, but the differences between \(\sqrt{n}\widehat{\delta}_1\) and the remaining estimators no longer hold.

\begin{figure}[htbp]
  \centering 
  \subfigure[Design-based \& $p=0.5$ \& $\text{ATE}=0$ \& $\text{HTE}=0$]
  {
   \includegraphics[width=0.225\textwidth]{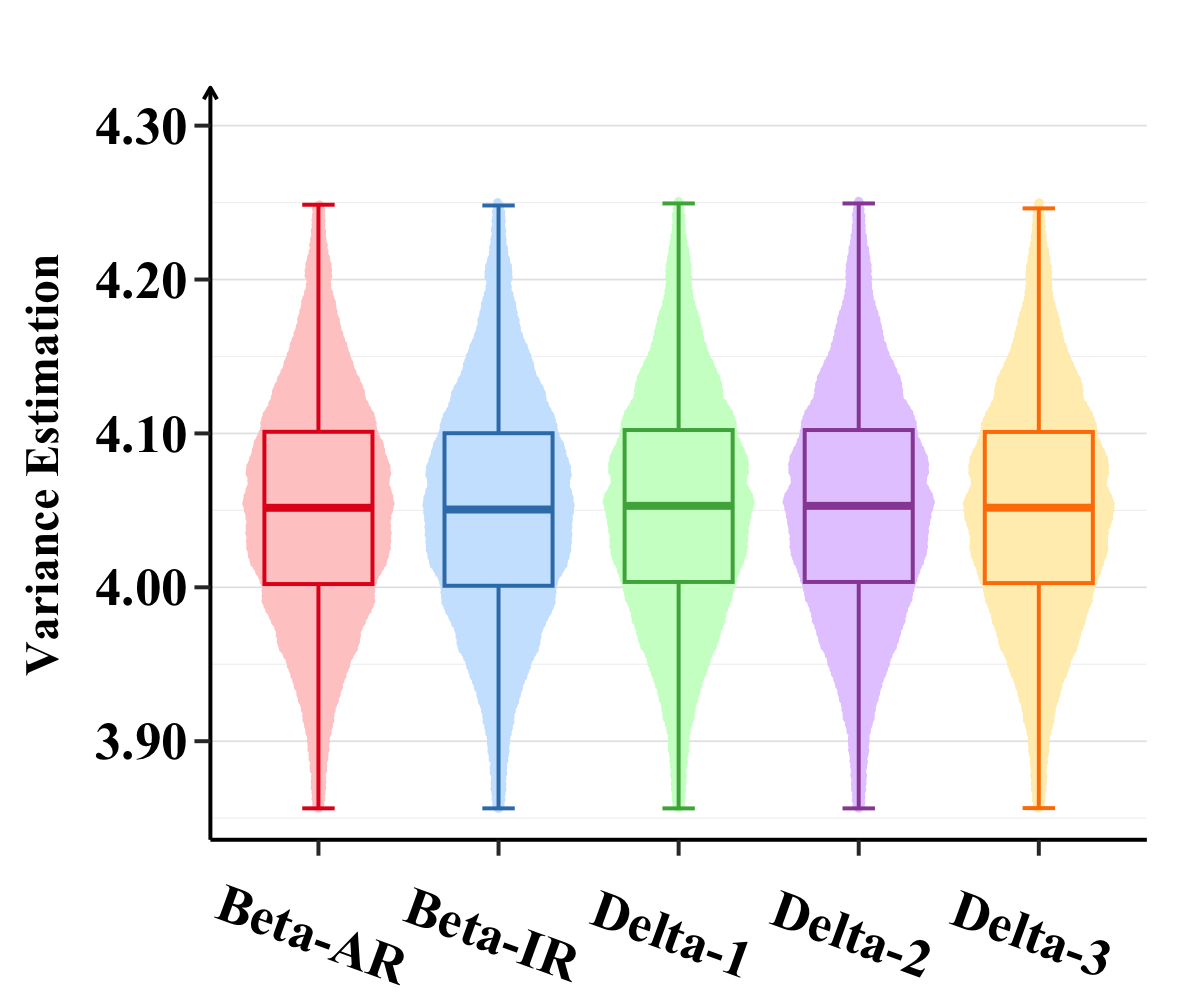}
  }
    \subfigure[Design-based \& $p=0.5$ \& $\text{ATE}=0$ \& $\text{HTE}=0.5$]
  { 
   \includegraphics[width=0.225\textwidth]{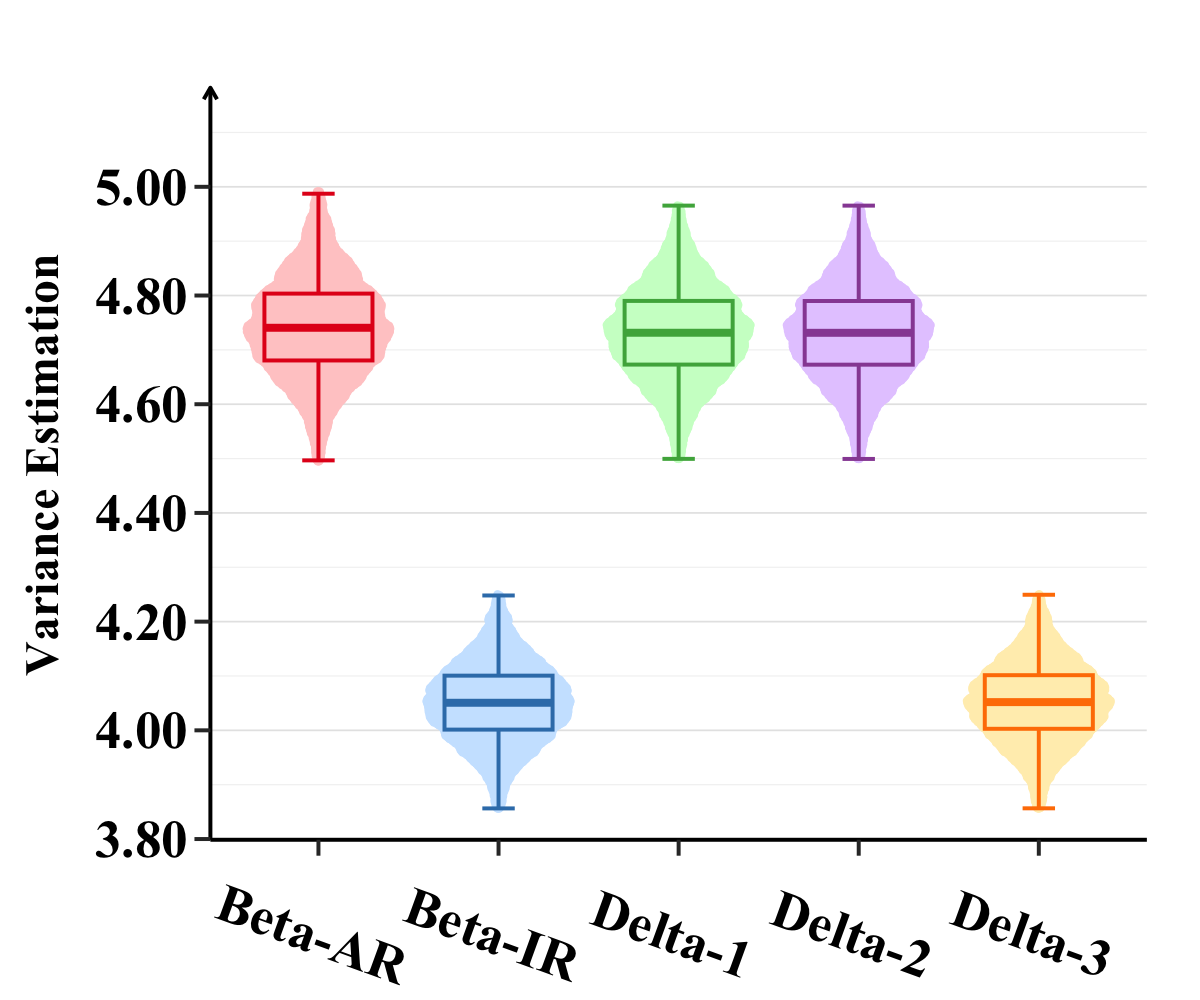}
  }
    \subfigure[Design-based \& $p=0.5$ \& $\text{ATE}=0.1$ \& $\text{HTE}=0$]
  {  
   \includegraphics[width=0.225\textwidth]{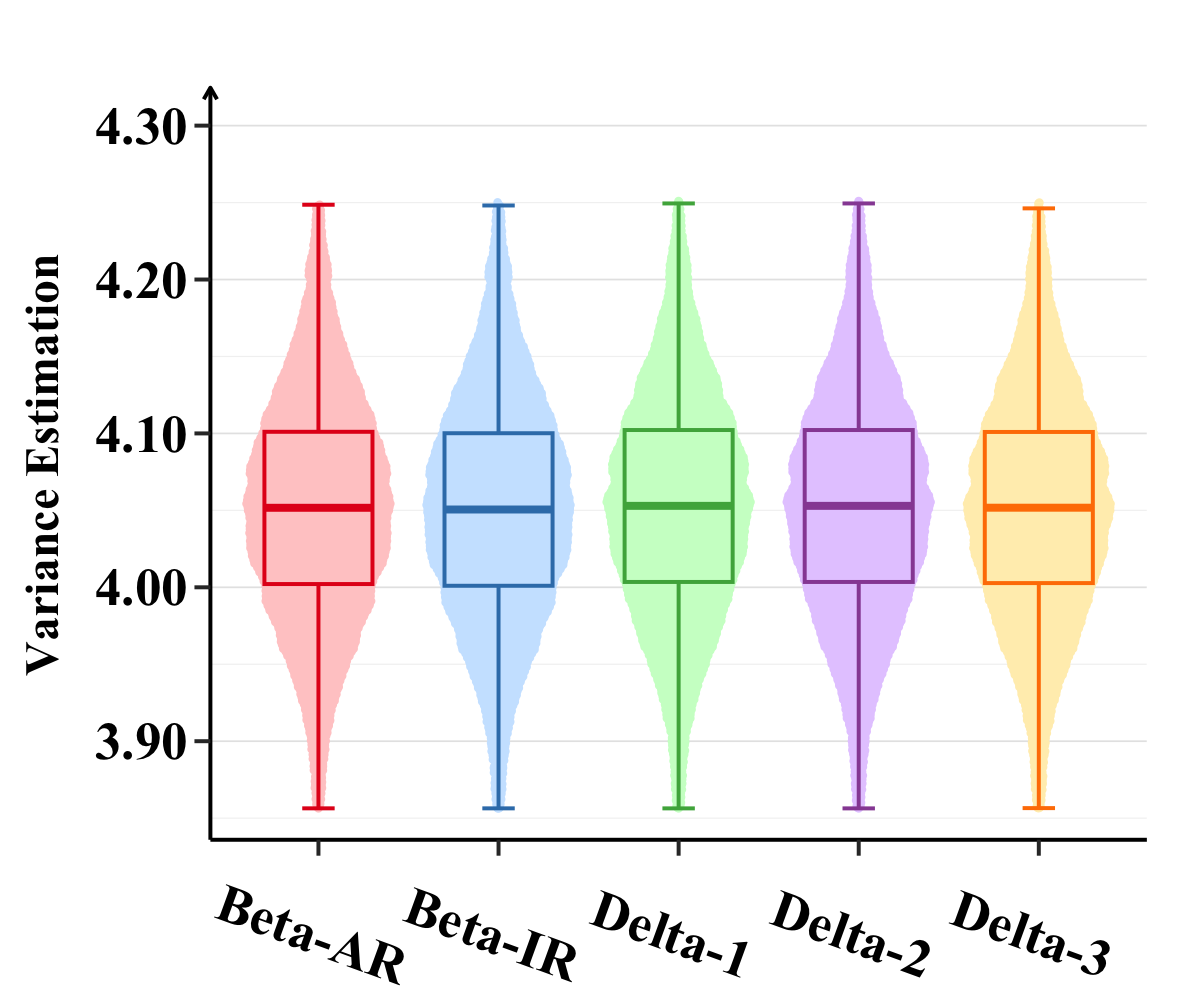}
  }
      \subfigure[Design-based \& $p=0.5$ \& $\text{ATE}=0.1$ \& $\text{HTE}=0.5$]
  { 
   \includegraphics[width=0.225\textwidth]{boxplot1122.png}
  }
  \subfigure[Design-based \& $p=0.4$ \& $\text{ATE}=0$ \& $\text{HTE}=0$]
  { 
   \includegraphics[width=0.225\textwidth]{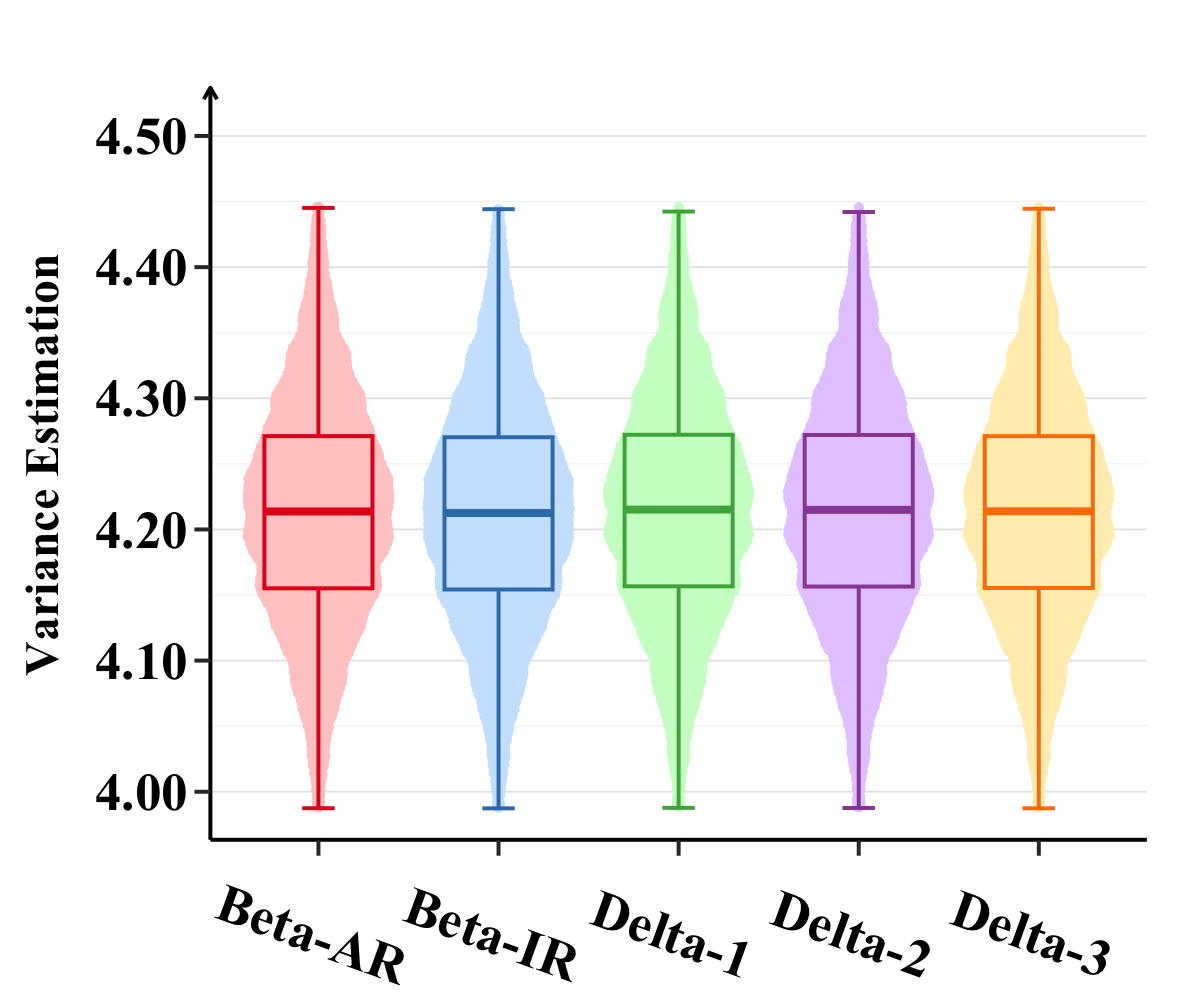}
  }
    \subfigure[Design-based \& $p=0.4$ \& $\text{ATE}=0$ \& $\text{HTE}=0.5$]
  {  
   \includegraphics[width=0.225\textwidth]{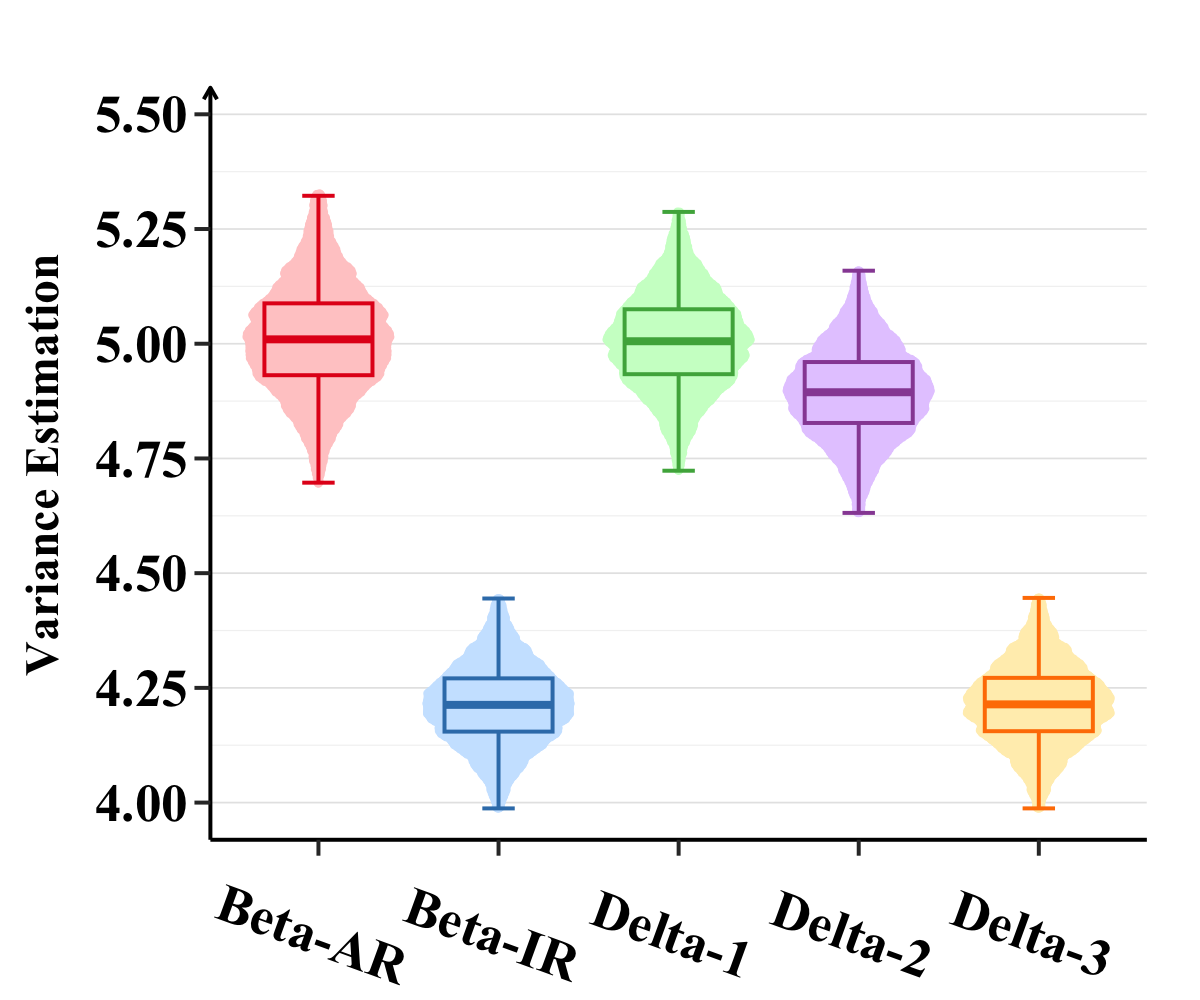}
  }
    \subfigure[Design-based \& $p=0.4$ \& $\text{ATE}=0.1$ \& $\text{HTE}=0$]
  {  
   \includegraphics[width=0.225\textwidth]{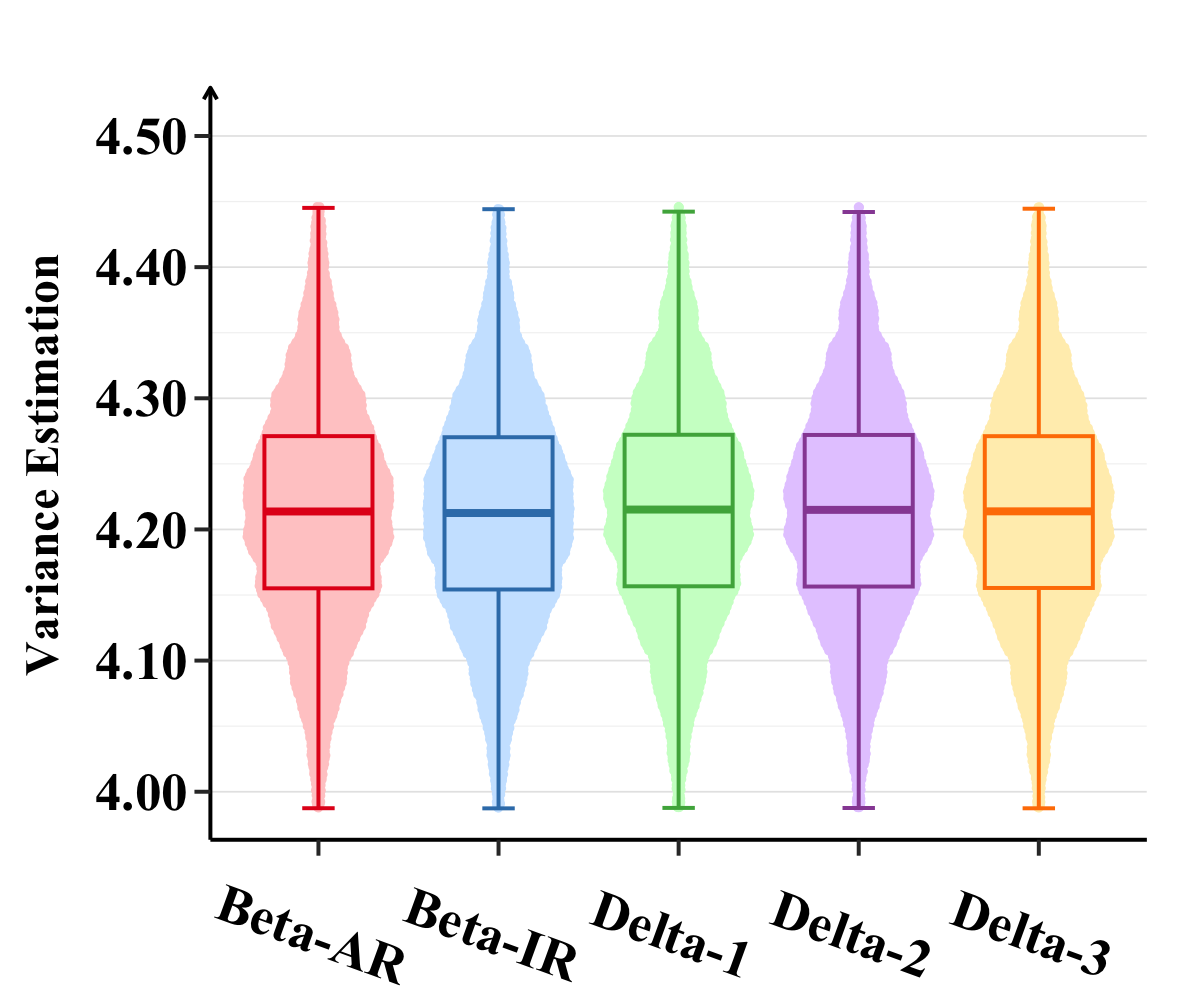}
  }
      \subfigure[Design-based \& $p=0.4$ \& $\text{ATE}=0.1$ \& $\text{HTE}=0.5$]
  {   
   \includegraphics[width=0.225\textwidth]{boxplot1222.png}
  }
  \subfigure[Model-based \& $p=0.5$ \& $\text{ATE}=0$ \& $\text{HTE}=0$]
  { 
   \includegraphics[width=0.225\textwidth]{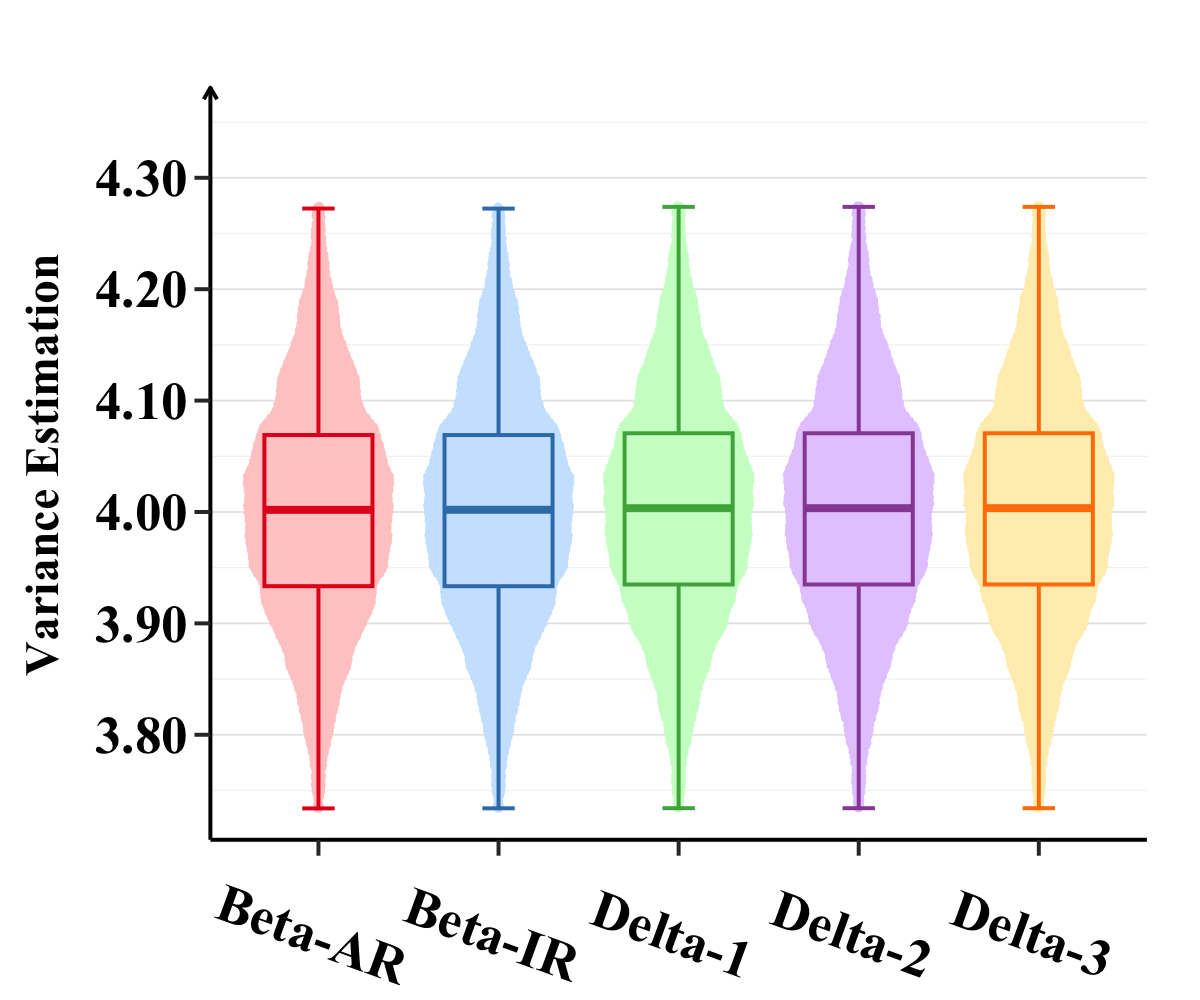}
  }
    \subfigure[Model-based \& $p=0.5$ \& $\text{ATE}=0$ \& $\text{HTE}=0.5$]
  { 
   \includegraphics[width=0.225\textwidth]{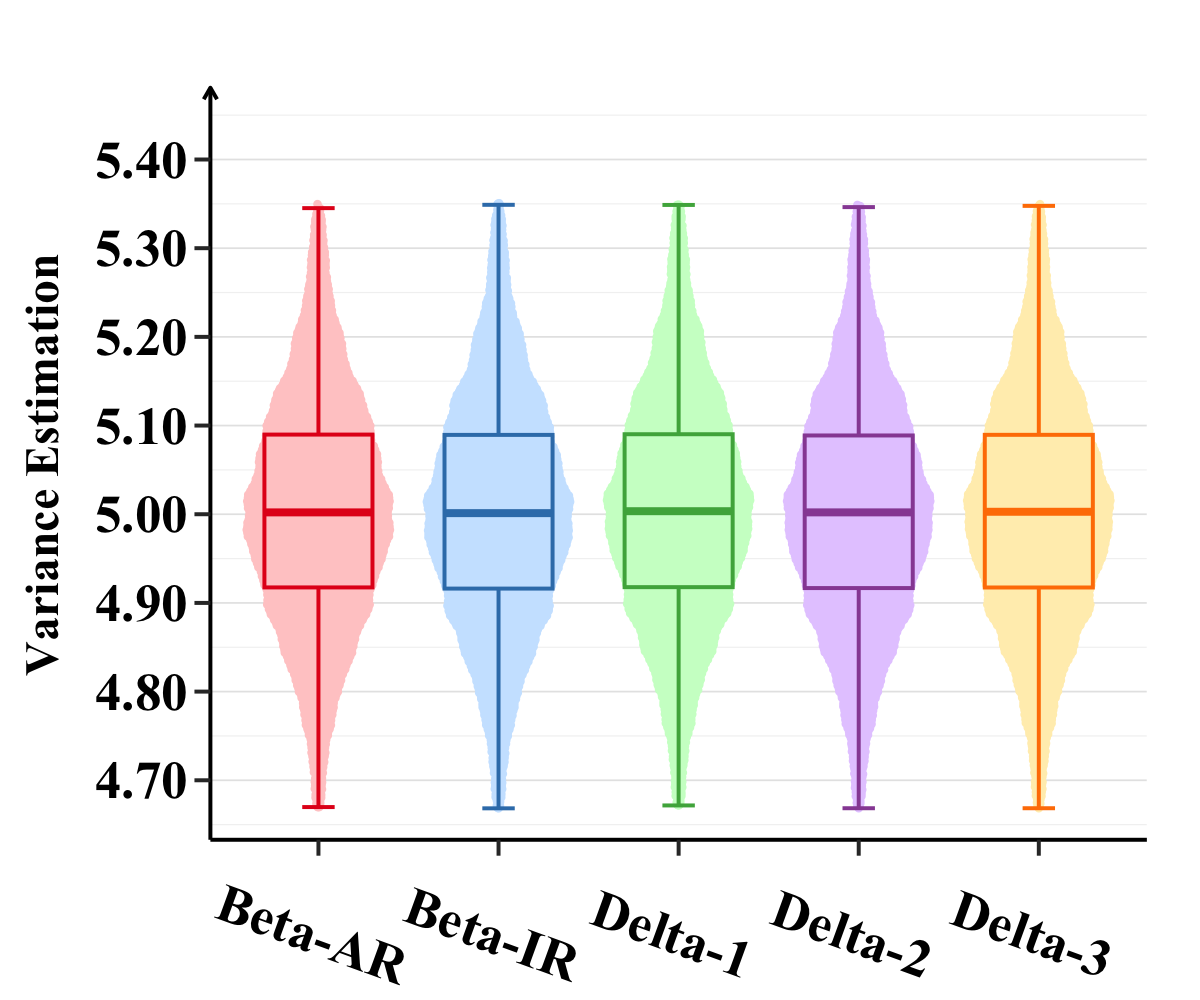}
  }
    \subfigure[Model-based \& $p=0.5$ \& $\text{ATE}=0.1$ \& $\text{HTE}=0$]
  { 
   \includegraphics[width=0.225\textwidth]{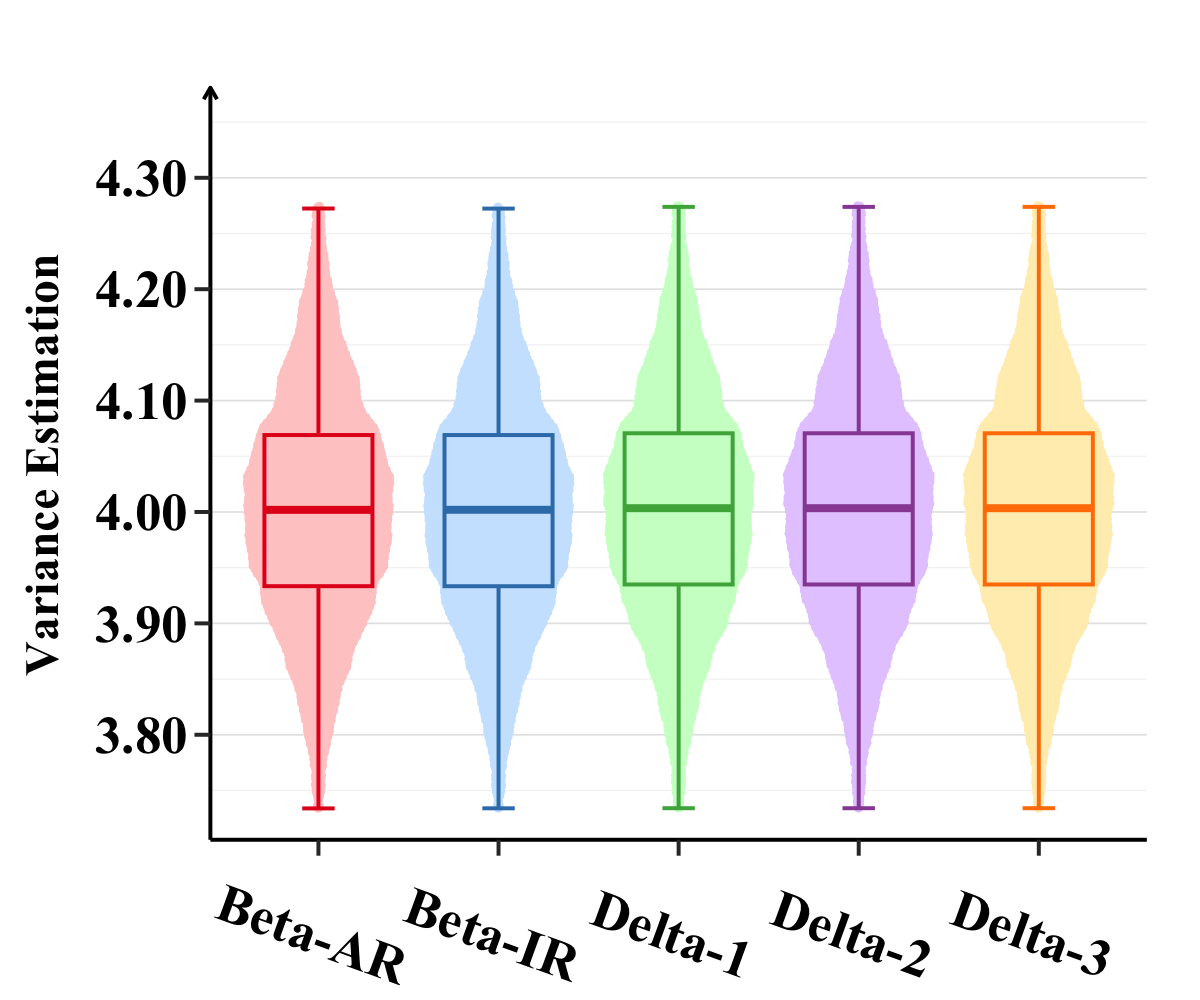}
  }
      \subfigure[Model-based \& $p=0.5$ \& $\text{ATE}=0.1$ \& $\text{HTE}=0.5$]
  { 
   \includegraphics[width=0.225\textwidth]{boxplot2122.png}
  }
  \subfigure[Model-based \& $p=0.4$ \& $\text{ATE}=0$ \& $\text{HTE}=0$]
  {  
   \includegraphics[width=0.225\textwidth]{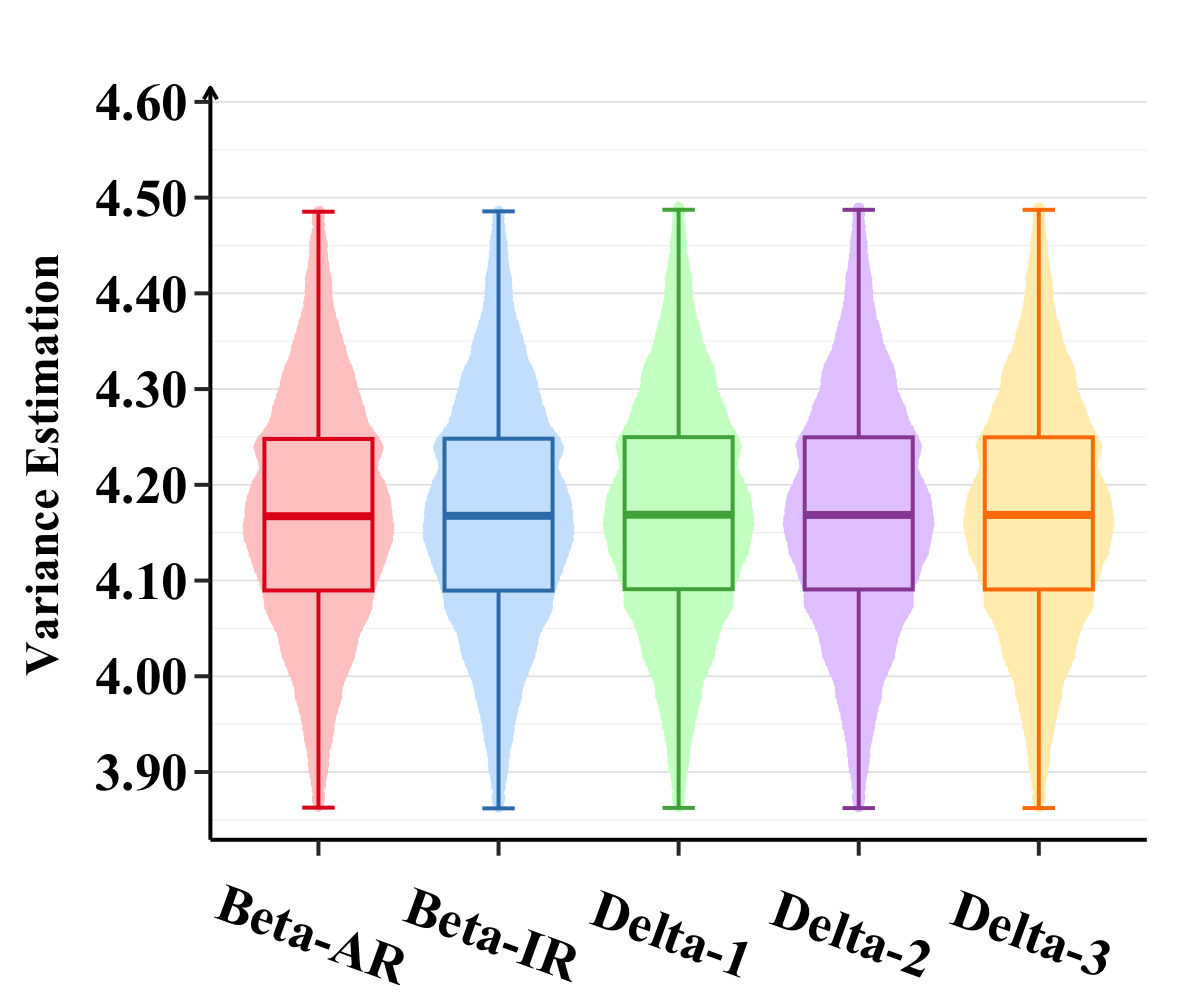}
  }
    \subfigure[Model-based \& $p=0.4$ \& $\text{ATE}=0$ \& $\text{HTE}=0.5$]
  {  
   \includegraphics[width=0.225\textwidth]{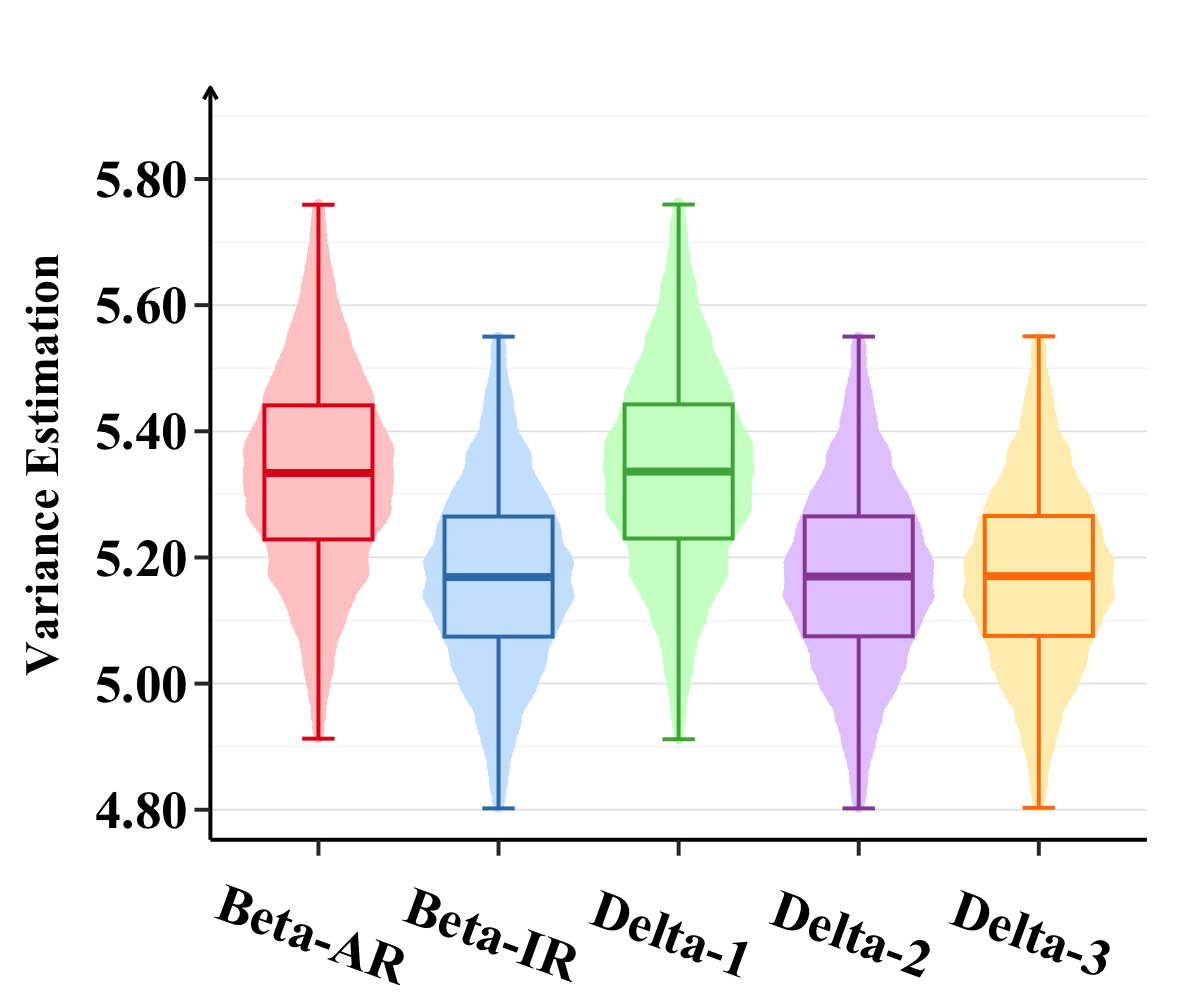}
  }
    \subfigure[Model-based \& $p=0.4$ \& $\text{ATE}=0.1$ \& $\text{HTE}=0$]
  { 
   \includegraphics[width=0.225\textwidth]{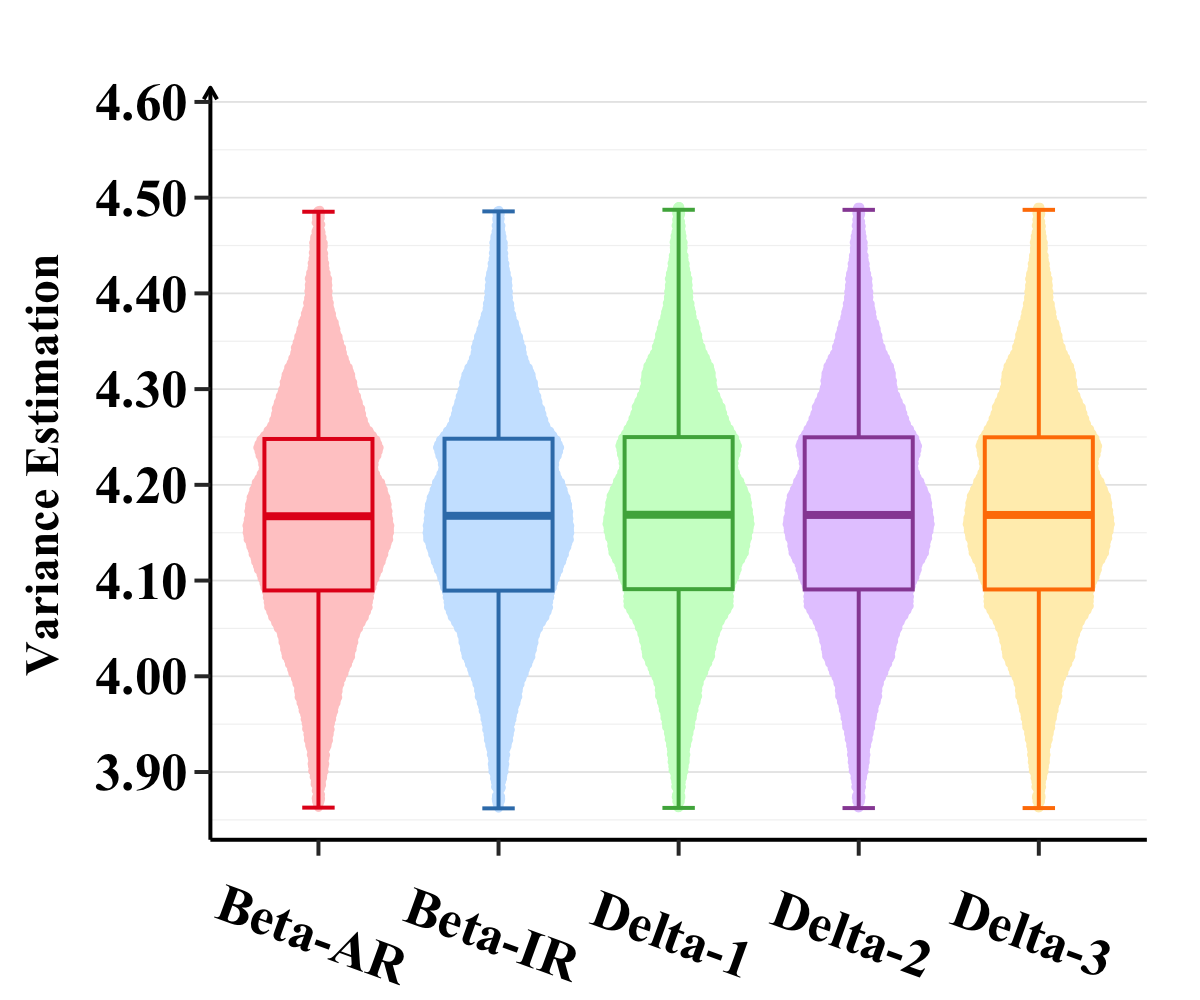}
  }
      \subfigure[Model-based \& $p=0.4$ \& $\text{ATE}=0.1$ \& $\text{HTE}=0.5$]
  { 
   \includegraphics[width=0.225\textwidth]{boxplot2222.png}
  }

  \caption{Box plots of the variance estimates of ATE estimators, scaled by \(n \times \widehat{\text{var}}  \). }
  \label{hatvar_all}
\end{figure}

We present box plots of variance estimates across all $16$ configurations in Figure \ref{hatvar_all}. The results reveal the following conclusions. Within the design-based Framework: 
\begin{itemize}

\item In the absence of HTE in the population, the variance estimates of all estimators are identical. 

\item When HTE is present in the population but group assignment probabilities are equal, the variance estimates of \(\wdd_3\) and \(\wb_T^{{IR}}\) are equivalent and lower than those of the remaining three estimators, whose variance estimates are identical to one another. 

\item When HTE is present in the population and group assignment probabilities are unequal, \(\wdd_2\) exhibits a slight advantage over \(\wdd_1\) and \(\wb_T^{{AR}}\), yet still lags behind \(\wdd_3\) and \(\wb_T^{{IR}}\). 

\end{itemize}

Within the model-based framework: Since \(\wdd_3\) and \(\wb_T^{{IR}}\) require variance correction, the variance estimates of all estimators are identical if either no HTE exists or group assignment probabilities are equal. Conversely, when neither condition holds, \(\wdd_2\), \(\wdd_3\), and \(\wb_T^{{IR}}\) outperform the other two estimators.

\end{document}